\documentclass[12pt]{article}

\usepackage{mathrsfs}
\usepackage[T1]{fontenc}
\usepackage{mathpazo}
\usepackage{setspace}
\usepackage{amsfonts}
\usepackage{amssymb}
\usepackage{epsfig}
\usepackage{latexsym}
\usepackage{color}
\usepackage{graphicx}
\usepackage{nicefrac}
\usepackage[latin1]{inputenc}
\usepackage{slashed}
\usepackage{multirow}

\usepackage{hyperref}

\def\hybrid{\topmargin -30pt    \oddsidemargin 0pt 
        \headheight 0pt \headsep 0pt
        \textwidth 6.25in       
        \textheight 9.5in       
        \marginparwidth .875in
        \parskip 5pt plus 1pt   \jot = 1.5ex}

\hybrid

\def\baselinestretch{1.2}

\catcode`\@=11

\def\marginnote#1{}
\newcount\hour
\newcount\minute
\newtoks\amorpm
\hour=\time\divide\hour by60
\minute=\time{\multiply\hour by60 \global\advance\minute by-\hour}
\edef\standardtime{{\ifnum\hour<12 \global\amorpm={am}%
        \else\global\amorpm={pm}\advance\hour by-12 \fi
        \ifnum\hour=0 \hour=12 \fi
        \number\hour:\ifnum\minute<10 0\fi\number\minute\the\amorpm}}
\edef\militarytime{\number\hour:\ifnum\minute<10 0\fi\number\minute}

\def\draftlabel#1{{\@bsphack\if@filesw {\let\thepage\relax
   \xdef\@gtempa{\write\@auxout{\string
      \newlabel{#1}{{\@currentlabel}{\thepage}}}}}\@gtempa
   \if@nobreak \ifvmode\nobreak\fi\fi\fi\@esphack}
        \gdef\@eqnlabel{#1}}
\def\@eqnlabel{}
\def\@vacuum{}
\def\draftmarginnote#1{\marginpar{\raggedright\scriptsize\tt#1}}

\def\draft{\oddsidemargin -.5truein
        \def\@oddfoot{\sl preliminary draft \hfil
        \rm\thepage\hfil\sl\today\quad\militarytime}
        \let\@evenfoot\@oddfoot \overfullrule 3pt
        \let\label=\draftlabel
        \let\marginnote=\draftmarginnote
   \def\@eqnnum{(\theequation)\rlap{\kern\marginparsep\tt\@eqnlabel}%
\global\let\@eqnlabel\@vacuum}  }

\def\draft2{
        \def\@oddfoot{\sl preliminary draft \hfil
        \rm\thepage\hfil\sl\today\quad\militarytime}
        \let\@evenfoot\@oddfoot \overfullrule 3pt
        \let\label=\draftlabel
        \let\marginnote=\draftmarginnote
   \def\@eqnnum{(\theequation)\rlap{\kern\marginparsep\tt\@eqnlabel}%
\global\let\@eqnlabel\@vacuum}  }

\def\preprint{\twocolumn\sloppy\flushbottom\parindent 2em
        \leftmargini 2em\leftmarginv .5em\leftmarginvi .5em
        \oddsidemargin -.5in    \evensidemargin -.5in
        \columnsep .4in \footheight 0pt
        \textwidth 10.in        \topmargin  -.4in
        \headheight 12pt \topskip .4in
        \textheight 6.9in \footskip 0pt
        \def\@oddhead{\thepage\hfil\addtocounter{page}{1}\thepage}
        \let\@evenhead\@oddhead \def\@oddfoot{} \def\@evenfoot{} }

\def\numberbysection{\@addtoreset{equation}{section}
        \def\theequation{\thesection.\arabic{equation}}}

\def\underline#1{\relax\ifmmode\@@underline#1\else
        $\@@underline{\hbox{#1}}$\relax\fi}

\def\titlepage{\@restonecolfalse\if@twocolumn\@restonecoltrue\onecolumn
     \else \newpage \fi \thispagestyle{empty}\c@page\z@
        \def\thefootnote{\fnsymbol{footnote}} }

\def\endtitlepage{\if@restonecol\twocolumn \else \newpage \fi
        \def\thefootnote{\arabic{footnote}}
        \setcounter{footnote}{0}}  

\catcode`@=12
\relax

\def\pl#1#2#3{ {\sl Phys. Lett.\/} {\bf#1}, #2(#3)}

\def\figcap{\section*{Figure Captions\markboth
        {FIGURECAPTIONS}{FIGURECAPTIONS}}\list
        {Figure \arabic{enumi}:\hfill}{\settowidth\labelwidth{Figure
999:}
        \leftmargin\labelwidth
        \advance\leftmargin\labelsep\usecounter{enumi}}}
 \relax
\def\tablecap{\section*{Table Captions\markboth
        {TABLECAPTIONS}{TABLECAPTIONS}}\list
        {Table \arabic{enumi}:\hfill}{\settowidth\labelwidth{Table
999:}
        \leftmargin\labelwidth
        \advance\leftmargin\labelsep\usecounter{enumi}}}
 \relax
\def\reflist{\section*{References\markboth
        {REFLIST}{REFLIST}}\list
        {[\arabic{enumi}]\hfill}{\settowidth\labelwidth{[999]}
        \leftmargin\labelwidth
        \advance\leftmargin\labelsep\usecounter{enumi}}}
 \relax
\makeatletter
\newcounter{pubctr}
\def\publist{\@ifnextchar[{\@publist}{\@@publist}}
\def\@publist[#1]{\list
        {[\arabic{pubctr}]\hfill}{\settowidth\labelwidth{[999]}
        \leftmargin\labelwidth
        \advance\leftmargin\labelsep
        \@nmbrlisttrue\def\@listctr{pubctr}
        \setcounter{pubctr}{#1}\addtocounter{pubctr}{-1}}}
\def\@@publist{\list
        {[\arabic{pubctr}]\hfill}{\settowidth\labelwidth{[999]}
        \leftmargin\labelwidth
        \advance\leftmargin\labelsep
        \@nmbrlisttrue\def\@listctr{pubctr}}}
 \relax
\makeatother

\def\be{\begin{equation}}
\def\ee{\end{equation}}
\def\ba{\begin{eqnarray}}
\def\ea{\end{eqnarray}}

\def\del{\partial}
\def\pl{\partial}

\def\k{\kappa}
\def\r{\rho}
\def\a{\alpha}

\def\G{\Gamma}
\def\d{\delta}
\def\D{\Delta}
\def\e{\epsilon}

\def\th{\theta}

\def\m{\mu}
\def\n{\nu}
\def\om{\omega}
\def\Om{\Omega}
\def\l{\lambda}
\def\L{\Lambda}
\def\s{\sigma}
\def\S{\Sigma}

\def\cL{{\cal L}}

\def\cN{{\cal N}}

\def\cV{{\cal V}}

\def\no{\noindent}

\def\qq{\qquad}

\def\IR{\relax{\rm I\kern-.18em R}}

\def\inv{^{\raise.0ex\hbox{${\scriptscriptstyle -}$}\kern-.05em 1}}

\def \ha {{\frac{1}{2}}}

\def \ov {\over}

\def\const{{\rm const.}}

\begin{document}

\renewcommand{\theequation}{\arabic{equation}}
\renewcommand{\theequation}{\thesection.\arabic{equation}}

\renewcommand{\theequation}{\thesection.\arabic{equation}}
\csname @addtoreset\endcsname{equation}{section}

\begin{titlepage}
\begin{center}

{}\hfill DMUS--MP--13/02

\phantom{xx}
\vskip 0.4in

{\large \bf Non-Abelian T-duality and the AdS/CFT correspondence:\\
new ${\cal N}=1$ backgrounds}

\vskip 0.45in

  {\bf Georgios Itsios}${}^{1,3}\,$\footnote{{\tt gitsios@upatras.gr} }, {\bf Carlos N\'u\~nez}${}^{2}\,$\footnote{{\tt c.nunez@swansea.ac.uk} },\\  {\bf Konstadinos Sfetsos}${}^{3,1}\,$\footnote{{\tt k.sfetsos@surrey.ac.uk}}\ and\ {\bf Daniel~C.~Thompson}${}^{4}$\footnote{{\tt dthompson@tena4.vub.ac.be}}

\vskip 0.2in

${}^1$Department of Engineering Sciences, University of Patras,\\
26110 Patras, Greece\\
 \vskip .1in
${}^2$Swansea University, School of Physical Sciences,\\
Singleton Park, Swansea, SA2 8PP,  UK\\
\vskip .1in
${}^3$Department of Mathematics, University of Surrey,\\
 Guildford GU2 7XH, UK\\
\vskip .1in
${}^4$  Theoretische Natuurkunde, Vrije Universiteit Brussel, and \\
International Solvay Institutes, \
Pleinlaan 2, B-1050 Brussels, Belgium. \\

 \vskip .2in
\end{center}

\vskip .4in

\centerline{\bf Abstract}

\no
We consider non-Abelian T-duality on ${\cal N}=1 $ supergravity backgrounds possessing well understood field theory duals.
For the case of D3-branes at the tip of the conifold, we dualise along an $SU(2)$ isometry. The result is a type-IIA geometry whose lift to M-theory is of the type recently proposed by Bah et. al. as the dual to certain ${\cal N}=1$ SCFT  quivers produced by M5-branes wrapping a Riemann surface.  In the non-conformal cases we find smooth duals in massive IIA supergravity with a Romans mass naturally quantised.  We initiate the interpretation of these geometries in the context of AdS/CFT correspondence. We show that the central charge and the entanglement entropy are left invariant by this dualisation.
The backgrounds suggest a form of Seiberg duality in the dual field theories which  also exhibit  domain walls and confinement in the infrared.

\end{titlepage}
\vfill
\eject

\def\baselinestretch{1.2}
\baselineskip 10 pt
\noindent

\tableofcontents

\def\baselinestretch{1.2}
\baselineskip 20 pt
\no

\newcommand{\eqn}[1]{(\ref{#1})}

 \def\beq{\be}
 \def\eeq{\ee}
 \def\bea{\ba}
 \def\eea{\ea}
 \section{Introduction}

Within the context of gauge/string duality, solution generating
techniques in supergravity are an extremely powerful tool.
Prominent examples include the use of
bosonic and fermionic T-dualities to show dual superconformal
symmetry at strong coupling  \cite{Alday:2007hr, Berkovits:2008ic};
the T-s-T transformations that are the string analogue
of $\beta$-deformations in gauge theory \cite{Lunin:2005jy}
and which can also be used to construct gravity duals for some
non-relativistic field theories
\cite{Maldacena:2008wh,Herzog:2008wg,Adams:2008wt}; the use of
$G$ structure rotations to obtain solutions with/without back-reacted source
branes in conifold related geometries
\cite{Gaillard:2010qg,Conde:2011aa,Conde:2011ab}.
Evidently some of these techniques, namely U-dualities,
are understood to be symmetries of the underlying string theory.
Fermionic T-duality, however, provides an example where the symmetry
is only valid at tree-level in string perturbation theory but
nonetheless has applications in AdS/CFT when considering just the planar limit.

Performing a T-duality with respect to a non-Abelian isometry
group is also a solution generating technique of supergravity.
Rather like the case of fermionic T-duality it is
not expected to be a full symmetry of string perturbation theory. But
it is nonetheless natural to ask what r\^ole it might have
within the gauge/string correspondence.
This study was initiated in \cite{Sfetsos:2010uq} in which
the dualisation of $AdS_5 \times S^5$ with respect to
an $SU(2)$ isometry group was carried out.
The result was somewhat suprising; the dual was found to be a
solution of type-IIA supergravity whose lift to M-theory
captures some universal properties of the solutions found by
Gaiotto and Maldacena in \cite{Gaiotto:2009gz},
as dual geometries to the generalised ${\cal N} =2$
quiver SCFT's proposed by Gaiotto in \cite{Gaiotto:2009we}.
Further progress and works in studying non-Abelian T-duality in this context
can be found
\cite{Polychronakos:2010hd,Lozano:2011kb,Itsios:2012dc,Lozano:2012au,Itsios:2012zv}
and a brief review of elementary aspects of non-Abelian T-duality in \cite{Sfetsos:2011jw}.

Motivated by this, in this paper we shall investigate non-Abelian T-duality
applied to solutions with minimal supersymmetry whose field theory dual is well understood.

The Klebanov--Witten (KW) solution \cite{Klebanov:1998hh} provides the first such example; this solution represents the $SU(N) \times SU(N)$ conformal field theory on D3-branes located at the tip of the conifold.   We are also interested in gauge theories which are not conformal but rather, have non-trivial RG flows.  The prototypical example is the Klebanov--Tseytlin (KT) solution \cite{Klebanov:2000nc} which incorporates fractional branes  (D5-brans
wrapped on the shrinking two-cycle of the conifold) and is a good model for the UV dynamics of a $SU(N)\times SU(N+M)$ theory.  As one flows towards the IR the theory under goes a sequence of Seiberg dualities to ever decreasing gauge group ranks.  In the IR the solution of KT is singular, a fact which is remedied
when $M$ is a multiple of $N$  ( $N=k M$), as it occurs in the Klebanov--Strassler (KS) solution \cite{Klebanov:2000hb},
wherein strong coupling effects take hold and remove
the singularity by replacing the conifold with its deformation.
In the IR the theory exhibits R-symmetry breaking
(or rather $\mathbb{Z}_{2n} \rightarrow \mathbb{Z}_2$),
confinement, domain walls and other interesting phenomena.
One level up in complexity  is the construction of the gravity dual
for the case in which the KS-field theory  is
exploring its
baryonic branch; in this case there exists a
one-parameter family of regular deformations
\cite{Butti:2004pk} interpolating between the
KS solution and the  wrapped D5-brane solutions in
\cite{Maldacena:2009mw,Gaillard:2010qg,Caceres:2011zn,Elander:2011mh}.

All of these examples possess rich isometry groups containing at least an $SU(2)$ factor along which we will dualise.  There is also a
$U(1)$  isometry of the metric (at least in the KW and KT solutions)  that may be understood  in the dual field theory as the R-symmetry.
The Killing spinors of the background are invariant under the $SU(2)$ action,
in a sense which we shall explain. This corresponds to the fact the super symmetries are uncharged under the global  (flavour-like) symmetries in the field theory and because of this performing the dualisation preserves supersymmetry.  This should be contrasted with performing an Abelian T-duality in the internal space which would either destroy supersymmetry or result in a singular background.  Since the $SU(2)$ isometry group has three generators one will arrive at solutions in (massive) IIA.
Let us now summarise what happens in each case in ascending order of complexity.

We find that in the case of KW the dual geometry can be lifted to M-theory and can be directly matched to some solutions recently proposed by
  \cite{Bah:2012dg,Bah:2011vv}, generalising the eleven-dimensional solutions of  \cite{Maldacena:2000mw},  as dual to the ${\cal N}=1$ SFCTs obtained
  from M5-branes wrapped on a Riemann surface.  Included in this class of SCFT are the so-called Sicilian quivers of  \cite{Benini:2009mz}.
  This is a direct ${\cal N}=1$ analogue of the dualisation of $AdS_5\times S^5$  to Gaiotto--Maldacena-like geometries that was performed in \cite{Sfetsos:2010uq}. Indeed,  one can obtain the KW theory by considering the ${\cal N}=2$ gauge theory dual to the
  orbifold $AdS_5\times S^5/\mathbb{Z}_2$ adding a relevant deformation and flowing to the IR.   We essentially find a T-dual complement of this relation.

For the dualisation of the KT solution one finds that the resultant geometry is a solution of massive IIA supergravity and the Romans mass is naturally quantised by the number of fractional branes.  The reason for this can be understood intuitively by the fact that there is a component of the RR three-form with legs along all the $SU(2)$ directions.
Upon dualisation, this then gets converted to a zero-form.  Since this is a solution of massive IIA it has no lift to M-theory; the fractional branes of the type-IIB solution represent some obstruction to this.

To get a better handle on this novel background we perform a number of checks.  The first is to look at the central charge
before and after the dualisation,  following the method of \cite{Freedman:1999gp} and \cite{Klebanov:2007ws}.  We find that, up to a subtlety that depends on the global properties of the geometry,  the central charge before and after the non-Abelian duality, matches.   As we will explain,
this can be understood by the fact that
the measure, $\sqrt{g}e^{-2\phi}$, is
an invariant of the duality (just as it is for Abelian T-duality).
The same invariance is present for the entanglement entropy.
By using probe branes one can define a gauge coupling.  A strange feature is that this suitably defined
gauge coupling does not behave like those of a renormalisable 4d QFT
where $\displaystyle g^{-2} \sim \ln r$ (as in the KT case),
 instead going like $\displaystyle g^{-2} \sim (\ln r)^{3/2}$
which hints at a rather unusual dual field theory (either that,
or the coupling so defined
does not represent the usual gauge interactions). Finally one can
consider the Page and Maxwell charges after duality.
Essentially what was D3-brane charge becomes D6-brane charge.
The Maxwell charge of D6-branes changes logarithmically. As we will
discuss, this is one among other similarities with the KS-cascade. We will
discuss below, a form of Seiberg duality that appears after the duality.

To probe the low energy physics,
one needs to look at the dual of the KS geometry.  In this case things are rather more involved, but  we verify that the IR signatures of confinement and domain walls are preserved after the dualisation.  The same pattern shows-up if we start with the solution describing $D5$-branes wrapping SUSY two-cycles \cite{Maldacena:2000yy} and dualise it.

Interestingly, for the Type IIB solution describing the baryonic branch of the KS-field theory, something qualitatively different happens. After the dualisation, we find that the large radius asymptotics of the metric is no longer (logarithmically) approaching $AdS_5$. We provide two suggestions as to the field theoretic interpretation of this; either this is due to presence of an irrelevant operator in the dual QFT or more conservatively that this theory ceases to have a baryonic branch.

Let us present now a "road map" that summarises the points above and lay-out the general idea behind this long and technical
paper.

\subsection{General Idea and Road Map}
\begin{figure}[h!]
  \centering
 \includegraphics[trim= 1.9cm 17cm 2cm 2.5cm, clip, scale=0.8]{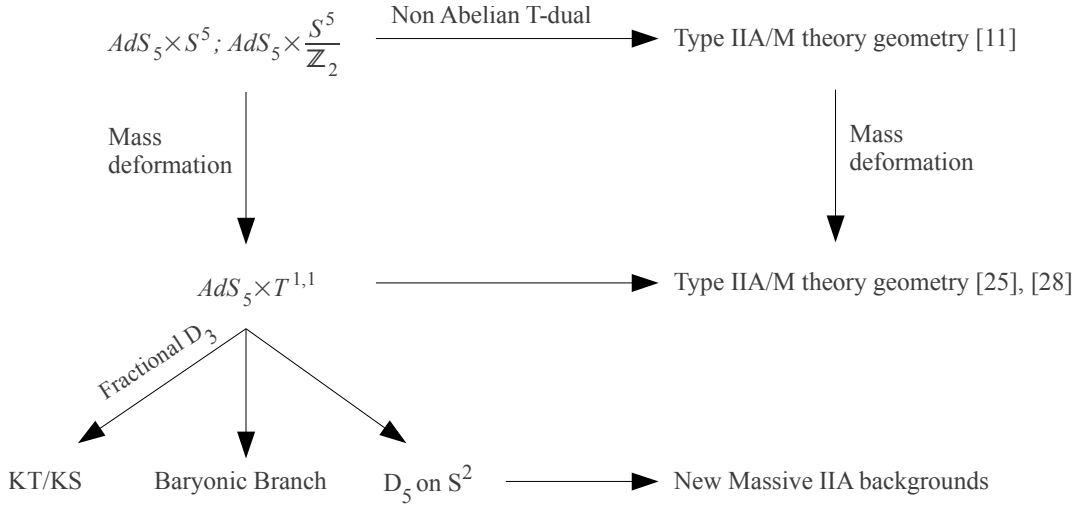}
 \caption{A road map for the paper. }
\end{figure}
We start with $AdS_5 \times S^5$ or better yet, with $AdS_5\times S^5/Z_2$,
with $N$ units of flux of the five form.
The field theories associated are ${\cal N}=4$ SYM or the ${\cal N}=2$ version
of the two groups quiver
$SU(N)\times SU(N)$ and adjoint matter. Following the paper \cite{Sfetsos:2010uq}
we can perform a non-Abelian T duality on the geometry to obtain a Type IIA/M-theory
geometry of the form proposed by Maldacena and Gaiotto \cite{Gaiotto:2009gz}
with the following characteristics (see \cite{Sfetsos:2010uq}  for details):
\begin{itemize}
\item{It contains a factor of $S^2$ instead of a hyperbolic plane $H^2$}
\item{The resulting geometry is singular. The dilaton field  diverges
at a given angular position. }
\item{In correspondence with the previous point, the 'charge distribution' in the language of
\cite{Gaiotto:2009gz} is $\lambda(\eta)=\eta$, which implies a quiver of the form
\beq
SU(2)\times SU(3)\times SU(4)\times....\times SU(N)\times SU(N+1)\times....
\eeq
}
\end{itemize}
A natural first step taken in this paper, is to apply the non-Abelian T-duality to examples preserving
${\cal N}=1$ SUSY.
We choose the Klebanov-Witten geometry \cite{Klebanov:1998hh} whose dual field theory is
the mass deformation of the $N=2$ field theory described above.
The non-Abelian duality is performed in Section \ref{section3www}. Some
interesting things are: that the solution is
{\it non-singular}, preserves ${\cal N}=1$ and falls within the class
of geometries proposed in \cite{Benini:2009mz} (originally found in \cite{Gauntlett:2004zh}).
These geometries have been proposed to be dual to the mass deformation
of the Maldacena-Gaiotto theories. Our geometry presented in Section \ref{section3www}
falls within this
class, for the case in which we have an $S^2$ factor instead of an $H_2$. The dual
field theory seems to be less understood in that case.

The following  step is  to continue with the known deformations of the
Klebanov-Witten theory/geometry. We then study the case of the
Klebanov-Tseytlin geometry, Klebanov-Strassler geometry,
Baryonic Branch and the background of D5 branes wrapping a
two cycle inside the resolved conifold.
We obtain in this case new backgrounds in {\it Massive IIA}
with a quantized mass parameter, proportional to the number of five branes $N_c$, the
'deformation' from the conformal point. We present arguments for the
non-singular behavior of these new solutions (the transformed of the KT-solution is
obviously singular as the seed solution is) and 'define' their field theory dual
calculating observables with the background.

In more detail, the structure of the paper is the following:
In Section \ref{section2www}  we develop the technology required to implement these non-Abelian duality transformations.
 In Section \ref{section3www} we apply this to the KW background.  In Sections \ref{secciondekt},
\ref{sectiondelosbackgroundscompletos} and \ref{seccion7deanalysis},
we turn our attention to the dualisations and field theory analysis of the non-conformal backgrounds described above.
 We conclude in Section \ref{section7www} , presenting some open questions
 and future topics for research.
 We provide generous appendices describing our conventions and generic Buscher-like rules for dualisation.

 %
 %
 %
 %

 %
 %
 %
 %

 \section{Non-Abelian T-Duality Technology }\label{section2www}

In this section we give details of the dualisation procedure used.
The hurried reader who simply wants to get the physical results should feel free to read the
following "Non-Abelian Duality 101" and skip past the rest of the section returning when he wishes to know more of the technicalities.

 \subsection{Non-Abelian T-Duality 101 }

 T-duality  states equivalence between string theories propagating on two different target spacetimes containing some abelian isometries. In its simplest form, it is the equivalence between strings on circle of radius R with those on a circle radius 1/R.   More generally T-duality provides a map, known as the Buscher rules, between one solution of supergravity and a second solution.   A powerful approach to deriving these rules is the path integral approach (or Buscher procedure)  \cite{Buscher:1987qj} .  This procedure is a three step recipe:  one begins with the string sigma model for the first spacetime and gauges a $U(1)$ isometry of this spacetime; second, one invokes a flat connection for this gauge field by means of a Lagrange multiplier; finally, one integrates by parts to yield an action with a non-propagating gauge field that can be eliminate by its equations of motion to produce the T-dual sigma model.

The Buscher procedure can be naturally generalised to the case of a target space equipped with a non-Abelian isometry group $G$. One follows exactly the same steps but in this case the gauge fields are valued in the algebra of $G$.  Doing so produces a map between one solution of supergravity and another. It is in this spirit of solution generating that we employ   non-Abelian T-duality in this paper.

Despite the fact the dualisation procedure is rather similar between the abelian and non-abelian cases there are some important differences.  Generically the isometry (and potentially supersymmetry) enjoyed by the starting geometry is, at least partially, destroyed.  However this lost isometry may be recovered as a non-local symmetry in the sigma model and the corresponding sigma models  are canonically equivalent.   A second point is the rather subtle effect of global issues that arise when performing the Buscher procedure on world sheets of arbitrary genera.  These global concerns mean one should not view non-abelian duality as a full symmetry of string (genus) perturbation theory but just a tree-level symmetry.  Nonetheless, if one's focus is on supergravity (as it will be in this paper) or the planar limit then one may  still harness its power as a solution generating symmetry.

 Early work on non-Abelian duality can be found in \cite{delaossa:1992vc,Giveon:1993ai,Curtright:1994be,Sfetsos:1994vz, Alvarez:1994zr,Lozano:1995jx,Sfetsos:1996pm} in context of purely Neveu-Schwarz backgrounds.   This subject has had something of a revival  following the work of  \cite{Sfetsos:2010uq}, in which this procedure was extended to geometries contain RR fluxes.  A particularly curious result came from performing a non-Abelian dualisation of an $SU(2)$ isometry group that acts within the sphere of $AdS_5 \times S^5$.  After dualisation the resultant geometry was a solution of IIA whose lift to M-theory bore a very close resemblance to the Giaotto-Maldacena geometries that come from considering M5 branes wrapped on Riemann surfaces.

We close this section by giving an example to get the reader in the spirit.  Consider the round metric on the $S^3$ which possess $SO(4) = SU(2)_L\times SU(2)_R$ isometry and may be written (in Euler angles) as
\be
ds^2 = d\th^2 + d\phi^2 + 2\cos \theta d\phi d\psi   + d\psi^2 \ .
\ee
After performing the dualisation with respect to say the $SU(2)_L$ isometry one finds a geometry that interpolates between $\mathbb{R} \times S^2$ and $\mathbb{R}^3$ with metric given by
\be
\widehat{ds}^2 = dr^2 + \frac{r^2}{1+ r^2} (d\th^2 + \sin^2 \th d\phi^2)\ .
\ee
In addition, in this example the dual geometry is supported by a NS two-form and dilaton given by
\be
\widehat{B} = \frac{r^3}{1+r^2} vol(S^2) \ , \qquad  \widehat{\Phi} = -\frac{1}{2} \ln (1+ r^2) \ .
\ee
This example serves to illustrates a two key features that we will encounter.  Firstly the $SO(4) = SU(2)_L\times SU(2)_R$ isometry gets reduced to just $SU(2)$ that is reflected by the presence of the $S^2$ in the dual.  Secondly there has been a serious topology change, indeed the dual geometry contains a non-compact direction.   Whilst this example does not represent, evidently, a full solution of supergravity on its own, it may be embedded into true supergravity solutions and indeed it is prototypical of the dualisations that we will perform.

An important ingredient in this paper will be the incorporation of  RR fields.  Lets us illustrate how this works by supposing that in the example above the initial geometry is supported by a RR three-form
\be
F_3 = vol(S^3) \ .
\ee
To extract the dual fluxes one may use the following formula,
\be
e^{\widehat{\Phi}} \widehat{\slashed{F}} =  \slashed{F}. \Omega^{-1} \ ,
\ee
where the slashes indicate the RR poly form (sum of RR forms)  contracted with gamma matrices to form a bispinor, i.e. $\slashed{F}= \Gamma^{123}$, and  $\Omega$  is a  matrix, the construction of which we describe in detail in the following section,  given by in this case
\be
\Omega^{-1}= \frac{1}{\sqrt{1+ r^2}} ( -\Gamma^{123} + r \Gamma^r) \ .
\ee
From this one ascertains that the dual geometry will contain a zero from and  two form:
\be
F_0 = 1 \ , \qquad F_{2} = \frac{r^3}{1+ r^2} vol(S^2) \ .
\ee
Notice that this would, when embedded into a true Type II supergravity background, lead to a solution in {\emph{ massive}} type IIA (the $F_{0}$ is the Romans mass and comes when the $\Gamma^{123}$ in $\Omega$ annihilate the same factor in $\slashed{F}_3$).    We shall see the same phenomenon happen in a number of the examples in this paper.  The fact the type of the supergravity changed from IIB to IIA is due to the fact the isometry group dualised had an odd dimension (if it were to be even dimensional the type would have remained the same).

We now present details of how to technically compute the dualisation rules for the non-abelian duality.

  \subsection{Non-Abelian T-Duality; some nuts and bolts }

We wish to consider backgrounds that support an $SU(2)$ isometry such that the metric can be cast as
\be
ds^2 = G_{\m\n}(x) dx^\m dx^\n + 2 G_{\m i}(x) dx^\m L^i + g_{ij}(x) L^i L^j\ ,
\ee
where $\m=1,2,\dots , 7$ and $L^i$ are the Maurer--Cartan forms.
Our group theory conventions can be found in Appendix A.
The NS sector comprises also the 2-form
\be
B = B_{\m\n}(x) dx^\m \wedge dx^\n + B_{\m i}(x) dx^\m \wedge L^i + \ha b_{ij}(x)  L^i \wedge L^j\
\label{bmnb}
\ee
and a dilaton
\be
\Phi = \Phi(x)\ .
\ee
Hence, all coordinate dependence on the  $SU(2)$ Euler angles $\th, \phi , \psi$ is
contained in the Maurer--Cartan
one-forms whilst the remaining data can all be dependent on the spectator fields $x^\mu$.
Notice that we could have taken $b_{ij}=0$,
\footnote{To see that note that in \eqn{bmnb} the relevant term becomes
\ba
& & B_{\m i} dX^\m \wedge L^i + \e_{ijk} b_k L^i \wedge L^j =B_{\m i} dX^\m \wedge L^i +  \sqrt{2} b_i dL^i
\nonumber\\
& &\ =  (B_{\m i} - \sqrt{2} \del_\m b_i) dX^\m \wedge L^i + \sqrt{2}\ d(b_i L^i)\ ,
\ea
where in the second line we have performed a partial integration.
Hence, the last term has no contribution to the field strength $dB$ and we may as well denote the
first term by $B_{\m i} dX^\m \wedge L^i$.}
however, we will not do that since in the specific examples we
will encounter it is necessary for
a clear presentation of the various results.

\no
In what follows it will be convenient to use a parametrisation of the frame fields given by
\be
e^A = e^A_\m dx^\m \ ,\qq e^a =  \kappa^a{}_j L^j + \l^a_\m dx^\m \ ,
\label{fraame}
\ee
where $A=1,2,\dots , 7$ and $a=1,2,3$.
By demanding that
\be
ds^2 = \eta_{AB}e^A e^B + e^a e^a \ ,
\ee
we obtain that
\be
\l^a_\m \l^a_\n =  K_{\m\n} \ ,\qq \eta_{AB}e^A_\m e^B_\n = G_{\m\n} - K_{\m\n}\ ,
\qq \kappa^a{}_i\kappa^a{}_j = g_{ij}\ ,\qq \kappa^a{}_i \l^a_\m = G_{\m i}\ ,
\ee
where $\eta_{AB}$ is the seven-dimensional Minkowski metric. Note that $\k^a{}_i$ and
$\l^a{}_\m$ depend in general on the $x^\m$'s.

\subsubsection{The non-Abelian T-dual of the NS-sector}

The corresponding Lagrangian density for the NS sector metric and antisymmetric fields is given by
\be
\cL_0 =Q_{\m\n} \del_+ X^\m \del_-X^\n
+  Q_{\m i} \del_+ X^\m L_-^i + Q_{i\m} L_+^i  \del_- X^\m + E_{ij} L_+^i L_-^j \ ,
\ee
where, in accordance with \eqn{litr}, $L^{i}_{\pm}=-i {\rm Tr}( t^i g^{-1}\partial_{\pm} g)$ and we have also defined
\be
Q_{\m\n} = G_{\m\n} + B_{\m \n} \ ,\qq
Q_{\m i} = G_{\m i } + B_{\m i } \ , \qq  Q_{ i\m} = G_{ i\m } + B_{ i\m } \ ,
\qq E_{ij} = g_{ij} +  b_{ij}\ .
\ee
To perform the non-Abelian T-duality we replace derivatives with covariant derivatives according to
\be
\del_\pm  g \to D_\pm g = \del_\pm g - A_\pm g \ ,
\ee
and add the Lagrange multiplier term
\be
-i {\rm Tr}(v F_\pm ) \  ,\qq F_{\pm} = \del_+ A_- - \del_- A_+ -[A_+,A_-] \ .
\ee
Then the total action is invariant under
\be
g\to h^{-1} g\ ,\quad v\to h^{-1} v h \ ,\quad
\quad A_\pm\to h^{-1}A_\pm h -h^{-1}\del_\pm h \ ,
\label{tranmult}
\ee
for a group element $h(\s^+,\s^-)\in SU(2)$. Under this transformation the fields $x^\m$ stay inert and
thus are called spectators.
After some partial integrations the Lagrange multiplier term takes the form
\be
{\rm Tr}(i \del_+ v A_- -i \del_-v A_+ - A_+ f A_-)\ ,\qq f_{ij} = f_{ij}{}^k v_k\ .
\ee

\no
We now can integrate out the gauge fields to produce a dual theory that still depends on $\th, \phi, \psi, v_i$ and the spectators.
One must now gauge fix the $SU(2)$ isometry to remove three of these variables.  The obvious way to proceed is to set
\be
g=\mathbb{I}\ ,
\label{gafig}
\ee
i.e.   $\th= \phi= \psi= 0$ in the notation of appendix A, 
which leaves an action in terms of just the $v_i$ and the spectators.
There are other gauge fixing choices that may be more 
revealing by, for instance, making manifest some residual isometries.
Different gauge fixing choices may be related, 
at least locally, through coordinate transformations as 
will demonstrate below in section 2.2.3.  For the time being 
we proceed with the gauge fixing choice $g=\mathbb{I}$.
Integrating out the gauge fields gives
\be
\label{Apmeq}
A_+^i =  i M^{-1}_{ji} (\del_+ v_j + Q_{\m j}\del_+ X^\m ) \ ,
\quad A_-^i = -i M^{-1}_{ij} (\del_- v_j - Q_{j\m} \del_- X^\m)\ ,
\ee
where we have defined the matrix
\be
M = E + f \ .
\label{mef}
\ee
Substituting back into the action gives the dual Lagrangian
\be
\widehat \cL = Q_{\m\n}\del_+ X^\m \del_- X^\n + (\del_+ v_i + \del_+ X^\m Q_{\m i})M^{-1}_{ij}
(\del_- v_j - Q_{j\m} \del_- X^\m )\ .
\label{tdulal}
\ee
From this we read off the background fields of the NS-sector for the T-dual theory as
\ba
&& \widehat Q_{\m\n} = Q_{\m\n} - Q_{\m i} M^{-1}_{ij} Q_{j\n}\ ,\qq \widehat E_{ij}  =M^{-1}_{ij}\ ,
\nonumber\\
&&
\widehat Q_{\m i} = Q_{\m j} M^{-1}_{ji} \ ,\qq \widehat Q_{i \m} = -M^{-1}_{ij} Q_{j \m} \ .
\ea
Additionally one finds that the dilaton receives a contribution at the quantum level  just as in Abelian duality
\be
\widehat \Phi(x,v)= \Phi(x) -\ha \ln (\det M)\ .
\label{dildiall}
\ee

It is clear from the above that the inverse of the matrix $M$ determines the dual geometry.
Since we are working with $SU(2)$ isometries it is simple enough to evaluate this explicitly.
In three-dimensions an antisymmetric matrix is dual to a vector, hence we may write
\be
b_{ij}= \e_{ijk} b_k \ .
\ee
Rescaling also $v_i\to v_i/\sqrt{2}$ we have to invert the matrix with elements
\be
M_{ij} = g_{ij} + \e_{ijk} y_k\ ,\qq y_i = b_i + v_i \ .
\ee
To compute the inverse define an antisymmetric density and a vector as
\be
\bar e_{ijk} = \sqrt{\det g}\ \e_{ijk}\ ,\qq z^i ={ y^i\ov \sqrt{\det g}}  ={ y^i\ov \det \k} \ .
\ee
In this way we may use the matrix $g_{ij}$ to lower and raise indices in $z^i$ since now
the index has been transformed into a curved index.
Then
\be
M_{ij} = g_{ij} + \bar \e_{ijk}z^k \ .
\ee
Then the inverse is found to be
\be
(M^{-1})^{ij} = {1\ov 1+z^2}\left( g^{ij} + z^i z^j -\bar \e^{ij}{}_k z^k\right)\ ,
\qq z^2 = z^i z^j g_{ij}= z^iz_i \ .
\ee
Returning to our original variables
\be
(M^{-1})^{ij} = {1\ov \det g + y^2} \left(\det g\ g^{ij} + y^i y^j - \e_{ijk} \bar y_k\right) \ ,
\ee
where $\bar y_i = g_{ij} y^j$ and $y^2 = y^i y^j g_{ij} = \bar y_i \bar y_j g^{ij}$.

\subsubsection{Computing the Lorentz transformation}

By making use of \eqn{Apmeq} one can establish that the worldsheet derivatives transform under the non-Abelian T-duality as
\ba
&& L^i_+ = -(M^{-1})_{ji} \left(\del_+ v_j + Q_{\m j}\del_+ X^\m \right)\ ,
\nonumber\\
&&
L_-^i = M^{-1}_{ij} (\del_- v_j - Q_{j\m} \del_- X^\m)\ ,
\label{dgkj1}
\\
&& \del_\pm X^\m ={\rm invariant}\ .
\nonumber
\ea
These relations, in fact, provide a canonical transformation in phase space between pairs of T-dual sigma models \cite{Curtright:1994be,Lozano:1995jx}.

Crucial to us will be that by virtue of \eqn{dgkj1},  left and right movers have different transformation rules and will define two different sets of frame fields.  However, since these frame fields will describe the same geometry they must be related by a Lorentz transformation.
Explicitly we find that the frames in \eqn{fraame}
transforms, using the "plus" and the "minus" transformations \eqn{dgkj1}, to the frames
\be
\label{Michu}
e\to \hat{e}_+ = -\kappa M^{-T} (dv + Q^T dX) + \l dX \ ,\qq
e\to \hat{e}_- = \kappa M^{-1} (dv - Q dX) + \l dX \ ,
\ee
where $(Q)_{i\m}=Q_{i\m}$.
Writing
\be
\hat{e}_+ = \L \hat{e}_-\ ,
\ee
where $\L$ is the Lorentz transformation matrix to be computed.
We find from equating the terms proportional to $dv$ in \eqn{Michu}
that
\be
\L = -\kappa M^{-T} M \kappa^{-1} = -\kappa^{-T} M M^{-T} \kappa^T\ .
\ee
The terms proportional to $dX$ equate identically with no extra condition.
To explicitly compute $\L$ note first that
\be
(\k^{-T} M \k^{-1})_{ab} = \d_{ab} + \e_{abc} \zeta^c\ ,
\ee
where
\be
\zeta^a = \k^a{}_i z^i \ ,
\label{kjsg1}
\ee
is the flat index coordinate. Then
\be
(\k M^{-1} \k^T)^{ab} ={1\ov 1+\zeta^2 }\left(\d^{ab} + \zeta^a \zeta^b -\e_{abc} \zeta^c\right)\ .
\ee
Then we compute that
\be
\L^{a b} ={\zeta^2-1\ov \zeta^2+1}\ \d_{ab}
- {2\ov  \zeta^2+1} (\zeta^a  \zeta^b +\e_{abc}\zeta^c)\ ,
\label{kjsg2}
\ee
where $\zeta^2 = \zeta_a \zeta^a$. Note that
this has exactly the same form as in \cite{Sfetsos:2010uq} for the case of the PCM.
Moreover, it is an $O(3)$ rotation as it has $\det \L=-1$.
The effect of the non-trivial extra couplings $g_{ij}$ and $b_{ij}$ is to "dress up"
the original Lagrange multipliers and is hidden into the the definition of $\zeta^a$.

\no
This Lorentz transformation also induces an action
on spinors given by a matrix $\Omega$ obtained by requiring that
\be
\Omega^{-1}  \Gamma^a  \Omega =  \Lambda^a{}_b \Gamma^b\ .
\label{spino1}
\ee
One finds that\footnote{The general expression for $\Om$ for a freely acting group $G$
can be found in \cite{Lozano:2011kb}.
}
\be
\Om = \G_{11} {-\G_{123} + \zeta_a \G^a\ov \sqrt{1+\zeta^2}}\ ,
\label{spino2}
\ee
where $\G_{11}$ is the product of all ten Gamma matrices,
that anticommutes with each one of them
and for Minkowski spacetime, it squares to unity. Note also that $\Om$
leaves invariant the
Gamma matrices $\G^A$ corresponding to the seven-dimensional
spectator spacetime and it is
of the same form as the corresponding matrix in \cite{Sfetsos:2010uq}.

\subsubsection{General gauge fixing and coordinate transformations}

\no
As noted above, gauge choices different
than \eqn{gafig} might be more convenient in certain applications.  To expound this point let us consider a more general situation where the target admits an isometry group $G$ and we dualise with respect to a subgroup $H$ (for the case at hand we dualise the full $SU(2)$ isometry and so $\dim G = \dim H = 3$).
Specific gauge choices among the original
$\dim G +\dim H $ variables are of the form $f_i(g,v)=0$, $i=1,2,\dots , \dim(H)$.
This leaves $\dim G$ variables for the T-dual model.
Nevertheless, one may show that the
different gauge choices are related by coordinate transformations. This can be done by defining the "dressed" Lagrange multipliers as
\be
\hat v_i = D_{ji} v^j\ ,
\label{vdhv}
\ee
where $D_{ij}$ denotes the components of the matrix defined in \eqn{adij},
then the results we   obtain for a general gauge fixing are given by just replacing
in the previous expressions $v_i$ by $\hat v_i$. The details of the derivation are given in Appendix B.
We also note that if the set of the $v_i's$ is non-compact the same is true for the $\hat v$'s since $D$ is
an orthogonal transformation.

\no
In the present paper where the symmetry group is a freely acting $SU(2)$, besides \eqn{gafig}
these are two other natural choices for gauge fixing:

\no
\underline{Case 1}:  One might choose to partially gauge fix the $SU(2)$ group element, by setting two of the
Euler angles $\th= \phi =0$, but leaving $\psi$ as a variable in the dual.  This choice is particularly
motivated for backgrounds in which $\pl_\psi$ is a Killing vector and corresponds, within the
gauge/gravity correspondence, to the $U(1)_{\rm R}$ symmetry of the $\cN=1$ supersymmetry in the field theory side.
To fix the remaining gauge freedom one may fix one of the Lagrange multipliers $v_2 = 0$ . The choice $v_1=0$
would work just as well, but $v_3 = 0$ is inadmissible since the fixing of $g$ has already used up that particular gauge freedom.
In this and similar manipulations one may use the transformations for the various variables given by \eqn{tran1}
and \eqn{tran2}.
Then
\ba
&& D =   \left(\begin{array}{ccc}
 \cos \psi & \sin \psi & 0\\
 -\sin \psi & \cos \psi & 0\\
 0 & 0 & 1
\end{array}\right )\ ,
\nonumber\\
&& v=( v_1,0,v_3 )\qq \Longrightarrow \qq \hat v = (\cos\psi v_1,\sin \psi v_1,v_3 )\ ,
\label{dvhatv1}\\
&&  v=( 0,v_2,v_3 )\qq \Longrightarrow \qq \hat v = (-\sin\psi v_2 ,\cos \psi v_2,v_3 )\ .
\nonumber
\ea

\no
\underline{Case 2}:
Another gauge choice is $\phi = v_1 = v_2 = 0$ leaving a dual depending on $v_3, \theta, \psi$ as coordinates.
In this case
\ba
&& D =   \left(\begin{array}{ccc}
 \cos\th \cos \psi & \cos \th \sin \psi & -\sin\th\\
 -\sin \psi & \cos \psi & 0\\
 \sin\th \cos \psi & \sin\th \sin\psi & \cos\th
\end{array}\right )\ ,
\nonumber\\
&& v=(0,0,v_3) \qq \Longrightarrow \qq \hat v = (\sin\th\cos\psi v_3,\sin\th \sin \psi v_3,\cos \th v_3  )\ .
\label{dvhatv2}
\ea
For the above gauge choices the $\hat v_i$'s are obtained by a transformation that
resembles the change of coordinates from Cartesian to either polar or spherical coordinates.
However, this is only a formal analogy since in order to be precise one has to specify the range of variables.
The above coordinate transformations imply that
the dual backgrounds for different gauge choices are locally diffeomorphic and one must be rather careful about global properties.
Indeed, the global properties of the dual coordinates should be established
by demanding that the gauged and ungauged  path integrals match.
In Abelian T-duality one finds that the periodicity of the
dual coordinates is determined by constraining the holonomies of the gauge connections to vanish.
In the present non-Abelian case this is an open problem.

\no
Finally we note that when we examine certain properties of the supergravity
backgrounds produced by a non-Abelian duality we will need to
perform certain integrations and thus will need the
information about the global properties of the T-dual coordinates.
In doing so, we note the following relation
\be
e^{-2\Phi} \sqrt{|g|}\Big |_{\rm original} \times ({\rm F.P.}) = e^{-2\Phi} \sqrt{|g|}\Big |_{\rm final}\ ,
\label{fadeevterry}
\ee
where the first factor in the left hand side is computed for the original background and for the specific gauge choice of the form $f_a(g,v)=0$, $a=1,2,3$, one has made.
The second factor is the Faddeev--Popov determinant of the $\dim H$-square matrix $\D_{ij}$ arising in the variation $\d f_i =\D_{ij} \e_j$ and in the
specific gauge slice $f_i=0$. Such a relation was first shown for gauged WZW models in \cite{Bars:1991pt},
but it is valid the context of non-Abelian duality as well. This is not a surprise given their close relationship established in
\cite{Sfetsos:1994vz,Polychronakos:2010hd}.

\subsubsection{Transformation of the RR flux fields}

For the RR fields, the transformation rules were
first realised for Abelian T-duality in \cite{Bergshoeff:1995as}
via the reduction and matching of type-IIA and
type-IIB supergravities in nine dimensions. These rules were also
obtained by considering how T-duality acts on spinors (or rather bispinors)
and was detailed in \cite{Hassan:1999bv,Hassan:1999mm}
from a space time perspective, in \cite{Cvetic:1999zs,Kulik:2000nr}
for the Green-Schwarz string and in \cite{Benichou:2008it,Sfetsos:2010pl}
for the pure spinor super string.
In the democratic formalism \cite{Bergshoeff:2001pv} RR fields
are combined with their Hodge duals to form a bispinor
\be
{\rm IIB}:\ P  = {e^{\Phi}\ov 2} \sum_{n=0}^4 \slashed{F}_{2n+1}\ ,
\qq
{\rm IIA}:\ \widehat{P} ={e^{\widehat \Phi}\ov 2} \sum_{n=0}^5 \slashed{\widehat F}_{2n}\ , 
\ee
where $\displaystyle \slashed{F}_{p}={1\ov p!}\G_{\m_1\dots m_p}F^{\m_1\m_2\dots m_p}_p$.
The higher $p$-forms are related to the lower ones by
\be
F_p = (-1)^{\left[ \frac{p}{2}\right]} \star F_{10-p}\ ,
\label{demoll1}
\ee
where our conventions for the Hodge dual are given by \eqn{hdfer4},
assuming Minkowski signature spacetimes.
The non-Abelian T-dual is simply obtained by
multiplication with $\Omega^{-1}$. If the transformation is
from type-IIB to massive type-IIA  the
transformation rules for the RR-fluxes are given
by comparing the two sides of \cite{Sfetsos:2010uq}
\be
\widehat{P} =  P \cdot \Omega^{-1} \ .
\label{ppom}
\ee
In the case of massive type-IIA to type-IIB the role of $P$ and $\widehat P$ is interchanged.
For a general ansatz one may read off the dual fluxes produced in this way and a systematic analysis is given in Appendix C.

\subsubsection{A comment on singularities}
One might wonder whether such a dualisation procedure can result
in singular geometries starting with smooth geometries. To address this let us
for a moment consider the case of abelian T-dualisation along a $U(1)$ isometry generated by a  vector field,
$\partial_\theta$, in adapted coordinates.  The duality acts by inverting the component of the metric
 $ g_{\theta \theta}  \rightarrow \frac{1}{  g_{\theta \theta} }$.
It is clear that the dual may become singular at points for which  $ g_{\theta \theta}$ vanishes,
in other words when the norm of the Killing vector about which we dualise vanishes.
Indeed,  this phenomenon occurs when dualising the polar angle of say $\mathbb{R}^2$ and
in such cases  non-perturbative effects would typically become important (an interesting
related example of this in the context of mirror symmetry is found in \cite{Hori:2000kt}
wherein the phase of  chiral superfields are T-dualised and the dual superpotential
receives vital instantonic corrections).  More generally one anticipates singularities to be formed when the action of the isometry has fixed points.

 In the examples considered in the remainder of the paper this is not the case; the norm
of the Killing vectors can be seen to be nowhere vanishing and singularities are not {{\it created}}
by the dualisation procedure.\footnote{However in our examples we will find some apparent singularities
but these will be only coordinate in nature. In particular we will find bolt singularities
that may be removed with an appropriate choice of ranges for dual coordinates.}

\section{Dualisation of the Klebanov-Witten Background }\label{section3www}

The system of D3-branes placed at the tip of the conifold was studied within the AdS/CFT
correspondence in \cite{Klebanov:1998hh}.
In this section we work out the non-Abelian T-dual of this background and study
various of its properties within the correspondence.
The result of the dualisation and some of the of the properties of the background have first
presented in \cite{Itsios:2012zv}.

\subsection{The KW Background }

The gauge theory on the branes for the background of \cite{Klebanov:1998hh} is an $\cN=1$ superconformal field theory with product gauge
group $SU(N)\times SU(N)$. There are two sets of bifundamental matter fields;
$A_i$ in the $(N, \bar{N})$
forming a doublet of an $SU(2)$ global symmetry
and $B^m$ in    $(\bar{N}, N)$ forming a doublet of a second  global $SU(2)$.
The superpotential is given by
\be
W = \frac{\lambda}{2} \e^{ij} \e_{mn} {\rm Tr}( A_i B^m A_j B^n) \ .
\ee
 This  gauge theory is dual to string theory on $AdS^5 \times T^{(1,1)}$ with $N$ units of RR flux on the $T^{(1,1)}$.
The geometry and the $5$-form self-dual flux form, are given by
 \ba
 ds^2 &=&  \frac{r^2}{L^2} dx_{1,3}^2 + \frac{L^2}{r^2} dr^2 + L^2 ds^2_{T^{1,1}} \ ,
 \nonumber \\
 F_{(5)} &=& {4 \ov g_s L} \left({\rm Vol}(AdS_5) - L^5 {\rm Vol} (T_{1,1})\right)\ .
 \ea
Here $T^{(1,1)}$ is the homogenous space $(SU(2) \times SU(2))/U(1)$ with the diagonal embedding of the $U(1)$.
It has an Einstein metric with $R_{ij} = 4 g_{ij}$ given by \cite{Page:1984ae,Candelas:1989js}
 \be
 ds^2_{T^{(1,1)}} = \frac{1}{6}( d\th_1^2 + \sin^2 \th_1 d\phi_1^2) +  \frac{1}{6}( d\th_2^2 + \sin^2 \th_2 d\phi_2^2) + \frac{1}{9} \left( d\psi + \cos \th_1 d \phi_1 + \cos \th_2 d\phi_2 \right)^2\ .
 \ee
Introducing the frame fields for the $S^2$
\ba
\s_{\hat 1} = \sin \th_1 d \phi_1  \ ,  \qq   \s_{\hat 2} = d\th_1
\ea
and the invariant Maurer--Cartan forms for $S^3$, that
up to an overall normalization factor coincide with \eqn{L113},
\ba
\label{eq:MaurerCartan}
&& \s_{1} =  \cos \psi \sin \th_2 d\phi_2 - \sin \psi d\th_2 \ ,  \qq
\s_{2}= \sin \psi \sin \th_2 d\phi_2 + \cos \psi d\th_2 \ ,
\nonumber\\
&&  \s_{3} = d\psi + \cos \th_2 d \phi_2 \ ,
\ea
allows one to recast the $T^{1,1}$ metric as
\be
ds^2_{T^{1,1}} = \l_1^2 (\s_{\hat 1}^2 + \s_{\hat 2}^2 ) +  \l_2^2 (\s_{1}^2 +\s_{2}^2 )  + \l^2 (\s_{3} + \cos\th_1d\phi_1 )^2 \ ,
\ee
with $ \l_1^2 = \l_2^2  = \frac{1}{6}$ and $\l^2 = \frac{1}{9}$.
The $SU(2) \times SU(2) \times U(1)$ isometries of this metric correspond to Killing vectors.
In particular, there are two commuting sets as in \eqn{k123} with $(\th,\phi,\psi)\to (\th_i,\phi_i,\psi)$, with $i=1,2$.
These can be labeled as $k_{(a)}$ and $k_{(a+3)}$, $a=1,2,3$.
There is also $k_{(0)} = \partial_\psi$ corresponding to the $U(1)_R$ symmetry in the dual field theory.

\no
In the various computations below we will use the following frame
\ba
&& e^{\m'} = {r\ov L} dx^\m\ ,\quad \m=0,1,2,3\ , \qq e^4 = {L\ov r}dr\ ,
\nonumber\\
&& e^{\hat 1,\hat 2} = \l_1 \s_{\hat 1,\hat 2}\ ,\qq 
e^{1, 2}= \l_2 \s_{ 1,2}\ ,\quad e^{3} = \l (\s_3 + \cos\theta_1 d\phi_1)\ .
\label{dhfkem}
\ea

  \subsection{Action of $SU(2)$ on Killing Spinors }

 The KW background has eight unbroken super symmetries, four of which correspond to Poicar\'e super symmetries and the other
 four corresponding to superconformal symmetries of the dual field theory.
 The Killing spinor equation coming from the gravitino variation in the $T^{(1,1)}$ directions reduces to
 \be
D_\mu \eta + \frac{i}{2 L} \G^{12 \hat{1}\hat{2}\hat{3}} \G_\mu  \eta = 0 \ , \quad  \mu = 1, 2,   \hat{1}, \hat{2},\hat{3} \  .
\ee
This is solved by a constant $\eta$ obeying the projectors\footnote{Acting on the column vector $\left(\begin{array}{c}
\eta_+ , \\
  \eta_-\\
\end{array} \right)$  where $\eta = \eta_+ + i \eta_-$ we have the projection
\be
\G_{12} \eta  = - ( i \s_2) \eta \ , \quad \G_{45} \eta = + (i\s_2) \eta \ .
\ee }
\be
\G_{12} \eta = i \eta \ , \quad \G_{\hat{1}\hat{2} }\eta = - i \eta \ .
\ee

 We will dualise with respect one of the $SU(2)$ isometries.  It is natural to ask what portion of supersymmetry is preserved by this and what is the behaviour of the Killing spinors under the $SU(2)$ action.  In  \cite{Sfetsos:2010uq} (and further developed in \cite{Itsios:2012dc}) it was shown that the criteria for whether Supersymmetry is preserved is provided by the spinor Lorentz-Lie/Kosmann derivative \cite{Kosmann,oFarrill:1999va,Ortin:2002qb}.   For a killing vector $k$ this derivative is well defined and is given by
\be
{\cal L}_k \eta = k^\mu D_\mu \eta + \frac{1}{4}\nabla_\mu k_\nu \G^{\mu \nu} \eta \ .
\ee
Inserting the form of the Killing vectors we see that
\ba
{\cal L}_{k^{(4)}}  \eta &=& 0 \  , \\
{\cal L}_{k^{(5)} } \eta &=& -\frac{1}{4} \sin(\phi_1) \csc(\th_1) \left( \Gamma_{12} + \Gamma_{\hat{1}\hat{2}} \right) \eta \ , \\
{\cal L}_{k^{(6)}  }\eta &=& \frac{1}{4} \cos(\phi_1) \csc(\th_1) \left( \Gamma_{12} + \Gamma_{\hat{1}\hat{2}} \right)  \eta \ .
\ea
Thus,
one sees that the Killing spinor has vanishing Kosman derivative
along the $SU(2)$.  This corresponds to the statement that in the dual
field theory the supersymmetry is not charged under the $SU(2)$ flavour symmetries.
Hence we anticipate that supersymmetry is preserved after performing
a T-duality along this $SU(2)$.  Moreover we anticipate that the
Killing spinor in the dual will have the form
\be
\hat {\eta} = \Omega \cdot \eta
\ee
where $ \Omega$ is the spinorial representation of the Lorentz transformation between the left and right moving  frames for the dual geometry.

Parallel to this discussion is the fact that the $U(1)_R$ symmetry commutes with the $SU(2)$ and hence one expects the corresponding isometry to be preserved after dualisation.

 \subsection{Dualisation of the NS sector and  the Lorentz transformation }

We will dualise with respect to the $SU(2)$ isometry group generated by $\{k^{(4)}, k^{(5)},k^{(6)}\}$
following the procedure outlined in section 2.
Since the metric is block diagonal in the $AdS_5 \times T_{1,1}$ spacetime, it is sufficient to focus our attention on the $T^{(1,1)}$ factor alone.
The $AdS_5$ just comes along for the ride as a spectator field.\footnote{In what follows we have set $L=1$;
this can be restored by rescaling $\lambda_i $ and $\lambda$ by a factor of $L$ and by dividing the RR fields by $L$.
}
Our gauge choice will be given by the first of the choices in \eqn{dvhatv1}, i.e. $v_2=0$.
Moreover we relabel $v_1= 2 x_1 $ and $v_3 = 2 x_2 $.

\no
Within $T_{11}$, two of the fields, i.e. $x^\m=(\th_1,\phi_1)$ are spectators.
The various matrices we have introduced in \eqn{fraame} and directly enter in our expressions
for the T-dual background, using the frame \eqn{dhfkem} take the form
\be
(\k)^a{}_i = \sqrt{2} \left(
               \begin{array}{ccc}
                 \l_2 & 0 & 0 \\
                 0 & \l_2 & 0 \\
                 0 & 0 & \l \\
               \end{array}
             \right)\ ,\qq (\l)^a{}_\m = \l \cos\th_1 \left(
                                         \begin{array}{cc}
                                           0 & 0 \\
                                           0 & 0 \\
                                           0 & 1 \\
                                         \end{array}
                                       \right)\ ,
\label{kailmi}
\ee
where $\m=1,2$ corresponds to $\th_1$ and $\phi_1$, respectively.

\no
The result of this procedure is a $\s$-model on a target space with NS fields given by
\ba
 d\hat{s}^2 &=& ds_{\rm AdS_5}^2 + \l_1^2 (\s_{\hat{1}}^2 + \s_{\hat{2}}^2 )+ {\l_2^2\l^2\ov \D} x_1^2 \s_{\hat 3}^2
   \nonumber \\
&&  + {1\ov \D} \left( (x_1^2 + \l^2 \l_2^2 )dx_1^2 + (x_2^2 + \l_2^4) dx_2^2 + 2 x_1 x_2 dx_1 dx_2   \right) \ ,
  \nonumber \\
\widehat B &=& - {\l^2\ov \D} \left[x_1 x_2 dx_1  + (x_2^2 + \l_2^4) dx_2 \right]\wedge \s_{\hat 3} \ ,
\label{T11dual}
\\
e^{-2 \widehat \Phi} &=& {8\ov g_s^2}\ \D\ ,
\nonumber
\ea
where $ \s_{\hat 3} = d\psi + \cos\th_1 d\phi_1$ and
\be
\D \equiv  \l_2^2 x_1^2 + \l^2 (x_2^2 + \l_2^4 ) \ .
\label{T11dualn}
\ee
The metric, besides the symmetries of the $AdS_5$ factor,
evidently has a $SU(2) \times U(1)_\psi$ isometry and for a fixed value of $(x_1 , x_2)$ the remaining directions give a squashed three sphere.
Although the geometry is regular and the dilaton never blows up we note that there is a
removable bolt singularity (for a standard review see \cite{Eguchi:1980jx}).
For small values of $x_1$ and fixed $x_2$ the metric on the internal space behaves as
\be
 \l_1^2 (\s_{\hat{1}}^2 + \s_{\hat{2}}^2 )  + \frac{\l_2^2}{x_2^2 + \lambda_2^4} \left(  dx_1^2 + x_1^2  \s_{\hat 3}^2   \right) \ .
\label{bonllr}
\ee
 For this to be removed we require the range of $\psi$ to be $2\pi$ (so that at fixed $\theta, \phi$
the apparent singularity at  $x_1=0 $ becomes just the coordinate singularity of $\mathbb{R}^2$
written in polar coordinates).\footnote{This assumes that $x_1$ takes values in the half-line,
if it is allowed to range over the full real line the range of $\psi$ should be further restricted to $\pi$.}
Rather curiously, before dualisation the coordinate $\psi$ had range $4\pi$ (it was the coordinate of the
fibre in $T^{1,1}$  viewed as a $U(1)$ bundle over $S^2 \times S^2$) so we see that  the dualisation
has effectively enforced a $\mathbb{Z}_2$ quotient on $\psi$.  This is illustrative of the point made
earlier that the global properties of these geometries after dualisation may be rather subtle.

\no
From \eqn{Michu} one can obtain expressions for the "internal" frame fields $\hat{e}_\pm^i$ for this metric and the corresponding  Lorentz transformation
relating the frames is given by  \eqn{kjsg2}. To explicitly compute that we take
into account \eqn{kjsg1}, \eqn{kailmi} and the transformation \eqn{dvhatv1}, so that
\be
\zeta^a =  {1\ov \l_2^2 \l} (\l_2 \cos\psi x_1\ , \l_2 \sin\psi x_1,\l x_2)\  .
\ee
From this one can  deduce the spinorial representation $\Omega$ of this Lorentz transformation by using \eqn{spino2}.
Explicitly, we find dependence on the gamma matrices $\G_1$ and $\G_2$ of the form $\cos \psi \G_1 + \sin\psi \G_2$.
This will be used to obtain the RR fluxes as detailed below.

\no
In fact, there is a more convenient and simpler choice for the frame fields which can be obtained by performing an additional rotation  in the 1--2 plane
\be
\label{eq:rotation}
\left(
  \begin{array}{c}
     \hat e'_{1\pm} \\
     \hat e'_{2\pm} \\
  \end{array}
\right)  = \left(
             \begin{array}{cc}
               \cos\psi & \sin\psi \\
               -\sin\psi & \cos\psi \\
             \end{array}
           \right)
           \left(
  \begin{array}{c}
     \hat e_{1\pm} \\
     \hat e_{2\pm} \\
  \end{array}
\right)\ .
\label{g55er2}
\ee

\no
Using the above and the relation between the world sheet
derivatives as in \eqn{lplm} the frame fields are given by
\ba
\hat e'_{1 \pm} &=& \mp {\l_2\ov \D}   \left[ ( x_1^2 + \l^2 \l_2^2 )  d x_1 + x_1 x_2 (d x_2 \pm \l^2  \s_{\hat 3 })  \right]\ .
\nonumber\\
\hat e'_{ 2 \pm} &=& {\l_2\ov \D} \left[\l^2 x_2  dx_1 -  \l_2^2 x_1( dx_2 \pm \l^2 \s_{\hat 3 } )  \right] \ ,
\label{s123}\\
\hat e_{ 3 \pm} &=&\mp {\l\ov \D}  \left[  x_1 x_2  dx_1 + (x_2^2 + \l_2^4) d x_2 \mp \l_2^2 x_1^2  \s_{\hat 3 }  \right] \ .
\nonumber
\ea
The Lorentz transformation $\L$ relating the $\hat{e}^i_\pm$ frames is given by \eqn{kjsg2}. For these rotated frames
we have that $\L'= D\L D^T$. One finds that
\be
\Lambda' =   -\mathbb{1} + {2\ov \D}  \left( \begin{array}{ccc}
    \l^2 x_2^2  & -\l_2^2 \l^2 x_2 &  -\l_2 \l x_1 x_2 \\
  \l_2^2 \l^2  x_2  & \l^2 x_2^2 + \l_2^2 x_1^2 & - \l_2^3 \l  x_1 \\
  -  \l_2 \l x_1 x_2  &  \l_2^3 \l x_1 \ & -\l_2^2  x_1^2   \\
  \end{array} \right) \ .
\label{lpro}
\ee

 \subsection{RR Field Transformation }

We may now use the rule for transforming the RR sector given in
\eqn{ppom} in which the spinorial representation of the Lorentz transform acts on the right  of the bispinor.
A change for both frames induces a corresponding change in the bispinor with an extra matrix $\Om_{\rm fr}$ which acts from
the left as well as from the right. It reads
\be
\widehat P = \Om_{\rm fr}\cdot P \cdot \Om^{-1} \cdot \Om_{\rm fr}^{-1} = P' \cdot {\Om}^{\prime -1}\ ,
\ee
where
\be
\Om' =\Om_{\rm fr}\cdot \Om\cdot \Om_{\rm fr}^{-1}\ ,\qq  P' =\Om_{\rm fr}\cdot P\cdot \Om_{\rm fr}^{-1}\ .
\label{ppompr}
\ee

\no
In our case using \eqn{spino1} with $\L$ the two-dimensional $SO(2)$ matrix in \eqn{g55er2} we easily find that
\be
\Om_{\rm fr} = \cos{\psi\ov 2} \mathbb{I} +  \cos{\psi\ov 2} \G_{12}\ .
\ee
Then in $\Om'$ the dependence on $\psi$ disappears and we have that
\be
\Om' = {\G_{11}\ov \sqrt{\D}}\left(-\l_2^2 \l \G_{123} + \l_2 x_1 \G_1 + \l x_2 \G_3\right)\ ,
\ee
which precisely corresponds to the Lorentz matrix \eqn{lpro}. On the other hand $P'=P$ since the dependence on the frames $e^1$ and $e^2$
is of the form $e^1\wedge e^2$. This amounts to the "naive" use of \eqn{ppom} without taking into account the effect of the rotation
of the bispinor $P$, with just renaming the combination $\cos\phi \G_1 + \sin\psi \G_2$ by $\G_1$.
Using the general formulae of Appendix C, in particular \eqn{fpfg},
we see that the only non-vanishing form is
$\displaystyle G_2^{(3)}= -{4 \l_1^2\ov g_s} \s_{\hat 1}\wedge \s_{\hat 2}$.
Then using eqs.\eqn{j3hg4or} and \eqn{j3hg4} we compute the forms supporting the type-IIA supergravity background as\footnote{We use thought this paper the
$e^a_+$ frame to display the results. }
\ba
 \widehat F_2 & = & {8\sqrt{2}\ov g_s}  \ \l_1^2 \l\ e^{\hat{1}}\ \wedge e^{\hat{2}}
\ = \ {8 \sqrt{2}\ov g_s}\ \l_1^4\ \l\ \s_{\hat 1}\wedge \s_{\hat 2}\ ,
\nonumber\\
 \widehat F_4 & = & -  {8\sqrt{2}\ov g_s} \ e^{\hat{1}} \wedge e^{\hat{2}} \wedge \hat e^{\prime 2} \wedge (\l_2 \ x_1 \ \hat e^3 - \l \ x_2 \ \hat e^{\prime 1})
\label{T11dualforms}
\\
 \phantom{xx} &= &  - {8 \sqrt{2} \ov g_s} \l_1^2\ \l_2^2\ \l {x_1\ov \D} \s_{\hat 1}\wedge  \s_{\hat 2}\wedge  \s_{\hat 3}
\wedge (\l_2^2 x_1\ dx_2- \l^2 x_2\ dx_1)\ .
\nonumber
\ea
These fluxes support the NS geometry as a solution of the supergravity equations of motion.
Moreover the solution has ${\cal N}= 1$
supersymmetry  as explained in detail in the Appendix E.
Using eqs.\eqn{j3hg4or} and \eqn{j3hg4} we also compute the higher forms
\ba
\widehat F_6 &  = &   {8\sqrt{2} \ov g_s} \ {\rm Vol}(AdS_5) \wedge (\l_2  x_1 \hat e^{\prime 1} + \l  x_2 \hat e^3)
\nonumber\\
& = & - {8\sqrt{2} \ov g_s} {\rm Vol}(AdS_5)  \wedge (x_1 dx_1+ x_2 dx_2)\ ,
\nonumber\\
 \widehat F_8  & = &    {8\sqrt{2}\ov g_s} \ \l_1^2 \ \l  {\rm Vol}(AdS_5) \wedge \hat e^{\prime 1} \wedge \hat e^{\prime 2} \wedge \hat e^3
\\
& = &  {8 \sqrt{2} \ov g_s} \l_1^2\ \l_2^2\ \l^2\ {x_1\ov \D} dx_1\wedge dx_2 \wedge \s_{\hat 3}\ .
\nonumber
\ea
These obey $\widehat F_6= -(\star \widehat F_4) $ and $\widehat F_8= \star \widehat F_2 $, in agreement with \eqn{demoll1},
as they should. In fact, we prove in Appendix C, that these consistency relations relating higher to lower rank
RR flux forms are preserved by non-Abelian T-duality.

\subsection{M-theory lift }

We are interested in lifting  and interpreting our solution to M-theory. To do that
we first read off, using \eqn{hdbf2f4}, the potentials corresponding to the fluxes $\widehat F_2$ and $\widehat F_4$
in \eqn{T11dualforms}.\footnote{In the rest of this section and in the next section we set besides $L=1$,
the string coupling $g_s=2\sqrt{2}$ and
$\l_1=\l_2={1\ov \sqrt{6}}$, $\l={1\ov 3}$, even if we keep the notation in some of the eqs.}
We find that
\be
C_1= {1\ov 27} \s_{\hat 3}\ ,\qq C_3 = {x_2\ov 27} \s_{\hat 1} \wedge \s_{\hat 2} \wedge \s_{\hat 3}\ .
\ee
The lift to eleven dimensions (along the circle with coordinate $x_{\sharp} $)
of the geometry we find after non-Abelian T-duality is given by,
\ba
ds^2 &=& \D^{1/3} \left( ds^2_{AdS_5} +  \l_1^2 (\s_{\hat{1}}^2 + \s_{\hat{2}}^2) \right)  +  \D^{-2/3} \left[(x_1^2 + \l^2 \l_1^2 )dx_1^2 \right.
\nonumber \\
&& \left. \quad  + (x_2^2 + \l_1^4) dx_2^2 + 2x_1 x_2 dx_1 dx_2
+  \lambda^2 \lambda^2_1 x_1^2 \s_{\hat{3}}^2  +  \left( d x_{\sharp} + \frac{ \s_{\hat{3}} } {27} \right)^2 \right]\ ,
\label{ds11we}
\ea
where $\D$ is given in \eqn{T11dualn}.
The four-form flux field is given by
\be
F_4 = d( C_3 + B \wedge dx_{\sharp} ) = {1\ov 27} dx_2\wedge  \s_{\hat 1} \wedge \s_{\hat 2} \wedge \s_{\hat 3} + H \wedge dx_{\sharp}\ ,
\label{ds11wefl}
\ee
where $H=dB$ is computed using the expression for $B$ in \eqn{T11dual}.

\no
In eleven dimensions
our solution preserves two commuting $U(1)$ isometries and eight supercharges.
Recently, a class of ${\cal N}=1$ (generically non-Lagrangian)
SCFT's, arising as the IR  fixed point of the dynamics
of M5-branes wrapped on a
genus $g$ surface $\S_g$, was engineered in \cite{Bah:2012dg,Bah:2011vv}.
These field theories enjoy not only a $U(1)_R$ global symmetry but also
an additional $U(1)$ global symmetry.
The geometrical dual to these theories fit within the ansatz
of $\cite{Gauntlett:2004zh}$, but now specialised
to the case that the internal six-dimensional manifold is a squashed $S^4$
fibration over  $\S_g$.
Our solution fits in this ansatz as we will soon demonstrate after
a brief review of the solutions of \cite{Bah:2012dg,Bah:2011vv}.

\subsubsection{Brief review of the solution in \cite{Bah:2012dg,Bah:2011vv}}
To give the metric and four-form flux in \cite{Bah:2012dg,Bah:2011vv},
we first introduce the constant curvature metric on the Riemann surface $\S_g$
\be
ds^2_2({\S_g}) = e^{2A(y_1,y_2)}(dy_1^2 + dy_2^2)\ ,
\ee
with scalar curvature $2 \k$, where $\k=1,0,-1$ for the two-sphere ($g=0$ for which we will be particularly interested),
for the torus and a hyperbolic surface, respectively. Hence the conformal factor obeys  Liouville's equation
\be
(\del_{y_1}^2 +\del_{y_2}^2)A + \k e^{2 A}= 0 \ .
\ee
It is also convenient to define the constants
\ba
&& a_1  = 2{2-2 g\ov \k} e^{2\n}\ ,\qq a_2 = -12{2-2 g\ov \k} e^{2\n}\left(1-{\k\ov 6} e^{-2 \n}\right)\ ,
\nonumber\\
&& e^{2\n}= {1\ov 6}\left(-\k \pm \sqrt{\k^2 +3\k^2 z^2}\right)\ ,
\label{a1a2an}
\ea
the functions
\be
k(q) = {a_2\ov a_1} q^2 + q-{1\ov 36}\ ,\qq f(y) = f_0+ 6{a_2\ov a_1} y^2\ ,\qq e^{6 \l(y,q)}= q f(y)+ 4 y^2\ ,
\ee
as well as the one forms $V$ and $\r$
\ba
&& dV = {\k\ov 2-2 g} e^{2A(y_1,y_2)} dy_1\wedge dy_2\ ,\qq \int_{\S_g} dV = 2\pi \ ,
\nonumber\\
&& \r = (2-2g)V -\ha \left(a_2+{a_1\ov 2q}\right)(d\chi + V)\ .
\ea
Then the metric takes the form
\be
\label{metricBBBW}
 ds_{11}^2 = e^{2 \lambda(y,q)} \left[ ds^2_{AdS_5} + e^{2 \nu} ds^2_2(\S_g) \right] + e^{- 4 \lambda(y,q)} ds^2_{M_4} \ ,
 \ee
 where $ds^2_{AdS_5}$ is the metric of $AdS_5$ with unit radius and
 \ba
 ds^2_{M_4} &  = &\left( 1+ \frac{4 y^2}{q f(y)}  \right)dy^2 + \frac{f(y) q}{k(q)} \left(dq + \frac{12 y k(q) }{qf(y) }d y \right)^2
\nonumber\\
&& + \frac{a_1^2}{4} \frac{f(y) k(q) }{q} (d \chi + V)^2  + \frac{q f(y)}{9} (  d\psi + \rho)^2 \ ,
 \label{wechtmetric}
\ea
 in which the $U(1) $ isometry generated by $\partial_\psi$ corresponds to the R-symmetry and is supplemented by an additional $U(1)$ generated by $\partial_\chi$.
 Demanding that the metric is positive definite places bounds on the coordinates $y$ and $q$ by imposing
 \be
 q f(y) \geqslant 0 \ , \qq k(q) \geqslant 0 \ .
 \ee
To support the geometry of the ansatz
\eqn{metricBBBW} a flux is turned one with structure\footnote{Some terms in the
expression below seem to be missing in
the corresponding expressions of the original literature.}
 \be
 \label{fluxBBBW}
 F_{4} = e^{2 A(y_1,y_2) + 2 \n} dy_1 \wedge dy_2 \wedge G_2 + V\wedge G_3 +   G_{\perp}\ .
\ee
The various functions are given by
\ba
&& G_{2} =  \frac{(24 q-1) (4 y^2 + q f(y)) - 144 y^2 k(q)}{ 18 q (4 y^2 + q f(y))} \  dy \wedge d\psi
\nonumber \\
 && - \frac{( 24 q-1) (a_1 + 2 a_2 q) [4 y^2 + q f(y)] -
 36 k(q) \left[4 \left(a_1 + 2 (6 a_1 + a_2) q\right ) y^2 + a_1 q f(y)\right]}{72 q^2 (4 y^2 + q f(y)) }\ dy \wedge d\chi
\nonumber  \\
 && -{1\ov 3} \frac{y f(y) }{4 y^2 + q f(y)} \ dq \wedge [ 2 d\psi - (6 a_1 + a_2) d\chi ]
 \label{g22}
 \ea
and
\ba
G_{3} &=& -{a_1 f(y)\ov 36} \  \frac{  96 y^2 + f(y)(1+12q)  }{ ( 4 y^2 + q f(y))^2} \  dq \wedge dy \wedge \left[ 2(g-1) d\chi + d\psi  \right]\ ,
\nonumber\\
 G_{\perp}  &=& - {a_1 f(y)\ov 36} \ \frac{  96 y^2 + f(y)(1+12q)  }{( 4 y^2 + q f(y))^2}\  dq \wedge dy \wedge d\chi \wedge d\psi \ .
 \label{g2p}
 \ea

\subsubsection{Explicit change of coordinates}

Let's consider the case where $g=0$, i.e.  $\S_{g=0} = S^2$ and also
take the numbers
$\k = z=1$. Then the various constants are given by
\be
e^{2 \nu} =\frac{1}{6}\ ,\qq a_1 = \frac{2}{3}  \ , \qq a_2 = 0\ .
\label{kolarov}
\ee
The remaining functions take the form
 \ba
 f(y) = f_0 \ , \quad k(q) = q - \frac{1}{36} \ ,  \quad e^{6 \lambda(y,q)} =  f_0 q + 4 y^2 \  .
 \ea
 The one-forms that determine the fibration are given by
 \be
 \rho = 2 V - \frac{1}{6 q } ( d\chi + V) \ , \qq dV = \frac{1}{2} e^{2A(y_1,y_2)} dy_1 \wedge dy_2 \ .
 \ee
 The conformal factor for the metric on $S^2$ is given by
 \be
e^{A(y_1,y_2)} =  \frac{2}{1+y_1^2 + y_2^2} \ ,
 \ee
 and the conversion to polar coordinates is given by
 \be
 y_1 = \cos \phi_1 \cot \frac{\th_1}{2}  \ , \qq  y_2 = \sin \phi_1 \cot \frac{\th_1}{2}  \ ,
 \ee
 whence
 \be
 V = \frac{1}{2} \cos \th_1 d\phi_1 \ .
 \ee
Then the metric \eqn{metricBBBW} can be written exactly in the form
\eqn{ds11we} if we make the coordinate change
\be
\label{coordchange}
q = \frac{1}{36} + \frac{3}{2} x_1^2 \ , \qq y = \frac{x_2}{6} \ ,  \qq x_{\sharp} = \frac{1}{9} \chi - \frac{1}{18} \psi \ , \qq f_0 = \frac{1}{9} \ .
\ee
 From the positivity requirement of the metric (and these coordinate redefinitions) we expect
{\it{ a priori}} that  $q\in [\frac{1}{36}, \infty ]$  and $y\in [-\infty , \infty]$.
There may be further conditions from quantisation of fluxes.

\no
For the forms in eqs.(\ref{g22}) and (\ref{g2p}) we find that,
 \ba
  G_{\perp} &=& -\frac{1}{54} \frac{(1+ 12 q + 864 y^2)}{(q + 36 y^2)^{2}}  dq \wedge dy \wedge d\chi \wedge d\psi \ ,
\nonumber\\
  G_{3} &=&  \frac{1}{54} \frac{(1+ 12 q + 864 y^2)}{(q + 36 y^2)^{2}}  dq \wedge dy \wedge ( 2 d\chi - d \psi) \ ,
\\
  G_{2} &=&  \frac{1}{18 ( q + 36 y^2 )  } \Big(  12 y dq\wedge( 2 d\chi - d \psi)    + 2 ( 1+ 1296 y^2) dy \wedge d\chi
\nonumber\\
&& + (24 q-1 - 432y^2 ) dy \wedge d\psi  \Big) \ .
\nonumber
 \ea
 Then, with the above coordinate change the four-form flux \eqn{fluxBBBW} becomes \eqn{ds11wefl}.

\subsection{Brief comments on the field theory}

Next we present a very quick synopsis the work of \cite{Bah:2012dg}
which interprets these geometries as coming from wrapped M5-branes.
Before back-reaction one considers the M5-brane on a genus $g$ curve
 ${\cal C}_g$ in a Calabi Yau threefold that is decomposable as
\be
\begin{array}{cc}
\mathbb{C}^2 \rightarrow  & {\cal L}_1 \oplus {\cal L}_2 \ , \\
& \downarrow\\
& {\cal C}_g
\end{array}
\ee
where ${\cal L}_i$ are line bundles, so that the space
has a natural $U(1)_1  \times U(1)_2$ symmetry acting
as phase rotations in the respective line bundle.
The Calabi--Yau condition restricts the Chern classes of the line bundles,
$c_1 ({\cal L}_1) = p $, $c_2 ({\cal L}_2)= q$, so that they cancel-off
against the curvature coming from the curve ${\cal C}_g$ i.e. one requires $p+q= 2g -2$.
In what follows it is helpful to encode the solution by the genus
$g$ and the {\it twisting} parameter $z$ defined through
\be
p = (1+ z) (g-1) \ , \qq q = (1-z) (g-1) \ .
\nonumber
\ee
The parameter $z$ is related to the warp factor $e^\nu$ entering into
the supergravity solutions above as in \eqn{a1a2an}. In our case recall that $\kappa=z= 1$,
which gives \eqn{kolarov}.
The construction in
\cite{Bah:2012dg}, allows one to calculate the four-dimensional central
charges $c, a$ using the anomaly polynomial
of the six dimensional ${\cal N} = (2,0)$ theory
(here specialised to the $A_N$ case)
\ba
\label{aandc}
& a = (g-1) (N-1) \frac{\zeta^3+\kappa \eta^3  -\kappa(1+\eta)(9+21 \eta + 9 \eta^2) z^2  }{48 (1+\eta)^2 z^2}\ ,
\nonumber\\
& c = (g-1) (N-1) \frac{\zeta^3+\kappa \eta^3 -\kappa(1+\eta)(6-\kappa \zeta+17 \eta + 9 \eta^2) z^2}{48 (1+\eta)^2 z^2}\ ,
\ea
where $\eta = N^2+1$ and
$\zeta = \left( \eta^2 + (1 +4\eta + 3\eta^2  )z^2 \right)^{1/2}$.
Generically, in the large $N$ limit the leading behaviour of these charges match
and scale as $N^3$ reflecting the six dimensional M5-brane origin of the theories.

\no
The corresponding field theories for $g>1$ were constructed in
\cite{Bah:2012dg}, starting from the theories discussed by
Gaiotto  and Maldacena \cite{Gaiotto:2009gz}, integrating out the
adjoint chiral multiplet inside the $N=2$ vector multiplets and
following a simple set of rules.
However this construction does not seem to extend easily for
genus zero and the field theory interpretation remains
somewhat mysterious in this case.
A surprising result is that
for the particular geometry that we obtained by dualising
KW we have $g=0, z^2=1$.
Then the anomaly polynomial result for central charges
in \eqn{aandc}  in the large $N$ limit gives,
\be
a= \frac{3}{8} (N -1 ) + {\cal O}\left(1\ov N\right) \ , \qq c =  \frac{1}{4} (N -1 ) + {\cal O}\left(1\ov N\right) \ .
\ee
The leading $N^3$ term cancelled out leaving just a linear dependence
on $N$. Since the leading term vanished one need not have
that $a$ and $c$ match.  A strange feature, however, is that the result
has no $N^2$ dependent piece as one might have expected.
As we shall see later this  does not match the expectation
from supergravity where we find the central charge is an invariant of
the T-dualisation.
The resolution of the puzzle probably lies on the fact that
a-maximization, used in \cite{Bah:2012dg} is breaking-down in this case
due to the presence of extra accidental symmetries.

\no
These features point to the fact that the theory obtained when M5-branes wrap
a two-sphere preserving conformality and minimal SUSY are apparently
out-of-line with the general ideas of \cite{Gaiotto:2009gz},
\cite{Benini:2009mz} and  \cite{Bah:2012dg}.

\no
On the other hand, the existence of a BPS operator corresponding to
a wrapped $M2$-brane seems guaranteed here. We will see below how to calculate its
dimension. In line with the differences pointed above, we will
also find qualitative differences regarding this operator
and that defined in  \cite{Gaiotto:2009gz},
\cite{Benini:2009mz} and  \cite{Bah:2012dg}.
\subsubsection{An operator associated with M2-branes}
Gaiotto and Maldacena \cite{Gaiotto:2009gz}
discussed the existence of an operator
associated with the
$T_{N}$ theories. They called this operator $O_{ijk}$ and
its characteristics are described in \cite{Gaiotto:2009gz}.
It is uncharged under the $U(1)$ symmetries of the field theory.
The dimension of this operator in the string dual is equated with
the energy of an M2-brane extending in time and wraps
a two-cycle (in their case an $H2$-space) which is {\it not}
fibered with the rest of the space.
In other words, the calculation in the putative type-IIA would give a D2-brane without
any worldvolume gauge field.

\no
We can proceed in analogy and {\it define} the dimension
of an operator by the volume of an M2-brane
that extends in time and wraps a two-cycle.
The main subtle difference is that in our case,
we will not be able to have this operator uncharged.
When we place our M2-brane on the manifold,
utilizing our M-theory geometry (\ref{ds11we}),
$M_3=[t, \theta_1,\phi_1]$ at constant values of $x_1, x_2, \psi, x_{11}$ and for the radial coordinate $r=r_0$,
we can then calculate the induced metric on this M2-brane to be
\ba
&& ds_{\rm ind, M2}^2 = \D^{1/3}\Big[ -r_0^2 dt^2 + \frac{1}{6} (d\theta_1^2 +\sin^2\theta_1 d\phi_1^2)
+ \frac{f(x_1,x_2)}{6} \cos^2\th_1 d\phi_1^2\Big] \ ,
\nonumber\\
&& f(x_1,x_2) = { x_1^2 + 2/27 \ov 9 \D}\ ,
\ea
where $\D$ is given by \eqn{T11dualn}.
We see here that we will  have non-zero $U(1)$ charge, as there
seems to be no way of getting rid of the fibration represented by the term
$\sigma_{\hat{3}}= d\psi + \cos\theta_1 d\phi_1$.
The energy of this M2-brane is given
in terms of the complete elliptic function of the second kind ${\bf E}$ as
\ba
 E &= & r_0 \frac{ \D^{1/2}}{6} \int_{0}^{\pi} d\theta_1 \sqrt{f(x_1,x_2)\cos^2\theta_1 + \sin^2\theta_1}
 \nonumber\\
 & = & r_0 \frac{ \D^{1/2} }{6}\left[ {\bf E}\left(\sqrt{1-f}\right) + \sqrt{f}\ {\bf E}\left(\sqrt{1-f^{-1}}\right)\right]\ .
\ea
After lengthly algebraic manipulations one can show that the minimum of the energy as a
function of $x_1$ and $x_2$ is at $x_1=x_2=0$ and has the arithmetic value $E_{\rm min}\simeq  0.039 r_0$ (note
our choice for $g_s$ in footnote 8).

\no
In a very similar way, we can calculate the
central charge at this conformal point, by defining it as proportional to
the volume of the internal six-manifold.
We will study the central
charge in more detail in Section \ref{centralentanglementxxx}.
Since the results there can be specialised to the constant value
obtained at the fixed point, we will postpone this study until
Section \ref{centralentanglementxxx}.


  \section{Dualisation of the Klebanov-Tseytlin Background }
  \label{secciondekt}

Let us now turn our attention to non-conformal backgrounds.
One can start with the KW solution and break conformal invariance
by adding $M$ fractional D3-branes i.e. D5-branes wrapping a contractible two cycle
of $T^{(1,1)}$ as in \cite{Klebanov:1999rd,Klebanov:2000nc,Klebanov:2000hb}.
This modifies the field theory to be $SU(N) \times SU(N+M)$,
hence no longer conformal. In fact this theory
has a rich RG dynamics undergoing a sequence of Seiberg dualities
to lower rank gauge groups as one proceeds to the IR (see
\cite{Strassler:2005qs} for a careful summary).
In the IR, strong coupling dynamics takes hold giving rise
to spontaneous $Z_{2M}$-symmetry breaking and confinement.

\no
For the time being let us concentrate not on the full solution of Klebanov and Strassler \cite{Klebanov:2000hb}
but rather on the simpler case of Klebanov and Tseytlin (KT) that appeared earlier \cite{Klebanov:2000nc}.
This background gives a good description of the UV of the duality cascade, but
is singular in the IR (where the strong
dynamics cures this pathology, replacing the singular conifold with the smooth deformed conifold).

\no
One of the purposes of this section then is to develop our 'intuitions'
on the effect of the non-Abelian duality on cascading geometries.
All of our results will be trustable away from the singularity, placed at the origin
of the radial coordinate (that will be labelled by $u$ below).
The philosophy adopted here, will be that the generated
background 'defines' a dual QFT, that we will start to understand with the
calculations proposed in this section.

\no
The geometry is given  \cite{Klebanov:2000nc} by\footnote{The dilaton is zero
so that there is no difference between string and Einstein frame.}
 \be
 ds_{10}^2 = e^{-\frac{2}{3} (B + 4 C) } ds^2_5 + ds^2_{5'}\ ,
 \label{klebatsey}
 \ee
 where
 \be
 ds^2_5 = du^2 + e^{2A} dx^2_{1,3}\ ,
 \label{klebatsey1}
  \ee
 is a deformation of $AdS_5$ and
 \be
  ds^2_{5'} = \frac{e^{2 C}}{6} \sum_{i=1}^2 ( d\th_i^2 + \sin\th_i^2d\phi_i^2 )
  + \frac{e^{2 B} }{9} ( d\psi + \cos \th_1 d\phi_1 + \cos \th_2   d\phi_2)^2 \ .
 \label{klebatsey2}
 \ee
 It is convenient to define the natural forms
\ba
&&  e^{\th_i} = \frac{1}{\sqrt{6}} d\th_i \ , \qq e^{\phi_i} = \frac{1}{\sqrt{6}} \sin \th_i d\phi_i\ , \quad i=1,2\ ,
\nonumber\\
&&  e^\psi = \frac{1}{3} ( d\psi + \cos \th_1 d\phi_1 + \cos \th_2   d\phi_2) \ ,
\\
&& \omega_2= \frac{1}{\sqrt{2} } ( e^{\th_1} \wedge e^{\phi_1} - e^{\th_2} \wedge e^{\phi_2} )\  .
\nonumber
\ea
Note that the above one-forms are not the frames defining the metric since they are missing the appropriate
exponentials.
The fluxes are given by
 \ba
&&  B_2 = - T \om_2 \ ,\qq F_3 = -P e^\psi \wedge \om_2\ ,
\nonumber\\
&&
F_5 = K e^\psi\wedge e^{\th_1}\wedge  e^{\phi_1}\wedge  e^{\th_2}\wedge e^{\phi_2} - K e^{4 A - {8\ov 3}(B+4 C)}du\wedge dx_{1,3}\
\label{fluxesklebatsey}
\ea
and the dilaton is, as explained, vanishing.
Since it is needed for the dualisation below, we also give, using \eqn{demoll1}, the expression for
\be
F_7 = -(\star F_3)=
-P e^{ 4 A + 4 f-  14 q} du \wedge dx_{1,3} \wedge \om_2\ .
\ee
The functions $A,B,C, K, T$  depend only on the radial direction $u$, whereas $P$ is a constant.
Introducing the functions
 \be
 f = - \frac{1}{5} (B- C)  \ , \qq q = \frac{2}{15} (B+ 4C).
\label{teve}
 \ee
one finds that the BPS conditions lead to\footnote{ there is a sign convention choice $P_{\rm here} = - P$ in \cite{Klebanov:2000nc}.}
\be
K = Q-P T \ ,\qq A =  q + \frac{2}{3} f - \frac{1}{P} T\ ,
\label{bpseqskt1}
\ee
as well as to a set of first order non-linear system of equations
\ba
&& {dT\ov du} = - P e^{4 f-4 q }\ ,
\nonumber\\
&&  {df\ov du} = - \frac{3}{5}  e^{4f-4 q }  \left(1- e^{-10f}\right)\ ,
\label{bpseqskt}\\
&&  {dq\ov du} = \frac{2}{15}   e^{-4 q + 4 f}  \left(3+ 2 e^{-10f}\right)  - \frac{1}{6}  (Q-P T) e^{-10 q}\ .
\nonumber
\ea
One may check explicitly that the flux equations and Bianchi identities
are satisfied on these BPS equations.  The dilaton equation is also
satisfied by virtue of the identity $H_3^2 = F_3^2$.

\no
We recall the special logarithmically running solution of KT.
This is constructed by setting the
function $f=0$, which is consistent with the system \eqn{bpseqskt}
and changing variables as
\be
e^{3 q}= r^2 h^{1/2}\ ,\qq du = e^{4 q} {dr\ov r}\ .
\ee
One finds the metric
 \ba
&&  ds^2 = h^{-1/2} dx_{1,3}^2 + h^{1/2} \left(dr^2 + r^2 ds^2_{T^{1,1}} \right) \ ,
\qq h = b_0 + \frac{P^2}{4 r^4} \ln (r/r_{\ast})\ ,
\nonumber\\
&& T = \tilde T -P \ln (r/r_\ast)\ ,\qq K=P^2 \ln(r/r_\ast)- {P^2\ov 4}\ ,
\label{aguero}
\ea
where $\tilde  T$, $r_{\ast}$ and $b_0$ are integration constants.
The latter should be set to zero in order to
decouple  the dual QFT from gravity. Then clearly, the gravitational description breaks
down at $r\sim r_\ast$. However, already at a larger radius at which the function $K(r)$
vanishes the gravitational force has changed sign.
This indicates that one needs a non-singular completion
of this solution towards the IR, which was achieved in \cite{Klebanov:2000hb}.
In what follows we shall keep the solution completely general and given
in terms of the functions entering into the BPS equations. Only final results
may use the explicit expression in \eqn{aguero}.

\subsection{Dualisation of the background}

We proceed now to indicate the result of performing a non-Abelian T-duality
on the geometry (\ref{klebatsey})-(\ref{fluxesklebatsey}).
We will use the general results of Section 2 and Appendix C.
Our gauge choice will be given by the second of the choices in \eqn{dvhatv1}, i.e. $v_1=0$.
We note that the result of this dualisation for the special solution \eqn{aguero}
have been already presented in \cite{Itsios:2012zv}.

\no
We implement the dualisation with the same gauge fixing as before using \eqn{dvhatv1}.
The matrices $\k$ and $\l$ are the same as in \eqn{kailmi}, but
\be
\zeta^a =   \left(-3\sqrt{3} \, e^{-B-C}  \sin\psi \, v_2\ , 3\sqrt{3} \, e^{-B-C}  \sin\psi \, v_2,  {1\ov \sqrt{2}} \, e^{-2 C}  {\cal V} \right)\  ,
\quad  {\cal V} =   6 v_3 -T \ .
\ee
The procedure leads us to define a set of frame fields
 \ba
&&  \hat{e}^{\m' } = e^{A - \frac{1}{3} (B + 4 C) }dx^{\m} \ , \quad  \m = 0, 1,2,3\ , \qq \hat{e}^4 =
e^{ - \frac{1}{3} (B + 4 C) }du\ ,
\nonumber\\
&& \hat{e}^\th = \frac{e^C}{\sqrt{6} }  d\theta_1 \ ,\qq
\hat{e}^\phi = \frac{e^C}{\sqrt{6} } \sin \theta_1 d\phi_1\ ,
\ea
We also have that
\ba
&& \hat{e}^{\prime 1}_\pm = -\frac{\sqrt{6} }{ 81 {\cal W} }  \left(    e^{2B +C} {\cal V} dv_2 - e^{3 C}v_2  (  9  d v_3  \pm   \sqrt{2} e^{2B} \s_3  )  \right)\ ,
\nonumber\\
&&  \hat{e}^{\prime 2}_\pm = \mp  \frac{\sqrt{3} }{ 81 {\cal W} } e^{ C} \left(  2  (e^{2B+ 2C}  + 27 v_2^2  ) dv_2
+   {\cal V} v_2  (  9  d v_3  \pm   \sqrt{2} e^{2B} \s_3  ) \right)\ ,
\\
&&  \hat{e}^3_\pm = \mp \frac{1}{ 54 {\cal W} } e^B \left(  \sqrt{2} ( 2 e^{4C} + {\cal V}^2) dv_3
+ 6 v_2  (\sqrt{2}{\cal V} dv_2  \mp  2 e^{2 C} v_2 \s_3   )   \right) \ .
\nonumber
 \ea
where we have made a frame rotation as in \eqn{g55er2}.
In the above we have defined the function
 \bea
{\cal W} = {1\ov 81}\left(2 e^{2 B+ 4 C} + 54 \,  e^{2C} v_2^2  +  e^{2 B}{\cal V}^2  \right)\ ,
 \eea
  \def\U{{\cal U}}
  \def\W{{\cal W}}
such that the dual metric is given by
\ba
&& d\hat{s}^2 =  e^{-\frac{2}{3} (B + 4 C) } ds^2_5   +\frac{ e^{2C}}{6} ( d\th_1^2 + \sin^2 \th_1 d\phi_1^2) + \sum_{a=1}^3 \hat e_\pm^a \hat e_\pm^a \ ,
\nonumber\\
&& \phantom{xx} (\hat e_\pm^{\prime 1})^2 + (\hat e_\pm^{\prime 2})^2 + (\hat e_\pm^{3})^2
= {1\ov 54 {\cal W}}\Bigg(4 (e^{2B+2C} + 27 v_2^2)dv_2^2
\label{metriccaseKT}
\\
&&\phantom{xxxxxxxxxxx} + 3 (2 e^{4 C}+ {\cal V}^2)dv_3^2  + 36 {\cal V} v_2 \, dv_2 \, dv_3  + 4 e^{2B+2 C} v_2^2 \s_3^2\Bigg)\ .
\nonumber
\ea
We can either choose the $e_+$ or
the $e_-$ triads. In the following we will prefer $e_+$ in accordance with footnote 7.
Note that there is also a removable bolt singularity at $v_2=0$ provided that
the range of $\psi$ is restricted to $2\pi$ or $\pi$ depending on whether $v_2$ takes values in the
half or entire real line, respectively, i.e. footnote 6.
This metric has, besides the obvious Poicar\'e symmetry of the $dx_{1,3}^2$ factor,
an $SU(2)\times U(1)_\psi$ isometry as in the case of the T-dualised Klebanov--Witten background.

\no
There is also a NS antisymmetric tensor given by
\be
 \widehat{B}_2=  -{T \sin\th_1 \ov 6\sqrt{2}} d\th_1 \wedge d\phi_1 +  {e^B\ov 3\sqrt{6} v_2}\left(
 -\sqrt{2} e^{C} \hat e^{\prime 1} \wedge \hat e^{3} +  e^{-C} {\cal V} \hat e^{\prime 2} \wedge \hat e^{3}\right)
\ee
and a dilaton
\be
\label{eq:dilatonKT}
e^{-2 \widehat{\Phi}}= {\cal W}\ .
\label{dilakt}
\ee
The above background does not get any more
singular than the original one. For instance the dilaton in \eqn{dilakt} never blows up. However, it
still has any singular behaviour inherited from the original background, e.g. when \eqn{aguero} is used.

\no
The Lorentz rotation is given by
\be
 \Lambda^\prime = - \mathbb{1} + \frac{1}{81 \W}
 \left(
   \begin{array}{ccc}
     2 (54 e^{2 C} v_2^2 + e^{2 B} {\cal V}^2) & -2 \sqrt{2} e^{2 B + 2 C}  {\cal V} & 12 \sqrt{3} e^{B+3C} v_2 \\
      2 \sqrt{2} e^{2 B + 2 C}  {\cal V}  & 2 e^{2 B} {\cal V}^2 & -6 \sqrt{6} e^{B + C} {\cal V} v_2 \\
     -12 \sqrt{3} e^{B+3C} v_2  & -6 \sqrt{6} e^{B + C} {\cal V} v_2 & 108 e^{2 C}v_2^2 \\
   \end{array}
 \right)\ .
 \ee
The spinorial counter part of this rotation is
 \be
 \Omega = \frac{1}{9 \sqrt{\W}} \Gamma_{11}  \left( - \sqrt{2} e^{B+2 C} \G_{123}  + 3 \sqrt{6} e^{C} v_2 \Gamma_2  +  e^{B} {\cal V} \Gamma_3  \right)\ .
\label{traomega}
 \ee
Using the diagonal combinations $f,g$ defined in (\ref{teve}),
the fluxes are (the Ramond fields are obtained
by the right-action of $\Omega$ in (\ref{traomega}) on the flux bi-spinor),
\ba
\widehat{F}_0 &=& {P \ov 9}\ ,
\nonumber\\
\widehat{F}_2 &=& \frac{ e^{-3q-2f}}{9 \sqrt{2}}  \Big[ (2K-P{\cal V}) \hat{e}^{\th_1} \wedge \hat{e}^{\phi_1} - P \mathcal{V} \hat{e}^{\prime 1} \wedge \hat{e}^{\prime 2}
+ 3\sqrt{6} P e^{5f}  v_2 \hat{e}^{\prime 1} \wedge \hat{e}^{3} \Big]\ ,
\nonumber\\
\widehat{F}_4 &=& {e^{-6q-4f} \ov 9} \hat{e}^{\th_1} \wedge \hat{e}^{\phi_1} \wedge \Big[ -(e^{6q+4f} P + K \mathcal{V}) \hat{e}^{\prime 1} \wedge \hat{e}^{\prime 2}
 \label{rrfluxescaseKT}\\
&& + 3 \sqrt{3} e^{5f} v_2 (\sqrt{2}  K  \hat{e}^{\prime 1} \wedge \hat{e}^{3} + e^{2f + 3q} P  \hat{e}^{\prime 2} \wedge \hat{e}^{3} ) )\Big]\ .
 \nonumber
\ea
Provided the BPS differential conditions \eqn{bpseqskt1} and \eqn{bpseqskt} are satisfied,
these fluxes obey the Bianchi identities and ensure that the Einstein equations are obeyed.
Notice that the mass $F_0$ is quantised naturally by $P$
which measured the number of fractional branes prior to dualisation.

\no
We also obtain the higher forms
\ba
 \widehat{F}_6 & = &  {e^{-6q-4f} \ov 9} \textrm{Vol}(\textrm{AdS}_5) \wedge \Big( 3 \sqrt{6} e^{5f} K v_2 \hat{e}^{\prime 2}
- 3 \sqrt{3} e^{3q+7f} P v_2 \hat{e}^{\prime 1} + (e^{6q+4f} P + K \mathcal{V}) \hat{e}^{3} \Big)\ ,
\nonumber\\
 \widehat{F}_8 & = & {e^{-3q-2f} \ov 9\sqrt{2}} \textrm{Vol}(\textrm{AdS}_5) \wedge \Big( -3 \sqrt{6} e^{5f} P v_2 \hat{e}^{\th_1} \wedge \hat{e}^{\phi_1} \wedge \hat{e}^{\prime 2}
-  P \mathcal{V} \hat{e}^{\th} \wedge \hat{e}^{\phi} \wedge \hat{e}^{3} \ ,
\nonumber\\
&& + (2 \, K - P \mathcal{V} ) \hat{e}^{\prime 1} \wedge \hat{e}^{\prime 2} \wedge \hat{e}^{3} \Big)
\label{rrfluxescaseKTho}\\
\widehat{F}_{10} & = & -{P \ov 9} \textrm{Vol}_{10}\ ,
\nonumber
\ea
which turn out to be the related to the lower ones as dictated by \eqn{demoll1}.
Not surprisingly,
the  whole structure is very similar to what we have already seen
in the Klebanov--Witten case. In that respect we mention that the dualisation has not
introduced any new singularities to the background in addition to those that might
be initially present, e.g. for the solution \eqn{aguero}. Indeed, notice that the function
$\cal W$ is nowhere vanishing and therefore the string coupling is clearly bounded.

\no
Also useful for us will be the (rather pleasingly simple) expressions for
the RR potentials which are found using \eqn{hdbf2f4}. They read
\ba
 C_{1} &= &  -\frac{1}{27\sqrt{2}} ( Q - 3 P v_3)\s_3 \ , 
\nonumber
\\
 C_{3} & = & 
-\frac{1}{324} \Big(( 6 Q v_3 + T Q - 3 P T v_3) - 9 P (v_2^2 + v_3^2) \Big) \s_1\wedge \s_2\wedge \s_3\ .
\label{oneandthreeforms}
\ea

\subsection{Probing the dual geometry}

In order to learn lessons about the new configuration
described in (\ref{metriccaseKT})-(\ref{rrfluxescaseKT}), we will perform some
calculations with its geometry and fluxes. In the Klebanov--Tseytlin case
these calculation provide an understanding of how field theory features are
encoded in the background. The goal here will be to gain a similar understanding
of the behaviour of a dual field theory using
the geometry and fluxes as a way of defining it.

\no
We will observe that various quantities, when calculated in the transformed
background present qualitatively
similar (or the same) behavior as in the original KT-case.
One may think then that
one is capturing the result of a correlation function
that is "uncharged" under the group of transformations used by the non-abelian T-duality.
Ideas of this sort were used in the context of other background-generating techniques.
See for example \cite{Gursoy:2005cn} and \cite{Maldacena:2008wh}.

\no
We will start by defining a couple of "geometric" cycles.

\subsubsection{Two and Three-Cycles}
In the original KT geometry (\ref{klebatsey})-(\ref{fluxesklebatsey}),
there is a two-cycle defined by
\be
\theta_1={\theta_2}\ ,\qq  \phi_1=2\pi-{\phi_2}\ ,\qq  \psi=\psi_0\ .
\label{xxxwww}
\ee
Notice that the definition above is such that the $U(1)$-fibration
coming from the term
\bea
d\psi+cos\theta_1 d\phi_1+\cos{\theta_2}d{\phi_2}\ ,
\nonumber
\eea
vanishes.
We use this criterion (absence of fibration) to identify a
two cycle in the T-dualised geometry. Let us
consider the sub-manifold defined by
\be
\Sigma_2=[\theta_1,\phi_1]\ , \qq  v_2=v_3=\psi=0\ .
\label{twocycle}
\ee
We can check that the fibration term that appears in the vielbein
$\hat{e}^{3}_+$ vanishes, together with any contribution
coming from $\hat{e}^{\prime 1}_{+}$ and $\hat{e}^{\prime 2}_+$.
Hence, after the T-duality, $\Sigma_2$ is
a well defined two manifold (actually we only need for that
$v_2=0$ and $v_3,\psi=\const$).
The two-cycle of (\ref{twocycle}), will be used below.

\no
Let us now define a three-cycle in the geometry. Consider  the submanifold
\be
\Sigma_3=[\theta_1,\phi_1,\psi]\ ,\qq v_2,v_3, u=\const\
\label{trescycle}
\ee
The three vielbeins $\hat{e}^{i}_+,$ when projected
to this submanifold read,
\ba
\hat{e}_+^{\prime 1}= \frac{\sqrt{12}  v_2  e^{3C+2B}}{81{\cal W}}\sigma_3\ ,
\qq
\hat{e}_+^{\prime 2}=-
\frac{\sqrt{6}  {\cal V}  v_2  e^{2B+C}}{81{\cal W}}\sigma_3\ ,
\qq
\hat{e}_+^3=
\frac{2  v_2^2  e^{B+2C}}{ 9{\cal W}}\sigma_3\ .
\nonumber
\ea
The induced
metric and the NS antisymmetric tensor on the three cycle are
\ba
&& ds_3^2= \frac{e^{2C}}{6}(d\theta_1^2 +\sin^2\theta_1 d\phi_1^2)+
\frac{2 e^{2B+2C} v_2^2}{27{\cal W}}(d\psi+\cos\theta_1 d\phi_1)^2\ ,
\nonumber\\
&&  \widehat{B}_2 \big |_{\S_3}=-\frac{T}{6\sqrt{2}}\sin\th_1 d\th_1 \wedge d\phi_1,
\ea
while the expression of the dilaton is given in \ref{eq:dilatonKT}.

\no
We propose that the cycles above will play an important role
in the study of this geometry and some of its field theoretical aspects.
To make this claim more solid, we will study possible field
theory observables, computed by brane probes
that partially wrap the cycles  above. We will start with domain walls.

\subsubsection{Domain Walls}\label{domainwallssectionzz}
All the material in this section should be taken with an important caveat:
 domain walls are
characteristic effects of the IR dynamics, while here
we have a singular geometry in IR (the singular behaviour inherited from KT).
We  will derive expressions that in principle, should be
evaluated at the origin of the radial coordinate, which  would lead to
ugly divergences. But that is not something of concern as the calculation
should be performed in the backgrounds we obtain once we consider the
non-singular geometries---see Section \ref{sectiondelosbackgroundscompletos}.
The point we want to understand what object should be identified with a domain wall
after the duality.
We will observe that in the simplified KT-case of this section, it is easy to
identify the probe whose interpretation in the dual QFT will be that of a domain wall.

\no
A domain wall in the KT-background (\ref{klebatsey})-(\ref{fluxesklebatsey})
is defined as a D5-brane extending along the manifold
\beq
\Sigma_6=[R^{1,2},{\theta}_2,{\phi}_2,\psi]\ .
\eeq
We calculate the corresponding Born--Infeld action for this D5-brane
and obtain
\bea
& & ds_{\rm ind,D5}^2= e^{2A-\frac{2(B+4C)}{3}}dx_{1,2}^2
+
\frac{e^{2C}}{6}(\s_1^2+\s_2^2) + \frac{e^{2B}}{9} \s_3^2 \ ,
\qq B_{\theta_2 \phi_2}=\frac{T}{6\sqrt{2}}\sin\theta_2\ ,
\nonumber\\
& & S_{\rm BI}=-T_{\rm eff} \int d^{2+1}x\ ,
\eea
where the effective tension is
\bea
T_{\rm eff}= T_{\rm D5} \left. \frac{ 8\pi^2 e^{3A-4C}}{9\sqrt{2}}\sqrt{2 e^{4C} + T^2} \right|_{u_0} \ .
\label{fkjeff}
\eea
In the computation we kept $\th_1,\phi_1$, $u=u_0$
as well as the extra Poincar\'e coordinate $x_3$ fixed.

\no
After the non-Abelian T-duality transformation we need to
specify what object will represent a domain wall.
We may reason as follows:
we started with a D5-brane and non-Abelian T-dualize in the three
directions of the brane $({\theta}_2,\phi_2,\psi)$, so we end up with a D2-brane.
We then propose that the domain wall in the new geometry is
represented by a D2-brane extended in $R^{1,2}$ and placed at a particular point
in the internal space.
We will set the rest of the coordinates
\bea
v_2=v_3=\theta_1=\phi_1=\psi=0\
\label{hart}
\eea
and keep the extra Poincar\'e coordinate $x_3$ and the holographic coordinate $u$ fixed.
Let us calculate the induced metric and BI action for this
putative probe D2-brane. For this we will need that
\beq
 {\cal W}|_{v_2=v_3=\psi=\phi_1=\theta_1=0}=\frac{e^{2B}(2 e^{4C} +T^2)}{81}\
\nonumber
\eeq
and to recall \eqn{dilakt}. We then calculate the corresponding BI action
\be
 ds_{\rm ind, D2}^2= e^{2A-\frac{2B+8C}{3}}dx_{1,2}^2 \ , \qq
S_{\rm D2}=-T_{\rm eff}\int d^{2+1}x \ ,
\ee
with
\be
T_{\rm eff}=
T_{\rm D2}\frac{e^{3A-4C} \ (2 e^{4C}+T^2)^{1/2}}{9}\ .
\label{fkjeff2}
\ee
The two effective tensions in \eqn{fkjeff} and \eqn{fkjeff2} are in agreement up to a constant factor!
This suggests that when we deal with the
whole non-singular KS/baryonic branch background--after transformed under
non-Abelian T-duality, see Section \ref{sectiondelosbackgroundscompletos}--
the D2-brane we studied here will be an actual domain wall in the QFT dual to the
transformed non-Abelian T-dual geometry.

\no
Another quantity that in a less subtle way
will behave similarly (will be "uncharged" under the non-Abelian duality)
is the Wilson loop. Let us comment briefly upon that.

\subsubsection{Asymptotics of $R^{1,3}\times R$ and Wilson loops.}
\label{asintoticseccion}
Let us study the asymptotic behavior of
the Minkowski and radial part of the space by specializing the internal coordinates as
in (\ref{hart}).
Then the metric
of the five-dimensional space i.e. (\ref{klebatsey})-(\ref{klebatsey1})
 reads,
 \be
 e^{-\frac{2}{3} (B + 4 C) } ( du^2 + e^{2A} dx^2_{1,3})\ .
  \label{5dimin}
 \ee
There is no mixing term coming from the "internal" manifold (the $\widehat B$ field does not induced
a new term in the $g_{uu}$ component of the metric).
This implies that asymptotically, the space will
behave like the Klebanov--Tseylin one and that a simple rectangular Wilson loop \cite{Maldacena:1998im},
calculated as a string on the configuration
\beq
x=\sigma\ , \qq t=\tau\ ,\qq u=u(\sigma)\ ,
\label{kompany}
\eeq
will proceed along the same lines as before the T-duality transformation is performed
(for general formulas see \cite{Brandhuber:1999jr,Sonnenschein:1999if}).
We could think that this particular Wilson
loop is  'uncharged' under the duality.
Hence, the short distance
behavior of the quark-antiquark potential will be
the same in the transformed dual field theory, exhibiting a
Coulombic behaviour with a logarithmical running charge.

\no
Let us analyse the radial behavior of the dilaton.
Using (\ref{hart}) the transformed dilaton (which---we remind--- is constant in the KT case) goes like,
\bea
e^{-2\widehat{\Phi}}= {\cal W}=
\frac{e^{2B}(2 e^{4C}+ T^2)}{81} \sim [\ln(r/r_*)]^{5/2}\ ,
\nonumber
\eea
where we have indicated the behaviour for the specific solution \eqn{aguero} (with $b_0=0$)
and similarly in the following two expressions.

\no
Let us finally propose an object that calculates the 't Hooft line (and the potential
between two magnetic monopoles).
In the KT geometry, this is usually identified with the 'Wilson loop' for an effective string associated
with a D3-brane probe that extends on $R^{1,1}$ and wraps the two-cycle in (\ref{xxxwww}).
The presence of the NS
B-field must be taken into account. We get an effective tension given by
\beq
T_{\rm eff}=\frac{4\pi}{3\sqrt{2}} \; T_{\rm D3} \; e^{2A-\frac{2}{3}(B+4C)}\sqrt{2 e^{4C}+ T^2} \sim r^2 [\ln (r/r_*)]^{1/2}\ .
\eeq
The last behaviour is valid for large values of the radial coordinate, so our previous comment on domain walls applies here too.
(we should actually be calculating these quantities in
backgrounds without IR singularity, like those in Section \ref{sectiondelosbackgroundscompletos}).

\no
After the non-Abelian T-duality, we propose that the same observable is computed by
extending a D4-brane on $R^{1,1}$ and wrapping
it over the three cycle $\Sigma_3$
in (\ref{trescycle}). The effective tension in this case is given by
\beq
T_{\rm eff} = \frac{2\pi^2 v_2}{9\sqrt{3}} \; T_{\rm D4} \; e^{2A-\frac{2}{3}(B+4C)}\sqrt{2 e^{4C}+ T^2} \; e^{B+C} \sim r^2 \ln (r/r_*)\ ,
\eeq
where, as discussed below (\ref{metriccaseKT}), we have used that the range
of the angle $\psi$ is $[0,\pi]$ to avoid a bolt-singularity.
In summary, for the 't Hooft loop (defined this way),
we are not obtaining exactly the same functional form.
It remains to study the functional form and values in the IR-smooth case,
which we will do in Section \ref{seccion7deanalysis}.

\no
We will now study two other quantities that when computed
before and after the non-Abelian duality show
the same qualitative behaviour in a quite interesting way.

\subsubsection{Central Charge and Entanglement Entropy}\label{centralentanglementxxx}

To assess the central charge we follow the procedure explained
in \cite{Freedman:1999gp}.  If it is implemented in full generality one should
reduce to a one-dimensional action depending on the unknown functions
entering into the solution.  In turn this should then be recasted as
a five-dimensional gauged supergravity from which the central charge function may be determined.
Fortunately \cite{Klebanov:2007ws} spares us of this technically challenging
reduction by giving some general results that after a small modification are applicable here as well.
For a ten-dimensional metric of the form
\bea
ds^2= \alpha \; dx_{1,3}^2+\alpha \beta \; du^2 + g_{ij} \; d\Theta^i d\Theta^j
\nonumber
\eea
and calling
\bea
V_{\rm int}=\int d\vec{\Theta} \; e^{-2\Phi} \sqrt{g_{ij}}\ ,
\nonumber
\eea
we can define the functions
\bea
H=  V_{\rm int}^2 \; \alpha^3\ ,\qq  \kappa=H^{1/3}\ ,
\nonumber
\eea
such that the Einstein frame five dimensional metric is
\bea
ds_5^2= \kappa( dx_{1,3}^2+\beta \; du^2)\
\nonumber
\eea
and this in turn, implies a central charge
\bea
c=27 \beta^{3/2}\frac{H^{7/2}}{(H')^3}\ .
\nonumber
\eea
Let us test this formula above for our KT-metric in (\ref{klebatsey}). We have
\bea
& & \alpha= e^{2A-2/3(B+4C)}\ ,\qq  \beta=e^{-2A}\ ,\qq V_{\rm int}^2=
\frac{(4\pi)^6 e^{8C+2B}}{36^2\times 9}\ ,
\nonumber\\
& &
c=\frac{2\pi^3}{27A'^3}\ \sim\ \frac{1}{A'^3} .
\eea
Now, let us analyze things {\it after} the non-Abelian T-duality.
The functions are
\bea
\alpha= e^{2A-2/3(B+4C)}\ ,\qq \beta=e^{-2A}\ .
\nonumber
\eea
After a short calculation one finds that
\bea
& & V_{\rm int}= 6\sqrt{2}\pi^2 \; e^{B+4C}\int dv_3\int dv_2 e^{-2\Phi}
\frac{v_2}{81 {\cal W}} \ ,
\nonumber\\
& & H=24^2 ( \frac{\pi^4}{2})
e^{6A}\Big[  \int dv_3\int dv_2 \frac{v_2}{81} \Big]^2
\nonumber
\eea
and this will in turn produce a central charge
\be
c\sim \Big[  \int dv_3\int dv_2 \frac{v_2}{81} \Big] \frac{1}{A'^3}\ .
\label{cdual}
\ee
At this point one can immediately
see that the central charges before and after duality
match up to a single RG scale {\it independent} coefficient (the integral appearing in the result above).
In fact this is not a coincidence. As discussed in (\ref{fadeevterry}), the "measure"
$e^{-2 \phi} \sqrt{g}$ is an invariant of the duality up to a factor arising from the Fadeev--Popov determinant.
Indeed the integral in \eqn{cdual} can be understood as the space time integral of the FP determinant and is   completely determined by the global properties of the dual coordinates.
This being so, follows the invariant of the central charge in the manner displayed above.

\no
In exactly the same way the central charge was analysed, we
can study the Entanglement Entropy \cite{Ryu:2006bv}.
Following equations (7)-(10) of \cite{Klebanov:2007ws}, we see that the integral
defining the entangled entropy (for the case of the Klebanov--Tseytlin background) is
\bea
\frac{S}{V_2}=\frac{1}{4G_{10}}\frac{(4\pi)^3 e^{-2\Phi_0}}{108}
\int_{-l/2}^{l/2} e^{3A}\sqrt{1+ e^{-2A}(\partial_x {u})^2} \ dx \ .
\eea
While if we compute things after the
non-Abelian T-duality, for the same reasons
as explained around (\ref{fadeevterry}) we will have
a nice cancellation of the transformed dilaton and the
involved combination we called ${\cal W}$.
We will then obtain, for the entanglement entropy after the non-abelian
T-duality,
\bea
\frac{S}{V_2}=\frac{24\pi^2
\Big[ \int dv_3 \int dv_2 \frac{v_2}{81}  \Big]}{16 G_{10}}
\int_{-l/2}^{l/2} e^{3A}\sqrt{1+ e^{-2A}(\partial_x {u})^2} \ dx \ .
\eea
We see again, that like with many other quantities described above
the non-Abelian T-duality preserves the dynamical
content of the central charge and the entanglement entropy.
These will behave equally in the original Klebanov-Tseytlin
cascading theory and in its non-abelian T-dual field theory.

\no
Let us close this sub-section
giving another argument
explaining why the central charge (and similarly the entanglement entropy)
should be invariant under the non-abelian transformation.
In the case of the flow between
$AdS_5 \times S^5/\mathbb{Z}_2$ and $AdS_5 \times T^{1,1}$ the corresponding central charges obey
\be
\frac{c_{\rm IR}}{c_{\rm UV}} = \frac{27}{32} \ .
\label{2732}
\ee
After non-abelian T-duality, the corresponding "UV"   geometry will just be some
$\mathbb{Z}_2$ quotient of the Gaiotto-Maldacena type geometry found
in \cite{Sfetsos:2010uq} and the "IR"  geometry the result provide in the preceding section (the Klebanov-Witten transformed).
One then can say that the ratio of  $c_{\rm IR}/c_{\rm UV}$ in \eqn{2732}
will also be preserved.  Indeed exactly this result was obtained directly
in   \cite{Bah:2012dg}. It is satisfying to see a
different interpretation for it.

\no
We will now focus on a quantity
that will display a qualitatively different behavior
before and after the non-abelian T duality. This is probably linked to the
changing behavior of the dilaton (respect to the
constant value in KT) obtained in (\ref{rrfluxescaseKT}).
We will move on to study a possible definition of the gauge coupling in the
dual QFT of the background of eqs.(\ref{metriccaseKT})-(\ref{rrfluxescaseKT}).

\subsubsection{Definition of a 4-d gauge coupling}\label{subsectionongaugecoupling}
We want to give a possible definition of the gauge
coupling of the field theory dual to the geometry of (\ref{metriccaseKT}).
We begin by reviewing how things work in the KT-background.

In the case of the Klebanov-Tseytlin background, one
defines two gauge couplings
in terms of the quantity $b_0$,
\bea
& & b_0=\frac{1}{4\pi^2}\int_{S^2} B_2\ ,
\label{couplingskt}
\nonumber\\
& & \frac{8\pi^2}{g_1^2}= \pi e^{-\Phi}[1 + b_0]\ ,\qq
\frac{8\pi^2}{g_2^2}= \pi e^{-\Phi}[1 - b_0]\ ,
\nonumber\\
& & \frac{4\pi^2}{g_+^2}= \pi e^{-\Phi}=\pi\ ,\qq \frac{4\pi^2}{g_-^2}=\pi e^{-\Phi}b_0 \ ,
\nonumber
\eea
where we have defined also the diagonal combinations
$g_{\pm}^2$ in the usual way.
These definitions arise when considering
String Theory on the conifold (actually for strings on $AdS_5 \times S^5/Z_2$).
In the context of the KT-background,
there are more practical ways of getting the information
encoded by the coupling $g_{-}^2$:
\begin{itemize}
\item{To consider the Action of an instanton
$e^{-S_{\rm inst}} = e^{-\frac{8\pi^2}{ g^2} + i\Theta}$ and equate it with
the Action of an euclidean D1 brane wrapping the two-cycle
of (\ref{xxxwww}). The presence of
the background $B$-field needs to be considered in the BIWZ Action.}
\item{
To consider a probe D5 brane that wraps the
two cycle in (\ref{xxxwww}) and that contains a gauge field
in the Minkowski part of its worldvolume.
It is again crucial
to take into account
the effect of the $B_2$-field with nonzero projection
on the two-cycle and worldvolume of the brane}
\end{itemize}
In summary, the calculation gives (in both cases described above),
\beq
\frac{1}{g^2}\sim 2\pi \; T_{D5} \; h(r)^{1/2} \; e^{2C}
\sqrt{1 + \frac{B_{\theta_2\phi_2}^2 }{h(r) e^{4C}}}
\sim\ln r\ .
\label{couplingKT}
\eeq
The high energy/large radius
logarithmic behavior is the expected one in a four dimensional QFT.

After the non-abelian T-duality have acted upon the KT-background,
we would like to define the gauge coupling
in the dual QFT.
We find again two possible definitions (that as above will agree):
one in terms of a D2 brane that wrapping
our three-cycle of (\ref{trescycle})
will behave as an instanton, the other in terms of
a D6 (with a gauge field in its Minkowski directions)
that wraps the same three-cycle.
In broad lines the calculations go as explained below.
The first definition considers an instanton and uses that
$
e^{-S_{\rm inst}}\sim e^{-\frac{8\pi^2}{g^2} +i\Theta}.
$
An instanton in our background is given by an
euclidean D2 brane
that wraps the three cycle described above. Its action will be
\beq
S_{\rm D2}=-T_{\rm D2}\int_{\Sigma_3} e^{-\widehat{\Phi}}
\sqrt{\det[g_{ab} +B_{ab}]} +
T_{\rm D2}\int_{\Sigma_3} C_3\ .
\eeq
So, we can associate
\bea
\frac{1}{g^2}\sim -T_{\rm D2}\int_{\Sigma_3} e^{-\widehat{\Phi}}
\sqrt{\det[g_{ab}+B_{ab}]}\ ,\qq
\Theta\sim \int C_3\ .
\label{couplingtheta}
\eea
The induced metric and the $B_2$ field that are relevant for this calculation are
\bea
& & ds_3^2= \frac{e^{2C}}{6}(d\theta_1^2+\sin^2\theta_1 d\phi_1^2)+
\frac{2 e^{2B+2C} v_2^2}{27 {\cal W}}(d\psi+\cos\theta_1 d\phi_1)^2\ ,
\nonumber\\
& & B_2=-\frac{T(u)}{6\sqrt{2}}\sin\theta_1 d\theta_1 \wedge d\phi_1\ .
\label{b2metric}
\eea
So, we can calculate
\beq
\frac{1}{g^2}\sim T_{\rm D2} \ \frac{\pi^2 e^{B+C} v_2}{9\sqrt{3}}
\sqrt{2e^{4C} + T^2}\ .
\label{gaugecoupling}
\eeq
Notice that the units are correct. The tension of the D2
together with the volume of the three cycle cancel to give something
dimensionless.
Similarly, with the $C_3$ field associated
with $\widehat{F}_4$---see (\ref{oneandthreeforms}), we can calculate the anomaly
associated with the changes in the theta-angle.

{\it The  definition in terms of D6's:}

Suppose that we consider a D6 brane wrapping the three cycle $\Sigma_3$
extended along $R^{1,3}\times (\theta_1,\phi_1,\psi)$.
We will also turn on  a gauge field on the Minkowski part of the D6
\footnote{It is enough to consider just the $F_{tx}$ component to see
the argument working.}.
We will now calculate
\beq
S_{\rm D6}=-T_{\rm D6}\int_{M7} e^{-\widehat{\Phi}}
\sqrt{\det{g_{ab}-(2\pi \alpha')F_{ab} + B_{ab}  } }
+ T_{\rm D6} \int_{R^{1,3}} \int_{\Sigma_3} C_7+ C_3\wedge F_2\wedge F_2\ .
\eeq
If the three-cycle is calibrated, then in an expansion for small
values of $\alpha'$, the first term coming from the square root and the one coming
from the WZ term proportional to $C_7$ should cancel.
It would be interesting to check if this is the case.
The second term in the expansion should give the gauge coupling
and  the anomaly,
\bea
\Theta\sim \int_{\Sigma_3} C_3\ .
\nonumber
\eea
Focusing on the gauge coupling,
the induced metric and
action are
\bea
\det[g_{ab}+B_{ab} +2\pi\alpha' F_{ab}]=
g_{xx}^4\Big(1-g^{tt} g^{xx} F_{tx}^2
\Big)\frac{ e^{2B+2C}v_2^2 \sin^2\theta_1}{972 {\cal W}}[2e^{4C}
+T^2]. \nonumber
\eea
Now we can expand for small values of $\alpha'$ or small
electromagnetic fields to get
\ba
S_{\rm D6} = \frac{1}{g_4^2} \int d^{3+1}x \ F^2_{tx} \ ,
\ea
where the prefactor of the Maxwell term is identified with the gauge coupling which is found to be
\beq
\frac{1}{g_4^2}\sim
T_{\rm D6}(4\pi^2)(\pi\alpha')^2
\frac{ e^{B+C} v_2}{9\sqrt{3}}
\sqrt{2e^{4C} + T(u)^2}\ .
\eeq
Notice that aside form the constant-factors
we have the same expression as in (\ref{gaugecoupling})
---using a different initial definition.
As with the previous definition
the units are correct.
To close this section, let us observe that
if we fix $v_2$ to a constant and we use the leading asymptotics,
we will have a behavior for the
gauge coupling
\beq
\frac{1}{g^2}\sim (\ln r)^{3/2}\ .
\eeq
This rather strange scaling perhaps suggests
that the field theory might be different from a conventional field theory.
Comparing with (\ref{couplingKT}),
it is clear that the new gauge coupling is
not behaving as a typical coupling in a 4d-theory. Another possibility
is that the quantity we have defined is not related
to the four dimensional gauge coupling of the QFT.

In hand with the non-conventional beta-function derived from
the above behaviour, we can study anomalies, associated
with changes in the $\Theta$-angle. Indeed, with our definition of the
gauge coupling, naturally comes a definition for the $\Theta$ angle, as we stressed in (\ref{couplingtheta}).
To calculate the integral of the RR-three form on the three-cycle defined in (\ref{trescycle}), we use the expression in
(\ref{oneandthreeforms}). We will focus in the case $Q=0$---well motivated, as this is the case in the smooth geometries
that IR-complete the KT-background---and we will also choose $v_3=0$.
The result of the $\Theta$ angle is
\beq
\Theta\sim \int_{\Sigma_3} C_3=\frac{\pi^2}{9}P v_2^2\ .
\eeq
Hence, changes of $\Theta$ in integer multiples of  $2\pi n$ imply a
periodicity in the $v_2$ coordinate or a quantisation on the changes
of $v_2$.
Similar reasoning applied to the KT-background gave a result for a anomalous breaking of the R-symmetry.
In this case, we  emphasise that the symmetry is not  the $U(1)_R$ R-symmetry (associated with translations in the angle $\psi$).

So, we have analysed different dynamical quantities with the goal of
narrowing or defining the possible field theory dual to our background in
eqs.(\ref{metriccaseKT})-(\ref{rrfluxescaseKT}).
It is of interest to analyse
the behavior of quantities that are either gauge invariant,
 conserved or quantized
like Maxwell and Page charges. We turn to this now.

\subsection{Maxwell and Page Charges}
Before the non-abelian T-duality,
we calculate the Maxwell and Page charges
\cite{Marolf:2000cb} of D3 and D5 branes,
\bea
& & Q_{\rm Max, D3}
=\frac{1}{(16\pi^4)}\int_{\theta_1,\phi_1,{\theta}_2,{\phi}_2,\psi}
F_5=\frac{K(r)}{27\pi}\sim  N_c\ln r\ ,
\nonumber\\
& &
Q_{\rm Max, D5}=\frac{1}{4\pi^2}\int_{{\theta}_2,{\phi}_2, \psi}F_3=
\frac{\sqrt{2}}{9}P\ .
\label{maxwellchargesd3d5}
\eea
This shows  the usual  logaritmic growth of the D3 brane charge, linked with
the logarithmic deviation of the geometry from the $AdS_5$
Klebanov-Witten fixed point. Also, the
Maxwell charge of D5 branes is quantized.
We can also compute the Page charges,
\bea
& & Q_{\rm Page, D5}=\frac{1}{4\pi^2}
\int_{{\theta}_2,{\phi}_2, \psi}F_3- B_2\wedge F_1=
\frac{\sqrt{2}}{9}P\ ,
\label{pagechargesktd3d5}\\
& & Q_{\rm Page, D3}=
\frac{1}{(16\pi^4)}\int_{\theta_1,\phi_1,{\theta}_2,{\phi}_2,\psi}
F_5- B_2\wedge F_3
=\frac{K+T P}{27 \pi}
=\frac{Q}{27\pi}\ .
\nonumber
\eea
The quantity $Q$ is usually taken to zero, it
indicates the number of "free/mobile" D3 branes on the conifold.
In the full solution with a good IR behavior
(the Klebanov-Strassler background or its baryonic branch counterpart),
one takes $Q=0$, precisely to avoid singularities.

All this analysis is valid and standard
before the non-abelian T-duality. Let us analyse things after the duality.
We follow this logic: we have D3 branes to begin with.
We will perform the non-abelian T-duality
in the directions $[{\theta}_2,{\phi}_2,\psi]$.
This will generate D6 branes. To calculate the Maxwell charge of
D6 branes we should integrate
the expression for $F_2$ specialized
on the cycle in (\ref{twocycle})\footnote{The nomalization factor
is chosen arbitrarily
to match with previous expressions},
\beq
Q_{\rm Max, D6}= \frac{1}{\sqrt{2}\pi^2}\int_{\theta_1\phi_1} F_2
=\frac{K+Q}{27\pi}\ .
\eeq
We should compare this with the Maxwell charge of D3 branes,
in (\ref{maxwellchargesd3d5}).
Following the same logic in the case of the D5 branes, we can  define
\bea
Q_{\rm Max, D8}=\sqrt{2}\int F_0= \frac{\sqrt{2}P}{9}\ ,
\nonumber
\eea
that can be put in correspondence with $Q_{Max, D5}$ in
(\ref{maxwellchargesd3d5}).
Let us now calculate Page charges\footnote{The Page charge of D4 branes is formally given by,
\bea
Q_{\rm Page, D4}=\frac{1}{\mu_{\rm D4}}\int F_4-B_2\wedge F_2+\frac{1}{2}B_2\wedge
B_2\wedge F_0;
\nonumber
\eea
but its interpretation is not clear, as we did not find a well defined
4-cycle to integrate on.
},
\bea
&& Q_{\rm Page, D6}=\frac{1}{\sqrt{2}\pi^2}\int_{\theta_1\phi_1}
F_2- F_0 B_2=\frac{2Q}{27\pi} \ ,
\nonumber \\
&& Q_{\rm Page, D8}= \frac{\sqrt{2}P}{9} \ .
\nonumber
\eea
Notice that we can make a correspondence also
between Page charges (before and after the duality),
if we choose $Q=0$ as above.

Using these Page charges, one can play a game similar to the one in
\cite{Benini:2007gx}, to get a hint on what is
the fate of Seiberg duality after the non-abelian T-duality.
To this we now turn.

\subsection{Maxwell, Page and Seiberg.}
In this section we study the fate of Seiberg duality, after the non-abelian T-duality.

We will define the geometric version of
"Seiberg duality" in the Klebanov-Tseytlin quiver
(before the non abelian T-duality), to be the operation
that changes in integer units, the Maxwell charges
(after a given change in the radial coordinate)
or equivalently, as the operation that changes the Page charge
after  suitable large gauge transformation in the NS-B field
(at a fixed value of the radial coordinate).
Both these equivalent definitions were introduced in
\cite{Benini:2007gx}. We will follow that logic after
the non-abelian T-duality and we will
learn that in the generated background/dual QFT,
there seems to be a Seiberg duality at work.
The result suggests some ideas for the generated quiver field theory.

\subsubsection{Seiberg duality before dualisation}

Let us start by summarising the approach to understanding the geometric realisation of Seiberg duality in the background prior to dualisation.  Before the T-duality we have the D3 charges given by
\bea
&& Q_{\rm Max, D3}=\alpha \int_{X_5}F_5= \frac{4^3\pi^3\alpha K(r)}{108} \ ,
\nonumber \\
&& Q_{\rm Page, D3}=\alpha \int_{X_5}F_5 - B_2\wedge F_3=
\frac{(4\pi)^3 \; \alpha }{108} \Big( K(r) + P T(r)\Big) \ .
\nonumber
\eea
We have left the coefficient $\alpha$
undetermined. From the previous subsection we know that
$
{16\pi^4} \alpha=1$ and $ K(r)= Q_{free}- P T(r)$.
The restriction of the NS field on the two cycle of (\ref{xxxwww}) gives
\bea
B_2|_{\Sigma_2}=\frac{T(r)}{3\sqrt{2}}\sin\theta_1 d\theta_1 \wedge
d\phi_1 .\nonumber
\eea
From which we can define
\bea
b_0=\frac{1}{4\pi^2}\int_{\Sigma_2} B_2=\frac{T(r)}{3\pi \sqrt{2}} \ . \nonumber
\eea
The logic to follow will be this:
we will consider two ways of changing $b_0$ in $n$ (integer) units,
\beq
b_0\to b_0 \pm n
\label{cambiob0}
\eeq
\begin{itemize}
\item{by changing $\Delta T(r)=\pm 3\pi\sqrt{2} n$}
\item{by changing \bea
B_2\to B_2 \pm \frac{n\pi}{2}(\sin\theta_1 d\theta_1 \wedge d\phi_1
- \sin\theta_2 d\theta_2 \wedge d\phi_2) \ .
\nonumber
\eea}
\end{itemize}
The change in $T(r)$ occurs as the radial
coordinate varies. We will apply this change to the Maxwell charge.
The large gauge transformation is performed at fixed
radial coordinate. We will apply this to the Page charge.
The idea is that both changes are the equivalent
of performing a Seiberg duality \cite{Benini:2007gx}.

Let us first apply $\Delta T=\pm 3\pi\sqrt{2} n$ to the
Maxwell charge. We obtain,
\bea
& &\Delta Q_{\rm Max}=
\alpha \frac{(4 \pi)^3}{108} \Delta K=
\alpha \frac{(4 \pi)^3}{108} (\Delta Q_{\rm free}- P \Delta T)=
-\alpha \frac{(4 \pi)^3 P}{108}   \Delta T \ ,
\nonumber\\
& &
\Delta Q_{\rm Max, D3}=
\pm \alpha P \frac{4^3 \pi^4}{18\sqrt{2}} n.\nonumber
\eea
Now, we calculate the change in the Page charge under a large
gauge transformation
in the $B_2$ field.
\beq
\Delta Q_{\rm Page, D3}=-\alpha \int_{X_5}\Delta B_2 \wedge F_3=\pm \alpha P n
\frac{4^3 \pi^4}{18\sqrt{2}}.\label{cambiopaged3}
\eeq
So, we see that under these two different changes,
the change in the Page charge equals that
of the Maxwell charge for D3 branes. Plugging specific values of $P,\alpha$
we see that this amount to a change in the charge of $\Delta Q= \pm n M$, this is the effect of $n$ Seiberg dualities on the
Klebanov-Tseytlin-Strassler quiver field theory.

\subsubsection{Calculations after the non-abelian T-duality}
Now, we will review this calculation, after the non-abelian duality.
The object of study are D6 branes whose Page and Maxwell charges we recall are given by
\footnote{If we use
\beq
K= - P T,\;\;\; P= 9 F_0,\;\;\; \hat{\alpha} \sqrt{2}=\alpha 16 \pi^2.
\label{relations}\eeq
we obtain that Maxwell and Page charges are invariants under the
non-abelian T-duality, as we saw in the previous section.},
\bea
& & Q_{\rm Max, D6}=\hat{\alpha}\int_{\Sigma_2} F_2 = \hat{\alpha}
\frac{4\pi}{54\sqrt{2}}(2K(r) + P T(r)). \nonumber\\
& &
Q_{\rm Page, D6}=\hat{\alpha}\int_{\Sigma_2} F_2 -F_0 B_2= \hat{\alpha}
\frac{4\pi}{54\sqrt{2}}(2K(r) + P T(r) + 9 F_0 T).
\eea
After the non-abelian duality, we have that
\bea
b_0=\frac{1}{4\pi^2}\int_{\Sigma_2}B_2= -\frac{T(r)}{6\pi\sqrt{2}}.\nonumber
\eea
For an integer $k$,
the two possible ways of changing
$\Delta b_0=\pm k$ are by moving radially such that
\[ \Delta T=\mp 6\pi\sqrt{2} k, \nonumber \]
or by performing a large gauge transformation
\[     \Delta B_2
= n \pi \sin\theta d\theta_1 \wedge d\phi_1.\nonumber
\]
We will now calculate, the change in the Maxwell charge,
under a change in $\Delta T$ and the
change in the Page charge, under a change in the $B_2$ field as specified.
We have
\bea
 \Delta Q_{\rm Max, D6}= -\hat{\alpha}\frac{4\pi}{54\sqrt{2}} \; P \Delta T(r)= \pm \hat{\alpha}\frac{4\pi^2}{9} P \; k \ .
\eea
We now calculate the change in the Page charge, under a large
gauge transformation in $B_2$ and we get,
\bea
\Delta Q_{\rm Page, D6}=\pm \hat{\alpha} \;  4\pi^2 F_0 \; k=
\pm \hat{\alpha} \frac{4\pi^2}{9} P \; k \ .\label{utilisima}
\eea
We used eq.(\ref{relations}). We then see that
\bea
\Delta Q_{\rm Max, D6}|_{\Delta T(r)}=
\Delta Q_{\rm Page, D6}|_{\Delta B_2}.\nonumber
\eea
Using the values for $\alpha, \hat{\alpha}$ in (\ref{relations}), we
observe that whenever the Page charge of
D3
changed in $2M$ units, the Page charge of D6 branes---after the
non-abelian duality---
changes by $M$ units.
Indeed, comparing (\ref{cambiopaged3}) with (\ref{utilisima})
\beq
\Delta Q_{\rm Page, D3}=\pm \alpha P n \frac{4^3 \pi^4}{18\sqrt{2}},\;\;\;\; \Delta Q_{\rm Page, D6}=
\pm \hat{\alpha}\frac{4\pi^2}{9} P \; k.
\eeq
Hence, since we associate changes in
the Page charge by $M$ units as a
Seiberg duality applied on the quiver---before
the non-abelian T-duality. We see that a change in $M$ units in the KT-quiver
reflects in a  change in $2M $ units in the transformed QFT.

This may suggest some ideas for what is the quiver
after the non-abelian T-duality.

\subsection{A  summary of this Section}
Let us summarise the results of this long section:
we have constructed the non-abelian T-dual of
the Klebanov-Tseytlin background.
We learnt about the QFT dual to our new background
in eqs.(\ref{metriccaseKT})-(\ref{rrfluxescaseKT})
by performing different calculations using the geometrical
description of this field theory. Among these, we learn about the
(would-be) domain walls, Wilson and 't Hooft loops, gauge coupling,
central charge, entanglement entropy,
conserved and gauge invariant charges. We also got a glimpse
at the existence of an operation like Seiberg duality. Our
transformed background
is probably dual to a QFT that presents a cascade similar to the one of the
Klebanov-Tseytlin-Strassler background.

The information we gather,
together with the one in the following sections
may help in narrowing or deciding for a given field theoretical description
of the background.

We will now move to study the full non-perturbative
dynamics of this putative QFT. We will do so by analising
the non-abelian T dual of geometries, like Klebanov-Strassler,
the baryonic branch of KS or the background produced by
D5 branes compactified
on the resolved conifold. These are smooth geometries that 'IR-complete'
the KT-case analysed above. As expected, the resulting backgrounds will
be quite involved (mostly due to the fact that we are
loosing the $U(1)_R$ associated with the Killing vector $\partial_\psi$
at the level of the metric). The contents in this section will
give an orientation
about the interesting observables to study.

\section{Dualisation of smooth geometries}\label{sectiondelosbackgroundscompletos}
The purpose of this section
is to apply the non-Abelian T-duality technique to three
trademark backgrounds
in type-IIB string theory, conjectured to be dual to $N=1$ SUSY QFT.
The non-perturbative dynamics (confinement, symmetry breaking, etc)
of the QFT is captured by the backgrounds and in this sense we will refer
to them as "IR-complete" solutions of type-IIB. They are the smoothed out
version of the KT background of the previous section.

\no
In Section \ref{seccion7deanalysis}, we will study QFT
aspects of the backgrounds
spelled out below.

 \subsection{Dualisation of Wrapped D5 solutions}
An important class of theories are the wrapped-brane models.
In what follows we will be interested in the the near brane geometry
of D5 branes wrapping a two-sphere with a twist in the normal bundle to
preserved ${\cal N}=1$ supersymmetry  \cite{Maldacena:2000yy}.
In the very far IR  the gauge theory on the brane,
described in \cite{Maldacena:2000yy}, reduces to pure (i.e. without matter)
${\cal N}=1$ SYM.
In fact this shares many similarities with the
"IR completed" geometry of the Klebanov-Strassler theory
which we turn to afterwards.

 \subsubsection{The wrapped D5 background}\label{wrappedd5seccion}
\def\tom{\tilde{\omega}}
\no First we give the vielbeins of this solution (as ever we are in string frame):
\ba
\label{eq:vielbs}
&& e^{x^{i}}=e^{\Phi/2}dx^{i} \ , \;\; e^{\r}=e^{\Phi/2+k}d\r \ , \;\; e^{\th_1}=e^{\Phi/2+h}d\th_1 \ , \;\; e^{\phi_1}=e^{\Phi/2+h}\sin \theta_1 d\phi_1\ ,
\nonumber\\
&& e^{1}=\frac{1}{2}e^{\Phi/2+g}(\s_{2} + a d\th_1) \ , \;\; e^{2}=\frac{1}{2}e^{\Phi/2+g}(\s_{1} - a \sin\th_1 d\phi_1)\ ,
\\
&& e^{3}=\frac{1}{2}e^{\Phi/2+k}(\s_{3} + \cos\th_1 d\phi_1) \ ,
\nonumber
\ea
%
\no in which the $\s$'s are the $SU(2)$ left invariant one-forms given in (\ref{eq:MaurerCartan}).
%
%
\no This background is supported by a RR three-form
\ba
F_{3} &=& e^{-3\Phi/2}\left( f_{1} \, e^{1} \wedge e^{2} \wedge e^{3} + f_{2} \, e^{\th_1} \wedge e^{\phi_1} \wedge e^{3} + f_{3} \, ( e^{\th_1} \wedge e^{2} \wedge e^{3} +  e^{\phi_1} \wedge e^{1} \wedge e^{3}) \right.
\nonumber\\
      && \left. + f_{4} \, ( e^{\r} \wedge e^{1} \wedge e^{\th_1} + e^{\r} \wedge e^{\phi_1} \wedge e^{2}) \right) \ .
\ea
\no The functions $f$'s above are defined by
\ba
&& f_1 = -2 N_c \, e^{-k-2g}, \;\; f_2 = \frac{N_c}{2} \, e^{-k-2h} \, (a^2 -2ab +1),
\nonumber\\
&& f_3 = N_c \, e^{-k-h-g} \, (a-b), \;\; f_4 = \frac{N_c}{2} \, e^{-k-h-g} \, b^{'} \ .
\ea
\no The dilaton $\Phi$ and the other functions $k, \, g, \, h, \, a, \, b $ depend on the coordinate $\r$ and obey BPS equations the details of which can be found for instance in the appendix of  \cite{Casero:2006pt}.

\subsubsection{The Wrapped D5 dual geometry }\label{TdualwrappedD5}
As before, we perform a T-duality along the $SU(2)$ isometry under which the   $ \s$'s \footnote{The $\s$'s here are related to the $\tilde{\om}'s$ of \cite{Casero:2006pt} in the following way: $\s_1 = \tilde{\om}_2, \s_2 = \tilde{\om}_1$ and $\s_3=\tilde{\om_3}$, while the angles $\th_1, \phi_1$ correspond to the angles $\th,\phi$ and also $\th_2, \phi_2$ here are the tilded angles of the previous paper .} are invariant. As in the previous section we will chose a gauge fixing that sets $\th_2= \phi_2= 0$ and introduces new coordinates $v_2$ and $v_3$ in the dual background.   The duality transformation leaves the frames $\{e^{x^i}, e^\r, e^{\th_1} , e^{\phi_1} \}$ invariant  and acts by sending $e^{i} \rightarrow  \hat{e}^{i}$.
To express compactly
the dual geometry we find it convenient to rotate the
frame fields in the 1-2 plane using the same rotation matrix as in  (\ref{eq:rotation}). Also to express the results it is very helpful to recombine the dependence on the angular coordinates $\theta_1, \phi_1, \psi$ into a new set of $SU(2)$ left invariant forms which we denote by $\omega^i$. \footnote{The $\om$'s here are defined in the same way as the $\s$'s in (\ref{eq:MaurerCartan}) but here instead of the angles $\th_2,\phi_2$ we have $\th_1,\phi_1$.}

 We then find that
\ba
\hat{e}^{\prime 1} &=& \frac{e^{ g+\frac{3}{2} \Phi} }{ 8 {\cal W} }  \Big[ 4 e^{2g} v_2 \left( dv_3 - v_2\, a\, \omega_1  \right)   + \sqrt{2} e^{2g+2k+ \Phi} \left(v_2 \omega_3 - a \, v_3 \omega_2 \right)
\nonumber\\
                   &&- 4 e^{2k} v_3 \left( dv_2 + a \, v_3 \omega_1 \right)  \Big]\ ,
\nonumber \\
\hat{e}^{\prime 2} &=&  -\frac{e^{ g+\frac{1}{2} \Phi} }{ 8 {\cal W} }  \Big[   4 e^{2k + \Phi} v_3  \left(v_2 \omega_3 - a \, v_3 \omega_2 \right)  +  \sqrt{2}  e^{2g+2k+ 2\Phi}   \left( dv_2 + a \, v_3 \omega_1 \right)
\nonumber\\
                   &&+ 8 \sqrt{2} v_2 \left(v_2 dv_2 + v_3 dv_3  \right)   \Big]  ,
\\
\hat{e}^3          &=& -\frac{e^{ k +\frac{1}{2} \Phi} }{ 8 {\cal W} }  \big[  \sqrt{2} e^{4g + 2 \Phi}  \left( dv_3 - v_2\, a\, \omega_1  \right)  - 4 e^{2 g + \Phi} v_2 \left(v_2 \omega_3 - a \, v_3 \omega_2 \right)
\nonumber\\
                   &&+  8 \sqrt{2} v_3    \left(v_2 dv_2 + v_3 dv_3  \right)      \big] \ ,
\nonumber
\ea
and the T-dual metric is given by
\be
d\hat{s}^2 = (e^{x^i})^2 + (e^\rho)^2 + (e^\theta)^2+ (e^\phi)^2  + (\widehat{e}^i)^2 \ .
\ee
The NS-two form field is
\be
\widehat{B}_2  = - \frac{e^{g+ k +\Phi}}{2 \sqrt{2} v_2} \hat{e}^{\prime 1} \wedge \hat{e}^3 +  \frac{e^{-g + k} v_3}{v_2} \hat{e}^{\prime 2} \wedge \hat{e}^3  + \frac{e^{g+ \Phi/2} a }{2} \hat{e}^{\prime 2} \wedge \omega_2 + \frac{e^{k +\Phi/2}  a v_3}{2 v_2 }  \hat{e}^3 \wedge{\omega_2}  \ .
\ee
 The dual dilaton is given by
 \be
 \qq \widehat{\Phi} =  \Phi - \frac{1}{2} \ln {\cal W}\ ,
 \ee
 in which we defined
\be
{\cal W} = \det\widehat{M} = \frac{1}{8} e^{4g+2k+3\Phi} + e^{2g+\Phi}v_2^2 + e^{2k+\Phi}v_3^2 \ .
\ee
This geometry is supported by a cornucopia of RR fluxes with $\widehat{F}_0, \widehat{F}_2$ and $\widehat{F}_4$ all activated,  given by

\ba
\widehat{F}_0 &=& {N_c \ov \sqrt{2}}
\nonumber\\
\widehat{F}_2 &=& e^{g-\Phi} \, f_4 \, v_2 \, (\cos\psi \, e^\rho \wedge e^{\th_1} + \sin\psi \, e^\rho \wedge e^{\phi_1}) - e^{k-\Phi} \, f_2 \, v_3 \, e^{\th_1} \wedge e^{\phi_1}
\nonumber\\
              && -\frac{1}{4} e^{k-\Phi} \, f_3 \, (\sqrt{2} \, e^{2g+\Phi}\sin\psi + 4\cos\psi \, v_3) (e^{\th_1} \wedge \hat{e}^{\prime 1} + e^{\phi_1} \wedge \hat{e}^{\prime 2})
\nonumber\\
              && +\frac{1}{4}e^{k-\Phi} \, f_3 \, (\sqrt{2} \, e^{2g+\Phi} \, \cos\psi - 4\sin\psi \, v_3)(e^{\phi_1} \wedge \hat{e}^{\prime 1} - e^{\th_1} \wedge \hat{e}^{\prime 2})
\nonumber\\
              && +e^{g-\Phi} \, f_3 \, v_2 \, (\cos\psi \, e^{\phi_1} \wedge \hat{e}^3 - \sin\psi \, e^{\th_1} \wedge \hat{e}^3) + e^{k-\Phi} \, f_1 \, v_3 \, \hat{e}^{\prime 1} \wedge \hat{e}^{\prime 2} - e^{g-\Phi} \, f_1 \, v_2 \, \hat{e}^{\prime 1} \wedge \hat{e}^3
\nonumber\\
\widehat{F}_4 &=& e^{g-\Phi} \, f_4 \, v_2 \, e^\rho \wedge (\sin\psi e^{\th_1} - \cos\psi e^{\phi_1}) \wedge \hat{e}^{\prime 1} \wedge \hat{e}^{\prime 2}
\\
              && -\frac{1}{4} e^{k-\Phi} \, f_4 \, (\sqrt{2} e^{2g+\Phi} \cos\psi - 4 \, v_3 \, \sin\psi) e^\rho \wedge (e^{\th_1} \wedge \hat{e}^{\prime 1} + e^{\phi_1} \wedge \hat{e}^{\prime 2}) \wedge \hat{e}^3
\nonumber\\
              && +\frac{1}{4} e^{k-\Phi} \, f_4 \, (\sqrt{2} e^{2g+\Phi} \sin\psi + 4 \, v_3 \, \cos\psi) e^\rho \wedge (e^{\th_1} \wedge \hat{e}^{\prime 2} - e^{\phi_1} \wedge \hat{e}^{\prime 1}) \wedge \hat{e}^3
\nonumber\\
              && +e^{g-\Phi} \, f_2 \, v_2 e^{\th_1} \wedge e^{\phi_1} \wedge \hat{e}^{\prime 2} \wedge \hat{e}^3 - \frac{e^{2g+k}}{2\sqrt{2}} \, f_2 \, e^{\th_1} \wedge e^{\phi_1} \wedge \hat{e}^{\prime 1} \wedge \hat{e}^{\prime 2}
\nonumber\\
              && +e^{g-\Phi} \, f_3 \, v_2 (\cos\psi e^{\th_1} + \sin\psi e^{\phi_1}) \wedge \hat{e}^{\prime 1} \wedge \hat{e}^{\prime 2} \wedge \hat{e}^3
\nonumber
\ea

Using $\tt{Mathematica}$ we have verified that indeed the equations of motion for fluxes, Bianchi identities, dilaton and Einstein's equations are all satisfied on the same set of BPS equations as the original geometry.

\subsection{Dualisation of Klebanov Strassler}\label{sectiononksdual}

\subsubsection{The KS background}
We now consider the Klebanov Strassler solution \cite{Klebanov:2000hb} in which the conifold is replaced with its smooth deformation so that the metric is given by
\be
ds^2_{10}  = h^{-\ha}(\tau) dx_n dx_n + h^{\ha}(\tau ) ds^2_6
\ee
where $ds^2_6$ is the metric of the deformed conifold.  The details of supergravity solution can be found in section 5 of  \cite{Klebanov:2000hb}.  The radial coordinate we now denote by $\tau$ in keeping with the notation of  \cite{Klebanov:2000hb}.\footnote{At large $\tau$ the KS solution asymptotes to the KT solution with a the radial coordinates related by $r^3 \sim  \varepsilon^2 e^\tau$.  }

So as to write the metric in a form compatible with our ansatz we find it convenient to introduce some functions which depend on the radial coordinate. The dictionary between the functions we use here and the original paper is as follows
 \be
 h_1= \varepsilon^{\frac{4}{3}} h(\tau)^\frac{1}{2} K(\tau)\ ,  \quad  h_2 = g_s M (f(\tau) + k(\tau) )\ , \quad  h_3 =  g_s M (f(\tau) - k(\tau) )\ .
 \ee
 The functions $h(\tau),  f(\tau)  , k(\tau) $ obey BPS equations given explicitly in \cite{Klebanov:2000hb} and $K(\tau)$ is a function fixed by the deformed conifold metric.  The appearance of the parameter $\varepsilon$ which describes the deformation of the conifold can be thought of the supergravity dual of dimensional transmutation.

As with the KT solution the geometry is supported by an NS two-form, an RR three form and a RR five form whose details may be found in  \cite{Klebanov:2000hb}.  Let us just remark that the RR three form interpolates between that of the KT solution at large $\tau$ and, to prevent the infinite charge density origin of the singularity, something with support only in the non-shrinking $S^3$ at $\tau = 0$.

As before we perform an non-abelian dualisation along the $SU(2)$ action that acts on the coordinates $\{\psi, \th_2, \phi_2 \}$ and we choose a gauge fixing choice in which $\th_2 = \phi_2 = v_1 = 0$ such that the coordinates of the dual theory are $\{ \tau, x^i , \psi, \th_1 , \phi_1, v_2 , v_3 \} $.  We express our results more
compactly by writing derivatives of the remaining Euler angles as
the left invariant Maurer-Cartan forms   $\omega_i[\psi, \th_1, \phi_1]$
as we did in Section \ref{TdualwrappedD5}.

\subsubsection{The KS dual geometry}
To express compactly
the dual geometry it is convenient to introduce the combinations
\def\cV{{\cal V}}
\def\cW{{\cal W}}
\def\cU{{\cal U}}
\ba
&\cV = h_2 + \sqrt{2} v_3 \ , \quad \cU = \cosh \tau h_3 - \cV \ , \nonumber \\
& \cW =  \det M = \frac{h_1}{12 K^3} \left( \cosh^2 \tau h_1^2 + 4 \cV^2 + 12 \cosh \tau K^3 v_2^2 \right)\ ,
\ea
and the one-forms
 \be
 \L_1 = -\cU \om_2 - \sqrt{2} \cosh \tau v_2 \om_3 \ , \quad \L_2  = \cU \om_1 - \sqrt{2} \cosh \tau dv_2 \ , \quad \L_3 = \cosh \tau dv_3 - v_2 \om_1 \ .
 \ee
Furthermore we perform a rotation of the dual frame in the 1-2 plane as we did in (\ref{eq:rotation}).

The frame fields in the dual are given by
\ba
\hat{e}^{\prime 1} &=&  \frac{h_1^{\frac{3}{2}} }{12 \cW K^3 \cosh^\ha \tau}  \big[ - h_1\cosh \tau \Lambda_1  + 2 \cV \Lambda_2 + 6\cosh \tau  K^3 v_2 \Lambda_3 \big]\ , \nonumber \\
\hat{e}^{\prime 2} &=&  \frac{h_1^{\frac{1}{2}} }{12 \cW K^3 \cosh^\ha \tau} \big[2h_1 \cV \Lambda_1 + h_1^2 \cosh\tau \Lambda_2 + 12 K^3 v_2 (v_2 \Lambda_2 - \cV \Lambda_3 ) \big] \ , \\
\hat{e}^3 &=&- \frac{h_1^{\frac{1}{2}} }{4\sqrt{3} \cW K^\frac{3}{2} \cosh  \tau} \big[ 2 h_1 v_2 \cosh \tau  \Lambda_1  -4 \cV (v_2 \Lambda_2 - \cV \Lambda_3 )  + h_1^2 \cosh^2 \tau \Lambda_3 \big]   \ ,\nonumber
\ea
so that the dual metric is given by
 \be
d\hat{s}^2 = h^{-\ha}(\tau) dx_n dx_n  + \frac{h_1}{6 K^3}d\tau^2+ \frac{h_1}{4}\sinh \tau \tanh \tau     (\omega_1^2 + \omega_2^2)   +  \sum_{i=1\dots3} (\hat{e}^i)^2 \ .
 \ee
 The NS 2-form is
 \ba
 \widehat{B} &=&  -\frac{h_1 \cosh^\ha\tau }{2\sqrt{3} K^{\frac{3}{2}} v_2 } \hat{e}^{\prime 1}\wedge \hat{e}^3 + \frac{\cosh^{-\ha}\tau \cV}{\sqrt{3}  K^{\frac{3}{2}} v_2} \hat{e}^{\prime 2} \wedge \hat{e}^3 + \frac{h_1^\ha \cosh^{-\ha}\tau   }{2 } \hat{e}^{\prime 2} \wedge \om_2  \\
 && - \frac{h_1^\ha \cosh^{-1}\tau \cU }{2\sqrt{3}K^{\frac{3}{2}} v_2   } \hat{e}^3 \wedge \om_2 + \frac{  h_3 \cosh^{-\ha}\tau}{h_1^\ha }\hat{e}^{\prime 1}\wedge\om_2 - \frac{ h_2 \cosh \tau  - h_3 }{2\cosh \tau} \om_1 \wedge \om_2
 \nonumber
 \ea
 and the dilaton is
 \be
 \widehat{\Phi} = - \frac{1}{2} \ln {\cal W}
 \ee
We will not quote the transformed RR fields.

 \subsection{Dualisation of the Klebanov-Strassler-baryonic branch }\label{sectionksbaryonicutil}
\subsubsection{The baryonic branch background}
  The geometry that describes the whole Baryonic branch
of the KS field theory
was constructed in \cite{Butti:2004pk}.
This background is given by the frame fields \footnote{Our convensions here are related to the convensions of \cite{Gaillard:2010qg}-\cite{Conde:2011aa}
in the following way:\\ $(\th_1,\phi_1)=(\th,\phi), \ (\th_2,\phi_2)=(\tilde{\th},\tilde{\phi})$ and $\s_1=\tilde{\om}_2, \ \s_2=\tilde{\om}_1, \ \s_3=\tilde{\om}_3$.}:
\bea
&   e^{x^i}          = e^{\frac{\Phi}{2}}\hat{h}^{-\frac{1}{4}} dx^i
\,,\;\;\;
    e^{\r}          =  e^{\frac{\Phi}{2}+k} \hat{h}^{\frac{1}{4}}d\r
\,,\;\;\;
    e^{\theta_1}  =  e^{\frac{\Phi}{2}+h} \hat{h}^{\frac{1}{4}}    d\theta_1
\,,\;\;\;
     e^{\phi_1}= e^{\frac{\Phi}{2}+h} \hat{h}^{\frac{1}{4}}    \sin\theta_1 \,
d\phi_1\,    ,\nonumber\\
&   e^{1}                =  \frac{1}{2}e^{\frac{\Phi}{2}+g}
\hat{h}^{\frac{1}{4}}   (\s_2 +a\, d\theta_1)\,  ,\qquad\qquad
    e^{2}               =       \frac{1}{2}e^{\frac{\Phi}{2}+g}
\hat{h}^{\frac{1}{4}}   (\s_1   -a\,\sin\theta_1\, d\phi_1)
\,,\nonumber\\
&   e^{3}                = \frac{1}{2}e^{\frac{\Phi}{2}+k}
\hat{h}^{\frac{1}{4}}   (\s_3 +\cos\theta_1\, d\phi_1)\,.
    \label{vielbeinafter}
\eea
Notice the factor ${\hat{h}}$
dressing up the frame fields in comparison to
the wrapped D5 background in (\ref{eq:vielbs}).

   The metric, RR and NSNS fields are
\bea
& & ds^2= \sum_{i=1}^{10} (e^{i})^2\,,\nonumber\\
& & F_3= \frac{e^{-\frac{3}{2}\Phi}}{\hat{h}^{3/4}}
\Big[f_1 e^{123}+ f_2 e^{\theta_1\phi_1 3}
+ f_3(e^{\theta_1 23}+ e^{\phi_1 13})+
f_4(e^{\r 1\theta_1}+ e^{\r\phi_1 2})   \Big]\,,\nonumber\\
& & B_2= \k\, \frac{e^{ \Phi}}{\hat{h}^{1/2}}\Big[e^{\r 3}-\cos\alpha
(e^{\theta_1\varphi_1}+ e^{12})-\sin\alpha(e^{\theta_1 2}+ e^{\varphi_1 1})
\Big]\,,\\
& & H_3=-\k\, \frac{e^{\frac{1}{2}\Phi}}{\hat{h}^{3/4}}
\Big[-f_1 e^{\theta_1\varphi_1 \r} - f_2 e^{\r 12}
- f_3(e^{\theta_1 2\r}+ e^{\varphi_1 1\r})+
f_4(e^{ 1\theta_1 3}+ e^{\varphi_1 2 3})   \Big]\,,\nonumber\\
& & C_4= -\k\,\frac{e^{2\Phi}}{\hat{h}}
dx^0\wedge dx^1 \wedge dx^2\wedge dx^3\,,  \nonumber\\
& & F_5= \k\, e^{-\frac{5}{2}\Phi -k}\hat{h}^{\frac{3}{4}}
\partial_\r \left(\frac{e^{2\Phi}}{\hat{h}}\right)
\Big[e^{\theta_1\phi_1 123 }- e^{x^0 x^1 x^2 x^3 \r}   \Big]\,. \nonumber
\label{configurationfinal}
\eea
We have defined
\beq
\cos\alpha= \frac{\cosh(2\r)-a}{\sinh(2\r)}\,,\qquad \sin\alpha= -\frac{2e^{h-g}}{\sinh(2\r)}\,,\qquad\qquad\hat{h}=1-\k^2 e^{2\Phi}\,,
\label{cosinh}
\eeq
where $\k$ is a constant that we will choose to be $\k= e^{-\Phi(\infty)}$,
requiring the dilaton to be bounded at large distances.

\subsubsection{The Baryonic Branch dual geometry }
We again perform the dualisation as outlined in section
(\ref{TdualwrappedD5}) however in this case the presence
of the NS two-form field renders the resulting geometry rather
complicated and in particular mixes the radial direction with the
internal space.

To express the dual geometry it is expedient to introduce a few combinations,
\def\cV{{\cal V}}
\def\cW{{\cal W}}
\def\cU{{\cal U}}
\def\hhat{{\hat{h}}}
\ba
\mu_1 &=&   a e^g \cos\a + 2 e^h \sin\a   \ , \quad \mu_2 = a e^g \cos\a + 4 e^h \sin\a  \ ,\nonumber\\
\cV &=& 2\sqrt{2}  v_3 +   e^{2 g+ 2 \Phi} \cos\a  \ , \nonumber \\
\cU &=& a \cV  + e^{g + 2 \Phi} \kappa \mu_1   \ , \\
\cW &=& det M = \frac{1}{8} e^\Phi \hhat^{\ha} \left( e^{4g + 2 k + 2 \Phi} \hhat  + e^{2k} \cV^2 + 8 e^{2g}v_3^2  \right)  \ , \nonumber
\ea
and to introduce a one-form
\be
\Lambda = dv_3 - \kappa \frac{e^{2k + 2 \Phi}}{\sqrt{2}} d\rho  \ ,
\label{oneformlambdaxxx}
\ee
 that will neatly encode the mixing of the radial and internal direction in the metric.

 Then T-dual vielbeins, after the rotation described in
section (\ref{TdualwrappedD5}), are given by
\ba
 \hat{e}^{\prime 1} &=& \frac{e^{ g+\frac{3}{2} \Phi} \hhat^{\frac{3}{4}} }{ 16 {\cal W} }
 \Big[ 8 e^{2g}v_2 (\Lambda - a v_2 \omega_1)    - e^{2k} a \cV^2 \om_1 - 2\sqrt{2} e^{2k} \cV dv_2 \nonumber\\
 && \qquad \qquad - e^{2k + g+ \Phi} \left( e^\Phi \cV \k \mu_1  \omega_1 + e^g \hhat ( \cU \om_2 - 2 \sqrt{2} v_2 \om_3)  \right)   \Big]\ , \nonumber \\
\hat{e}^{\prime 2} &=&  -\frac{e^{ g+\frac{1}{2} \Phi} \hhat^{\frac{1}{4}} }{ 16 {\cal W} }
 \Big[
e ^{2g+2k +2\Phi} \hhat (\cU \om_1 +2 \sqrt{2}dv_2 )  - e^{2k +\Phi} \hhat \cV ( \cU\om_2 - 2\sqrt{2} v_2  \om_3 )   \nonumber\\
&& \qquad \qquad + 8 v_2 (\cV \Lambda + e^{g+2\Phi} \kappa\mu_1v_2 \omega_1 + 2\sqrt{2} v_2 dv_2 )
 \Big]  ,  \\
\hat{e}^3 &=& -\frac{e^{ k +\frac{1}{2} \Phi} \hhat^{\frac{1}{4}} }{ 8 {\cal W} }  \Big[
\sqrt{2} e^{2g + \Phi} \hhat v_2 (\cU \om_2 - 2 \sqrt{2} v_2\om_3 ) + \sqrt{2}  \cV ( \cV_3 \Lambda + 2 \sqrt{2} v_2 dv_2)  \nonumber\\
&& \qquad \qquad + 8 \sqrt{2} e^{g + 2\Phi} (\cV_3 \kappa \mu_1 v_2 \om_1 + e^{3g} \hhat (\Lambda - a v_2 \om_1)  ) \ ,
  \Big] \ . \nonumber
 \ea
 The metric is then given by
 \be
  d\hat{s}^2 = (e^{x^i})^2 + (e^\rho)^2 + (e^{\theta_1})^2+ (e^{\phi_1})^2  + (\hat{e}^i)^2  ,
 \ee
  the NS two-form by
  \ba
  \widehat{B} &=&  - \frac{e^{g+ k +\Phi} \hhat^\ha}{2 \sqrt{2} v_2} \hat{e}^{\prime 1} \wedge \hat{e}^3 +  \frac{e^{-g + k} \cV}{2 \sqrt{2} v_2} \hat{e}^{\prime 2} \wedge \hat{e}^3  + \frac{e^{g+ \Phi/2} a \hhat^{\frac{1}{4}} }{2} \hat{e}^{\prime 2} \wedge \omega_2 + \frac{e^{k +\Phi/2} \hhat^{\frac{1}{4}}   a \cU }{4 \sqrt{2} v_2 }  \hat{ e}^3 \wedge{\omega_2} \nonumber\\
  && - \frac{\kappa e^{2\Phi} }{4}( a (\mu_1+ \mu_2) e^g  - 4 e^{2h} \cos\a ) \om_1 \wedge \om_2 - \frac{\kappa e^{2 k +2 \Phi} }{2}  \om_3 \wedge d\rho - \frac{\kappa \mu_1 e^{3\Phi/2} }{2\hhat^{\frac{1}{4} }} \hat{e}^{\prime 1} \wedge \om_1 \nonumber \\
  \ea
  and the dilaton by the usual formula
  \be
  \widehat{\Phi} =  \Phi - \frac{1}{2} \ln {\cal W} \ .
  \ee

As a consistency check one can
readily verify that these reduce to the results
of the wrapped D5 system in the limit where $\hhat \rightarrow 1$.
As before this geometry is supported by fluxes $F_0,F_2$ and  $F_4$
the exact details of which we do not require
at this stage and so omit for concision.

\section{Analysis of the dual field theories.}\label{seccion7deanalysis}
In this section, we will complete the analysis initiated
in Section \ref{secciondekt}.
As we repeatedly emphasized, the study of many of the observables
considered in Section \ref{secciondekt} was
incomplete, due to the fact
that we were analysing certain low-energy effects in backgrounds that do
not contain the information of the IR dynamics
(as it is the Klebanov-Tseytlin background and
its non-abelian T-dual transformed). The logic was that we were
{\it defining} the way to calculate these observables in the simpler setting
of the KT-transformed background.

In this section we remedy this deficiency. We will study IR effects
in backgrounds that are fully non-singular, containing all the information
on the non-perturbative dynamics of the dual QFT. We organise
this section by presenting different observables and quoting
the result when calculated in the non-abelian transformed of the
following:
\begin{itemize}
\item{(wD5): The wrapped D5 background, that as explained
in \cite{Conde:2011aa} could be thought
of as the "low energy effective theory" of the KS/baryonic branch
backgrounds}
\item{(KS): The Klebanov-Strassler background. A specially symmetric case, with a
four-dimensional UV dynamics.}
\item{(KS+bb): The baryonic branch, interpolating between both previous cases.}
\end{itemize}
To add clarity to the presentation, in what follows we mark the formulas by  (wD5), (KS), (KS+bb) to indicate the case to which they refer.

We start by writing explicitly the induced metrics and IR behavior of
the two-cycle and three-cycle we defined in eqs.(\ref{twocycle})
and (\ref{trescycle}), respectively.
Then, we will go down our list of non-perturbative effects,
drawing conclusions on the different characteristics of
the associated dual field theories.

\subsection{Two-cycle and Three-cycle}

The two cycle $\Sigma_2 = [\theta_1, \phi_1]$ with $v_2 = v_3 =\psi = 0 $  defined previously in the dual
of KT, (\ref{twocycle}) is also
well defined in each of the three cases that occupy us here.
The induced metric on the two cycle is given by
\ba
 (wD5):&\quad & ds^2 |_{\Sigma_2} = e^{2 h + \Phi}
\left(d\theta^2_1 + \sin^2 \theta_1 d\phi_1^2 \right) \  , \nonumber \\
(KS): &\quad & ds^2 |_{\Sigma_2} = \frac{h_1 \big( -8 h_2 h_3 + \cosh^3 \tau h_1^2 + \cosh \tau(4 h_2^2 + 4 h_3^2 - h_1^2 ) \big)}{4 ( 4 h_2^2 + \cosh^2 \tau h_1^2) }
\nonumber\\
      & & \qquad \qquad \qquad \times \left(d\theta_1^2 + \sin^2 \theta_1 d\phi_1^2 \right)\  ,  \\
(KS+bb):&\quad & ds^2 |_{\Sigma_2} = \frac{e^\Phi \hhat}{\hhat + e^{2\Phi} \k^2 \cos^2 \a} \Big( e^{h} (e^h \hhat + e^{h+2\Phi} \k^2 \sin^2 \a + a e^{g+ 2 \Phi} \k^2 \sin 2\a  )
\nonumber\\
      & & \qquad \qquad + e^{2\Phi} ( a^2 e^{2g} + e^{2h}) \kappa^2 \cos^2 \a  \Big)
\left(d\theta_1^2 + \sin^2 \theta_1 d\phi_1^2 \right) \ .
\nonumber
\ea
The volume of the cycle follows immediately.
The IR asymptotics are (making use of   \cite{Klebanov:2000hb}
 and \cite{Conde:2011aa}
\footnote{The careful reader will find useful the following expansions (at
$\tau\to 0$ and $\r\to 0$)
\bea
& & h_1\sim K(\tau)\sim h(\tau)\sim 1,\;\;\; h_2\sim h_3\sim \tau
\label{davidstar}\\
& & \hat{h}\sim e^{g}\sim e^{k}\sim e^{\Phi}\sim \sin\alpha\sim 1,\;\;\;
e^{h}\sim \mu_1\sim\mu_2\sim a \sim \cos\alpha\sim \sin 2 \a \sim \r.\nonumber
\eea
}),
\ba
(wD5):   && Vol[\Sigma_2]  \sim \rho^2+.... \ , \;\;\; (KS): Vol[\Sigma_2] \sim\tau^2+.... \ , \nonumber\\
(KS+bb): && Vol[\Sigma_2] \sim \rho^2+.... \  .
\ea
The two cycle vanishes in the origin of the radial coordinate.

Let us move to the study of the three cycle defined in (\ref{trescycle}).
The three cycle is defined by $\Sigma_3 = [\theta_1, \phi_1 , \psi_1 ]$
at constant values of the other internal coordinates.
The induced metric on this cycle is given by,
\ba
\label{threecycle}
(wD5):&& ds^2 |_{\Sigma_3} = \big(e^{2h+\Phi} +
\frac{a^2 e^{2g+2\Phi}}{4{\cal W} } (  e^{2g} v_2^2 + e^{2k}v_3^2 )  \big)
(\omega_1^2 + \omega_2^2)
+ \big(\frac{  v_2^2 e^{2g +2k +2 \Phi} } {4 {\cal W}}  \big)
\omega_3^2   \nonumber \\
&& \qquad \qquad  - \big( \frac{ a^2 v_2^2 e^{4g+2\Phi}}{4 {\cal W}} \big)
\omega_2^2  - \big(\frac{ a v_2 v_3e^{2g +2k +2 \Phi} } {2 {\cal W}}   \big)
\omega_2 \ \omega_3 \ ,    \nonumber \\
(KS):&&   ds^2 |_{\Sigma_3} = \big(\frac{h_1}{4}\sinh\tau \tanh\tau  + \frac{h_1^2 \cU^2}{12 \cosh\tau K^3 \cW } \big)
(\omega_1^2 + \omega_2^2)  +  \big( \frac{\cosh \tau h_1^2 v_2^2 }{6 K^3 \cW}  \big) \om_3^2 \nonumber \\
&& \qquad \qquad  + \big(\frac{(4 h_3^2 + h_1^2) v_2^2}{3\cW} \big) \om_1^2 + \big(\frac{  h_1^2 \cU v_2 } {3\sqrt{2} K^3 {\cal W}}   \big) \omega_2 \ \omega_3 \ , \\
(KS+bb):&& ds^2 |_{\Sigma_3} = \big( e^{2h+\Phi}\hhat^\ha + \frac{e^{2g+ 2k + 2\Phi}  \cU^2}{32 \cW}   \big) (\omega_1^2 + \omega_2^2)   + \big(\frac{  v_2^2 e^{2g +2k +2 \Phi} } {4 {\cal W}}  \big)   \omega_3^2    \nonumber \\
&& \qquad \qquad  +\big(\frac{(a^2 e^{4g+2\Phi} \hhat  + e^{2g+ 4\Phi} \kappa^2 \mu_1^2) v_2^2 }{4 \cW} \big)  \om_1^2   - \big(\frac{  v_2 \cU e^{2g +2k +2 \Phi} } {4 \sqrt{2} {\cal W}}   \big)
\omega_2 \ \omega_3 \ . \nonumber
\ea
The volume of these three cycle behaves in the far IR as
\ba
 (wD5):    Vol[\Sigma_3]  \sim \rho +....  \  , \quad
 (KS):   Vol[\Sigma_3] \sim \tau....\nonumber \  , \quad
 (KS+bb):   Vol[\Sigma_3] \sim \rho+....\nonumber \  . \nonumber
\ea
and in all cases the functional dependence of $v_i$ is shared in the IR limit with the volume behaving as
\be
\sqrt{g|_{\Sigma_3}}  \sim \rho v_2 \ \frac{\sqrt{v_3^2 + v_2^2} }{ {\rm const} + v_3^2 + v_2^2}  + \dots
\ee
reflecting the underlying similarities of these geometries.

We see that the volume of the three cycle also vanishes in the far IR.
Though both the two and three cycle defined above vanish, this does not
imply that we will get trivial results for all observables.
Indeed, in most cases, as we will see, it is the "stringy volume", that is
$\det [g+B]$ what plays a role in calculations.

We will move now into calculating observables of the dual QFT, using the three
new geometries we have found.

\subsection{Domain walls}
In the original backgrounds, domain walls are defined as
$D5$ branes that extend along $R^{1,2}$
and $(\th_2, \phi_2, \psi)$. These objects have finite
tension when placed in the
origin of the radial coordinate. This indicates the objects
"exist" and are formed due to non-perturbative dynamics.
From the induced Action one can calculate the
tension of the domain wall (before the non-abelian T-duality) to be,
\be
T_{DW} = T_{D5} \, 2\pi^2 e^{2\Phi+2g+k}\sim 2\pi^2 e^{2\Phi(0)} .
\ee
We used the IR expansions of the functions, quoted, for example in
Appendix A of
\cite{Conde:2011aa}---see also our (\ref{davidstar}).
\no In the T-dualised background we
have defined the Domain walls as $D2$ branes extending along $R^{1,1}$ with
vanishing $(v_2, \, v_3, \, \th_1, \, \phi_1, \, \psi)$.
Such configurations have vanishing B
field on their world-volume.
The Born Infeld action which in this case is for the non-abelian
transformed of the wrapped D5 system
\be
(wD5): \quad  S = - T_{\rm D2} \; e^{-\widehat{\Phi}+2\Phi/2} \; \int d^{2+1}x
\ee
\no Using the expression for the dual dilaton we
obtain for the effective tension,
\be
(wD5): \quad  T_{\rm eff} = T_{\rm D2} \left. {e^{2g+k+2\Phi} \ov 2\sqrt{2}} \right|_{\rho=0}
\ee
again, we observe a finte tension object. This indicates that the
field theory dual to the transformed
background contains different vacua, separated by these walls, formed
non-perturbatively.

In the cases of the KS-Baryonic Branch non-abelian T dual backgrounds
the expressions read,
\ba
(KS ):& \quad&  T_{\rm eff} = T_{\rm D2}  \left.  \frac{ \e^{2/3} h(\tau)^{-\ha}  (4 h_2^2 + \cosh^2 \tau K(\tau)^2 \e^{\frac{8}{3}} )^\ha  }{2\sqrt{3} K(\tau) }  \right|_{\tau=0} \nonumber \ , \\
(KS+bb) & \quad&  T_{\rm eff} = T_{\rm D2} \left. {e^{2g+k+2\Phi} (\hhat + e^{2\Phi} \k^2 \cos^2 \a)^\ha  \ov 2\sqrt{2} \hhat^\ha } \right|_{\rho=0}
\ea
Using the small radius expansions in \cite{Klebanov:2000hb}
 and \cite{Conde:2011aa}, we find that the
domain walls have constant tension in
the three analysed cases.

\subsection{The 't Hooft loop and gauge coupling}
In Section \ref{secciondekt},  we defined
the 't Hooft loop as
the non-local operator calculated with a
$D4$ brane extending in $R^{1,1}$ and
wrapping the three-cycle $\Sigma_3$ of (\ref{trescycle}).
In the case of the background
we obtain when transforming the wrapped D5 system---
see Section \ref{TdualwrappedD5}.
The induced metric on the probe D4 takes the form,
\ba
ds^2 &=& e^{\Phi}dx_{1,1}^2 \, + \Big(e^{2h+\Phi} + \frac{a^2 e^{2g+2\Phi}}{4{\cal W} } (  e^{2g} v_2^2 + e^{2k}v_3^2 )  \Big) (\omega_1^2 + \omega_2^2)  + \big(\frac{  v_2^2 e^{2g +2k +2 \Phi} } {4 {\cal W}}  \big) \omega_3^2
\nonumber \\
     &&  - \big( \frac{ a^2 v_2^2 e^{4g+2\Phi}}{4 {\cal W}} \big) \omega_2^2  - \big(\frac{ a v_2 v_3e^{2g +2k +2 \Phi} } {2 {\cal W}}   \big) \omega_2 \ \omega_3 \ .
\ea
\no while the B field induced on the brane is,
\be
B = {a \, e^{4g+2k+3\Phi} \ov 8 \, \sqrt{2} \, W}
( v_2 \, \om_1 \wedge \om_3 - a \, v_3 \, \om_1 \wedge \om_2 )
\label{cccuuu}
\ee
Now one calculates the
Born-Infeld action and the effective tension for the above configuration---for
the non-abelian T-dual of the wrapped D5 background,
\ba
(wD5): && S = T_{\rm eff} \int d^{1+1}x \ ,
\\
       && T_{\rm eff} = T_{\rm D4} \, 4\pi^2 \, e^{g+h+k+2\Phi} \, v_2 \, \sqrt{a^2 \, e^{2g} + 4 \, e^{2h}}.
\nonumber
\ea
\no Note that the above result is independent of $v_3$. Also, as already
observed, it is the stringy volume what plays a role in this calculation.

Using the small radius expansion from Appendix A of \cite{Conde:2011aa} or
(\ref{davidstar}),
we can see that this object becomes tensionless in the far IR
as $T_{\rm eff}\sim \r$.
This indicates that
monopoles are "screened", which is a typical signature of a confining
field theory.
Let us briefly analyse the analog result on the KS/Baryonic Branch
non-abelian T-dual backgrounds where the effective tensions are
\ba
(KS): && T_{\rm eff} =  T_{\rm D4} \, \frac{\pi^2 \e^{\frac{4}{3}}}{2\sqrt{3} K(\tau)^\ha } v_2\, \left( 16 g_s M ( (-1 + \cosh \tau)f(\tau)^2 +(1 + \cosh \tau)k(\tau)^2 ) \right. \nonumber \\
&& \qquad \qquad \qquad \qquad \qquad \qquad \qquad  \left. + 2 \e^\frac{8}{3} h(\tau)K(\tau)^2 \cosh\tau \sinh^2 \tau    )^\ha  \right)    \nonumber \ , \\
(KS+bb): && T_{\rm eff} =  T_{\rm D4} \, \pi^2 \, e^{g+h+k+2\Phi} \hhat^\ha \,
v_2 \, \sqrt{a^2 \, e^{2g} + 4 \, e^{2h}}   \ .
\ea
Using the IR expansions of \cite{Klebanov:2000hb}
 and \cite{Conde:2011aa}, we see that monopoles are
screened in the three backgrounds/dual QFTs.

\no Finally, in Section \ref{secciondekt},
we defined the gauge coupling in terms of
a $D2$ brane that wraps the three-cycle in (\ref{trescycle}).
The induced NS B field in this case is the same as that
we found in (\ref{cccuuu}).
The Born Infeld action for this euclidean "instantonic" D2 brane gives
\be
(wD5): \quad S = T_{\rm D2} \, \pi^2 \,
e^{g+h+k+\Phi} \, v_2 \, \sqrt{a^2 \, e^{2g} + 4 \, e^{2h} }
\label{manadax}
\ee
Notice that we have again considered the range of $\psi$ to be
$[0,\pi]$, to avoid the bolt-singularity.
Note also that this result is $v_3$ independent.
Using this, we obtain (in the case of the non-abelian T-dual to the
wrapped D5 system), that a logic similar to the one spelled around
(\ref{gaugecoupling}) tells us that the inverse gauge coupling
vanishes in the far IR as
\be
(wD5): \quad  \frac{1}{g_4^2}\sim \r ,
\ee
which is again a characterisc sign of a confining theory.
The definition based on D6 branes is expected to give the same result.

Let us briefly analyse the analog result on the KS/Baryonic Branch
non-abelian T-dual backgrounds.  The euclidean D2 actions are
\ba
(KS):& \quad & S = T_{\rm D2} \, \pi^2 \frac{h_1}{K^{3/2} 4 \sqrt{6} }  \big(- 8 h_2h_3 + \cosh^3 \tau h_1^2 + \cosh \tau ( 4h_2^2 + 4h_3^2 - h_1^2  )  \big)^\ha \,  \nonumber \\
(KS+bb):  & \quad &S = T_{\rm D2} \, \frac{\pi^2}{2	} e^{g+ k + \Phi} \hhat^\ha  v_2 \left( 4 e^{2h} \hhat (a^2 e^{2g} + 4 e^{2h} ) + e^{2\Phi}\k^2 ( a^2 e^{2g} (\mu_1+ \mu_2 )^2 \right.\nonumber\\
&& \qquad \qquad  \qquad \left. + 16 e^{4h} \cos^2\a  + 4 e^{2h} ( \mu_1^2 - 2 a e^{g} (\mu_1 + \mu_2) \cos \a ) \right)^\ha \ .
\label{mamabax}
\ea
Correspondingly we obtain the behaviour in the deep IR ( and at fixed $v_2$),
\ba
(KS): \quad  \frac{1}{g_4^2}\sim  \tau,  \qquad
(KS+bb): \quad  \frac{1}{g_4^2}\sim   \r  \ .\nonumber
\ea
Notice that the expression in (\ref{mamabax})
reduces to (\ref{manadax}) in the un-dressing limit
$\kappa\to 0,\; \hat{h}\to 1$.

In conclusion, the gauge coupling defined as in
Section \ref{subsectionongaugecoupling},
is such that in all three backgrounds/QFTs grows unbounded for
small energies. As explained, this is in
good correlate with  confinement of quarks and
the screening of monopoles discussed above.

\subsection{Central charge and Entanglement Entropy}
In Section \ref{secciondekt}, we found expressions
that hinted at the fact that the central charge and the
entanglement entropy were invariants of the non-abelian T-dual operation.
indeed, we found that up-to a factor depending on
the volume of the $v_2,v_3$ space, expressions before and after the
non-abelian duality were identical.

In the cases of the non-abelian T-dual of the wrapped D5 and the
KS  we can readily obtain the expressions for the central charge using the formalism of \cite{Klebanov:2007ws}.  Since in these cases the space spanned by $\mathbb{R}^{1,3}$ and the radial coordinate is unaltered it is sufficient just to look at the (dilaton adjusted) volume of the internal manifold which are given by
\ba
(wD5):&\quad & V_{\rm int}  = \int e^{-2 \Phi} \sqrt{g}  =  \sqrt{2}  \pi^2 e^{2g +2 h + k + \Phi/2}  {\cal I}   \ , \nonumber \\
(KS): &\quad & V_{\rm int}  = \int e^{-2 \Phi} \sqrt{g}  =\frac{ \pi^2}{2\sqrt{3}} \e^{\frac{10}{3}} h(\tau)^\frac{5}{4} K(\tau) \sinh^2 \tau   \   {\cal I} \ ,
 \ea
 where in both cases
\be
{\cal I} = \int dv_3 \int dv_2 \, v_2 \ ,
\ee
 is the RG scale independent integral left over from
the Fadeev-Popov determinant.  Up to this constant
factor these precisely match
the corresponding expressions before dualisation and
hence the central charges
agree.
Interestingly, notice that $V_{int}$ vanishes in the far IR.
The situation is rather more subtle when it comes to the
Baryonic branch; the `radial' coordinate $\rho$ mixes with the
internal directions, this means a direct use of \cite{Klebanov:2007ws}
is not viable. We will
 comment more on this unusual feature in the next section.

It may be interesting to push these calculations a bit more and
analyse as the authors of \cite{Klebanov:2007ws} did
the possible first order (quantum) phase transition observed when plotting
the entanglement entropy in terms of the separation between regions
$l$. We leave this demanding numerical work for the future.

\subsection{Wilson loops, Asymptotics
of the Dilaton and five dimensional metric}
In Section \ref{asintoticseccion}, we considered
the "radial" behavior of the dilaton and the metric of the
$R^{1,3}\times R$ of the metric for the non-abelian T-dual of the
Klebanov-Tseytlin background.
We observed that when fixing the internal directions as in (\ref{hart}),
the metric of the five-space received no "contributions"
from the internal manifold. The conclusion of this  was the fact that
the UV-asymptotics of the Wilson loop---computed with a string hanging from
the far UV in the configuration of (\ref{kompany})---was not
supposed to present changes. We can quickly verify which is situation
in the non-abelian T dual transformed of our wrapped D5,
KS and Baryonic branch backgrounds.

In the transformed solutions
corresponding to the case in which the "seed" background are those of wrapped
D5s and KS, we check that the $R^{1,3}\times R$ remains the same
before and after the non-abelian T-duality, in the case of
the Baryonic Branch background something uncanny occurs.
Inded, the one form $\Lambda$ defined in (\ref{oneformlambdaxxx}),
generates a new term in the component of $g_{\r\r}$ even when
the internal directions are fixed to be constants.

In particular we find that the  $R^{1,3}\times R$ part
of the metric is given by
\ba
ds^2_{R^{1,3}\times R} & = & e^{ \Phi} \hhat^{-1/2}
( - dt^2 + dx_1^2 + dx_2^2 + dx_3^2)\ +\  e^{2k + \Phi}\ \hhat^{1/2}\ d\rho^2
\nonumber\\
&& \
+\ \frac{ e^{4 k+ 4\Phi } }{ 8 \cW} ( e^{4g + 2\Phi} \hhat + \cV^2 )\ d\rho^2 \ .
\label{desviacion}
\ea
In addition to this there is an off-diagonal mixing
\ba
g_{\rho a}dx^a & = &
- \frac{e^{2k+ 2\Phi} \kappa }{4 \cW} \Big[ 4 \cV v_2 dv_2 + (e^{4g + 2\Phi}\hhat + \cV^2 )\ dv_3
\nonumber\\
&&\ - \sqrt{2} e^{g + 2\Phi} v_2 ( Ae^{3g} \hhat - \cV \kappa \mu_1 ) \omega_1     \Big] \ .
\ea

What distinguished the baryonic branch from the KS system and
the D5 wrapped background (all considered {\it before}
the non-abelian T-duality) is that in the baryonic branch case
we have a new component in the NS-$B_2$ field, that is ultimately
responsible for this mixing expressed by the one form $\Lambda$.
Field theoretically, the baryonic branch contains a non-zero
VEV for an operator of dimension two. The operator is called
${\cal U}$ in \cite{Dymarsky:2005xt} and is roughly indicating
the differences in VEVs of baryon and antibaryon fields. Here,
it seems that
what was a difference of VEV's before the non-abelian T-duality gets mapped in to an
irrelevant operator that deviates the geometry away from the
"logarithmically approaching AdS"
characteristic of KS and the baryonic branch. This in indicated
by the last term in (\ref{desviacion}). The numerology
points to an irrelevant of dimension six.  Another more conservative interpretation is that there simply does not exist a baryonic branch of the field theories obtained after T-dualisation; indeed since some global symmetries are lost it might not be possible to from an appropriate baryonic operator.
We leave for a more dedicated study
the understanding of this feature of the QFT.

Let us summarize the results of this section. We performed an analysis
that complements that of Section \ref{secciondekt}, for the
observables sensitive to the non-perturbative IR dynamics of the QFT.
The information obtained here is key in
deciding about the quiver/lagrangian of the dual QFT.

We have presented a set of three geometries
in massive IIA String theory, that aside from being smooth geometries,
are dual to
a minimally SUSY field theory that confines,
generates domain walls (hence different vacua), presents a mechanism
similar to Seiberg duality
(for the $KS$ case )
when
flowing down from an approximate fixed point.

Lots of observables that have been calculated with the trademark
backgrounds of Type IIB, could be computed in our new geometries to learn
more about these QFT.

  \section{Conclusions and Outlook}\label{section7www}

In this work we have demonstrated the utility of non-abelian T-duality as a solution generating technique of supergravity backgrounds, in particular in the context of the AdS/CFT correspondence. We began by considering dualisation of the Klebanov Witten geometry with respect to an SU(2) isometry group.  The result of this procedure was to find an $N=1$ supersymmetric background in type IIA supergravity whose lift to M-theory  coincides with a special case of the geometries dual to  certain N=1 quiver gauge theories described in
\cite{Bah:2012dg}.  However,  in our situation the dual geometry contains a genus zero surface and the corresponding gauge theories are, perhaps surprisingly, less   understood than those of higher genus.  Understanding more precisely the theories of  \cite{Benini:2009mz}, \cite{Bah:2012dg} in the genus zero case and their connection to the geometry present here is an interesting question that we hope will be the subject of further study.

We also looked at dualisation of  geometries dual to non-conformal theories, beginning with the Klebanov Tseytlin geometry and then considering its completion in the IR.  Here we find an immediate departure to the conformal case; the backgrounds we generate are solutions of massive type IIA supergravity with the mass parameter naturally quantised by the number of fractional branes before dualisation.  The action of T-duality has a natural consequence for the Page charges; the D5 and D3 charges get mapped on to D8 and D6 charges respectively.  Under a large gauge transformation we saw that these charges display the signatures of a Seiberg duality cascade encoded in the geometry.  A puzzling feature is that whereas a Seiberg duality before the T-duality
changed charges (or the ranks of gauge groups)
by M units, after duality it has the effect of
changing charges by 2M units.  A fundamental question to ask
is then what is the would be cascading  field theory corresponding to this solution.

 We hope that by performing a number of probes of this geometry we have assembled some facts that will be of use in future attempts to answer this question.   Many features that were present before the dualisation seem to persist. Features that are preserved (and which can thus be thought of as 'neutral'
under the isometry group dualised) are the presence of well defined two and three cycles; the aforementioned Page charges and indications of duality cascade; the constant tension of domain walls and the fact that the $R^{1,3}\times R $ directions, logarithmically approach $AdS_5$.    Perhaps the most compelling similarity is the fact that the central charge is preserved under the dualisation (up to a single RG scale independent integral that depends on the global properties of the geometry).   There however are some notable changes after dualisation.  Firstly the geometry is supported by a non-constant dilaton . Somewhat related to this is that a suitable  gauge coupling, defined in terms of a brane probe wrapping the three cycle, has a rather peculiar asymptotic behaviour and not what is typical in 4d gauge theories.
 On the same vein, we find an associated anomaly, that is not the usual one breaking the $U(1)_R$ symmetry of the seed backgrounds.

In order to probe the IR physics we also addressed IR smooth
N=1 dual geometries, namely the Klebanov Strassler, the wrapped five brane and the baryonic branch that interpolates between them.  In the UV the story is qualitively similar to the Klebanov Tseyltin.  In the deep IR we find that although the cycles are shrinking the presence of the dilaton and NS flux ensures finite tension for the domain walls.  The 't Hooft loop, defined as a probe D4 wrapping the three cycle, has an effective tension that vanishes
 in the deep IR indicating screening of monopoles, a signature of confinement.
  In alignment with this we find that the rectangular Wilson loop,
uncharged under global symmetries, displays an area law.
Also, the  gauge coupling described above diverges at low energies.
Again the central charges are preserved by the dualisation.

For the case of the baryonic branch 
the dualisation has rather more severe consequences; the $R^{1,3}\times R$ space actually mixes with the internal space (this is due to the activation of a certain component of the NS two form in baryonic branch background). This strongly deviates the geometry from the logarithmically approaching AdS.  One interpretation of this could be that the dimension two operator that takes a VEV on the baryonic branch has been converted into an irrelevant operator in the dualisation.

Let us now describe some unresolved puzzles that we hope will be the topic of future study.

{\it{What are the dual field theories?}}

The dual of the conformal case seems to have at least superficial
relation to the theories coming from wrapped M5 branes on Riemann surfaces
\cite{Benini:2009mz}, \cite{Bah:2012dg}.
This corresponds well to the fact that a similar dualisation applied
to $AdS_5 \times S^5$ gave a geometry rather similar to
that of Gaiotto and Maldacena \cite{Gaiotto:2009gz} which are
an $N=2$ counter part to the Sicilian theories of  \cite{Benini:2009mz}.
However the case of genus zero that seems relevant to us is
the least  understood.

Moreover, it is  puzzling  that in the non-conformal case the
geometries are in massive IIA which does not lift to M-theory
in the conventional way.  One is then led to ask how to modify the
picture of wrapped M5 branes to induce a mass.
One might wonder if there is some non-geometric compactification
involved especially given the connection between non-abelian duality and
gauged supergravity \cite{Avramis:2009xi}.   Equally it is know that certain supergravities, which don't have a higher dimensional origin in the conventional geometric sense, can be obtained by reduction in the T/U-duality symmetric generalised geometry \cite{Grana:2012rr,Berman:2012uy}.  It is interesting to ask whether these massive IIA theories can be obtained from an appropriate reduction of the generalised M-theory considered in \cite{Berman:2010is}.

{\it{Clarifying and developing the underlying geometry}}

At first sight, the solutions presented here, seem rather complicated.
However certain similarities emerge due to the simplicity
of the $SU(2)$ group dualised.  One might hope to find
a better language with which to describe the geometries
presented within this work.   In fact all of the initial
geometries studied can be nicely phrased in terms of $SU(3)$-structures
(i.e. by the data in a globally defined two form $\cal J$
and holomorphic  three form $\Omega_{hol}$).   The action
of abelian T-duality on the $SU(3)$-structure has been studied before
and it seems very likely the situation would be similar for the
non-abelian duality.  It is then natural to conjecture that
these geometries may be described by non-local $SU(2)$ structures
\cite{Grana:2008yw}.
With such technology one might even hope to find a more general
class of cascading solutions in massive type IIA
by using the solutions found here as the starting point
for an intelligent ansatz.

{\it{Compatibility with S-duality}}

In this paper one system considered was the wrapped D5 brane whose geometry is supported by quantised RR 3-forms flux.
In general one can consider a $p-q$ system of wrapped NS5 and D5 branes whose geometry has both NS and RR 3-forms. It is natural to ask what becomes of this system under the duality.  A similar question stands for the D1-D5 system considered in  \cite{Sfetsos:2010uq}.

{\it{Developing the technology of non-abelian T-duality}}

One of the challenges in non-abelian T-duality is the difficulty
(at the very least) to demonstrate it to be an exact symmetry
in string (genus) perturbation theory.  One reason for this is
that it is hard to understand how to constrain the periodicities
and global properties of the Lagrange multipliers so that partition
functions match exactly before and after duality.
Although one may not be concerned about this difficulty when thinking
about the dualisation as a solution generating technique of supergravity
or in the context of large N AdS/CFT where string genus corrections
are suppressed, it would be desirable to understand better the global properties
of the backgrounds we have presented within.

{\it{Wider applications}}

It seems that here, and in recent works, we have only started seeing the utility of these duality transformations. In principle one could apply these procedures whenever a space-time admits some non-abelian isometry group.  There are of course many examples of this and we hope that further study will prove fruitful.

\section{Acknowledgements}
We would like to thank Francesco Benini, 
Davide Forcella, Tim Hollowood,
Prem Kumar, Yolanda Lozano, 
Alberto Mariotti, Niall MacPherson, Eoin \'O Colg\'ain, Maurizio Piai, 
Diego Rodr{\'{\i}}guez-G\'omez,
Brian Wecht and especially Kostas Siampos for interesting 
discussion and correspondence. We also acknowledge related 
work \cite{LOCRGS} on probe brane analysis in the
background found in \cite{Lozano:2012au}.
The research of G. Itsios has been co-financed by the ESF
and Greek national funds through the Operational Program "Education and Lifelong Learning" of the NSRF - Research Funding Program:
``Heracleitus II. Investing in knowledge in society through the European Social Fund''.
This research is implemented (K.S.) under the "ARISTEIA" action
of the "operational programme education and lifelong learning" and is co-funded by
the European Social Fund (ESF) and National Resources. D. Thompson is supported in part by the Belgian Federal Science Policy
Office through the Interuniversity Attraction Pole P7/37, and in part by the
"FWO-Vlaanderen" through the project G.0114.10N and through an "FWO-Vlaanderen"
postdoctoral fellowship project number 1.2.D12.12N.

 %
 %
 %
 %

  \begin{appendix}
  \section{Group theory conventions }

We give details of the conventions used in performing dualisation with respect to $SU(2)$ isometry groups.
The Pauli matrices are
\be
\tau_1 =  \left(
\begin{array}{cc}
 0 & 1 \\
 1 & 0
\end{array}
\right)  \ , \qq
\tau_2 = \left(
\begin{array}{cc}
 0 & -i \\
 i & 0
\end{array}
\right)  \ , \qq
\tau_3 =  \left(
\begin{array}{cc}
 1 & 0 \\
 0 & -1
\end{array}
\right) \ ,
\ee
and obey $
\tau_i \tau_j =  \delta_{ij} \mathbb{I}  + 2 \e_{ijk} \tau_k
$.
We define $SU(2)$  generators
\be
t^{i} = \frac{1}{ \sqrt{2}} \tau_i \ ,
\ee
such that
\be
Tr(t^i t^j) = \delta^{ij}  \ ,  \quad  [t^i, t^j] = i f^{ij}{}_k t^k = i \sqrt{2} \e_{ijk} t^k \ .
\ee
Left invariant one-forms are defined in general by
\be
L^i = - i {\rm Tr}(t^i g^{-1} d g)
\label{litr}
\ee
and obey
\be
dL^i = \frac{1}{2} f^{i}{}_{j k } L^j \wedge L^k\ .
\ee
In the Euler parametrisation a group element is given by
\be
g = e^{\frac{i}{2} \phi \tau_3}\cdot e^{\frac{i}{2} \theta \tau_2} \cdot e^{\frac{i}{2} \psi \tau_3}\ ,\qq
0 \leqslant \th \leqslant \pi \ ,\quad 0 \leqslant \phi \leqslant 2 \pi\ ,\quad 0 \leqslant \psi \leqslant  4\pi \
\ee
and the left-invariant Maurer--Cartan forms by
\ba
L_1 &=& \frac{1}{\sqrt{2} } \left(-  \sin \psi d\theta + \cos \psi  \sin \theta  d \phi   \right) \ ,
\nonumber \\
L_2 &=& \frac{1}{\sqrt{2} } \left( \cos \psi  d\theta + \sin \psi  \sin \theta  d \phi \right) \ ,
\label{L113}
\\
L_3 &=&  \frac{1}{\sqrt{2} } \left( d\psi +  \cos \theta  d\phi  \right) \ .
\nonumber
\ea
A useful matrix is the adjoint action
\be
D^{ij} = {\rm Tr} \left( t^i g t^j g^{-1} \right)\ .
\label{adij}
\ee
which is an orthogonal matrix, i.e. it obeys
\be
D^{ij} D^{i k} = \delta^{jk} \ .
\ee
Its explicit expression in terms of the Euler angles is not necessary for our purposes.

\no
Finally note that the $SU(2)$ transformations on the group element have the following action on the Euler angles
\ba
&& \d  \th = \e_1 \sin\phi + \e_2 \cos\phi \ ,
\nonumber\\
&& \d \phi = \cot\th \left(\e_1 \cos\phi - \e_2 \sin \phi \right) + \e_3 \ ,
\label{tran1}
\\
&& \d \psi = {1\ov \sin\th} \left(-\e_1 \cos\phi + \e_2 \sin\phi\right )\ .
\nonumber
\ea
The corresponding Killing vectors are
\ba
k_{(1)} &=& - \cos \phi \partial_{\th} + \cot \th \sin \phi \partial_{\phi}  - \csc \th \sin \phi \partial_\psi \quad\ ,
\nonumber \\
k_{(2)} &=& - \sin \phi \partial_{\th} - \cot \th \cos \phi \partial_{\phi}  + \csc \th \cos \phi \partial_\psi \quad \ ,
\label{k123}
\\
k_{(3)}&=& - \partial_{\phi}  \ ,
\nonumber
\ea
and obey $[k_{(i)},k_{(j)}] = - \e_{ijk} k_{(k)}$, i.e. the $su(2)$ algebra.

\no
Similarly the Lagrange multipliers are infinitesimally transforming as, i.e. \eqn{tranmult}
\be
\d v_a = \e_{abc} \e_b v_c\ .
\label{tran2}
\ee

\section{General gauge fixing }

In this appendix we consider the general case in which we gauge fix some variables
of the group group element and use the residual symmetry to gauge fix in addition some of the
Lagrange multipliers.
Let $g$ and $v$ denote the set of $\dim G$ variables left after gauge
fixing $\dim H $ among the original $\dim G + \dim H$ ones.
Then we have the replacement
\be
L^i_\pm \to L^i_\pm + i D_{ji}A^j_\pm\ .
\label{lipm}
\ee
In addition we denote
\be
Q_{\m i}\del_\pm X^\m = Q_{\pm i }\ ,\qq Q_{ i\m}\del_\pm X^\m = Q_{i \pm}\ .
\ee
Then we have the Lagrangian (we suppress group indices)
\ba
\cL & = &  Q_{\m\n}\del_+X^\m \del_- X^\n + Q_+ L_- + L_+ Q_- + L_+ E L_-\
\nonumber\\
&& i A_+ (D Q_- + D E L_- -\del_- v) + i (Q_+ D^T + L _+ E D^T + \del_+ v) A_-
\\
&& - A_+ (D E D^T + f) A_-\ .
\ea
Integrating out the gauge fields we find that
\ba
&& A_+ = i \widehat M^{-T}\left(Q_+ D^T + L_+ E D^T + \del_+ v\right)\ ,
\nonumber
\\
&& A_- = - i \widehat M^{-1}\left(\del_- v - D Q_- - D EL_-\right)\ ,
\label{apamons}
\ea
where
\be
\widehat M = D E D^T + f \ .
\ee
The dual action is
\be
\cL_{\rm dual} = \cL_0 + (Q_+ D^T + L_+ E D^T +\del_+v)\widehat M^{-1} (\del_-v - D Q_- - DE L_-)\ ,
\label{ldlliL}
\ee
where $\cL_0$ the original action.

\no
The transformation of the worldsheet derivatives can be written as
\ba
&& L_+ \to L_+ + i D^T A_+ = -D^T \widehat M^{-T}(D Q_+ + f D L_+ + \del_+ v)\ ,
\nonumber\\
&& L_- \to L_- + i D^T A_- = D^T \widehat M^{-1}(\del_- v + f D L_- -DQ_-) \ ,
\ea
where we have used \eqn{lipm} and utilized \eqn{apamons}.
Next we perform the coordinate transformation  \eqn{vdhv}
that defines the dressed Lagrange multipliers in \eqn{vdhv}. Also let that
\be
\hat f_{ij} = f_{ijk} \hat v^k\ ,\qq M\equiv E + \hat f \ ,
\label{Mnew}
\ee
where note that $M$ above is defined slightly different than \eqn{mef}.
Then using the identity arising from group theory considerations
\be
D^T f D = \hat f\ ,
\ee
we obtain
\be
\widehat M  = D M D^T \ .
\ee
Using that and in addition the identity
\be
\del_\pm D^T v = \hat f L_\pm \ ,
\ee
we find that
\be
L_+ \to -M^{-T}(\del_+ \hat v + Q_+ )\ ,\qq L_- \to M^{-1}(\del_- \hat v - Q_- )\ ,
\label{lplm}
\ee
which are the same expressions as those found with the gauge choice $g=\mathbb{I}$,
but with the $v$'s replaced with the $\hat v$'s as in \eqn{vdhv}.

\no
Similar manipulations give for the dual action \eqn{ldlliL} the form \eqn{tdulal}, i.e.
\be
\tilde \cL_{\rm dual} = Q_{\m\n}\del_+ X^\m \del_- X^\n + (\del_+ \hat v_i + \del_+ X^\m Q_{\m i})M^{-1}_{ij}
(\del_- \hat v_j - Q_{j\m} \del_- X^\m )\ ,
\label{tdulal9}
\ee
with $M$ given by \eqn{Mnew}, plus the term
\be
L_+ \del_- \hat v - \del_+ \hat v L_- - L_+ \hat f L_-\ ,
\ee
which is however a total derivative.

\no
Finally, we note that the dilaton is still given by \eqn{dildiall}, but with
$M$ given by \eqn{Mnew}.

\section{The general form of the transformation on RR fields }

Our conventions on Hodge duality are such that on a $p$-form in a $D$-dimensional
spacetime is
\be
(\star F_p)_{\m_{p+1}\cdot \m_D} ={1\ov p!} \sqrt{|g|}\ \e_{\m_1\dots \m_D} F_p^{\m_1\cdots \m_p}\ ,
\label{hdfer4}
\ee
where $\e_{01\dots 9} = 1$. With this we have the useful identity $\star\star F_p = s(-1)^{p(D - p)}F_p$, where
$s$ is the signature of spacetime which we take to be mostly plus.

\no
We would like to see how one works out the details concerning \eqn{ppom} from which the transformation of the flux fields arises.
Its inverse (which is actually what we need) will necessarily have the form
\be
\Om^{-1} = (A_0\G^1 \G^2 \G^3 + A_a\G^a)\G_{11}\ ,
\label{ommone}
\ee
where $A_0$ and $A_a$ are some coefficients that may depend one fields and $a=1,2,3$ runs over the directions that use to be a,
generally squashed, $S^3$ that has the $SU(2)$ symmetry w.r.t. which we performed the non-Abelian T-duality.
These coefficients are given in our case by
\be
A_0={1\ov \sqrt{1+\zeta ^2}}\ ,\qq A^a= {\zeta^a\ov \sqrt{1+\zeta ^2}}\ .
\ee

\no
We will denote by $e^a$ the frames "containing" the original $SU(2)$ directions.

\no
In general we have the following decomposition for a $p$-form
 \be
 F_p = G_p^{(0)} + G_{p-1}^{a}\wedge e^a + \ha G_{p-2}^{ab}\wedge e^a \wedge e^b + G_{p-3}^{(3)} \wedge e^1 \wedge e^2 \wedge e^3 \ ,
 \label{fpfg}
\ee
where $ G_{p-1}^{ab} =-G_{p-1}^{ba}$.
Then
\be
\slashed{F}_p = \slashed{G}_p^{(0)} \mathbb{1}  + \slashed{G}_{p-1}^{a} \Gamma^a + \ha  \slashed{G}_{p-2}^{a b}\Gamma^{ab} + \slashed{G}_{p-3}^{(3)} \Gamma^{123} \ ,
\ee
 where $\slashed{G}_q$, $q=p-1,\dots , p-3$ is defined with the Gamma matrices $\G^A$ corresponding to the seven-dimensional spectator spacetime.

\no
If there is more symmetry some of the above forms are actually zero.
For instance, if there is an extra $SU(2)$ symmetry so that the symmetry group is actually $SO(4)$, then
\be
G_{p-1}^a= G_{p-2}^{ab}= 0 \ .
\ee
The following identities are needed
 \ba
&& \G^a \G^{123} = \frac{1}{2} \e^{ab c} \G^{bc}  \ , \qq \G^{ab} \G^{123} = - \e^{abc} \G^c \ , \qq \G^{123}\G^{123} = -1 \ ,
\nonumber\\
&& \G^a \G^b = \delta^{ab} + \G^{ab} \ , \qq \G^{ab} \G^c = \delta^{bc} \G^a - \delta^{ac} \G^b + \e^{abc} \G^{123}  \ ,
\\
&& \Gamma^{123} \G^a=   \frac{1}{2} \e^{ab c} \G^{bc} \ .
 \nonumber
 \ea
 Then we have that
\ba
&&  \slashed{F}_p   \Gamma^{a}   = \slashed{G}_p^{(0)} \G^a   +   \slashed{G}_{p-1}^{a} +  \slashed{G}_{p-1}^b \G^{ba} - \slashed{G}_{p-2}^{ab}  \G^b
+  \ha \slashed{G}_{p-2}^{bc}\epsilon^{abc} \G^{123} +  \frac{1}{2} \slashed{G}_{p-3}^{(3)} \e^{abc} \G^{bc} \ ,
\nonumber\\
 && \slashed{F}_p   \Gamma^{123}   = \slashed{G}_p^{(0)}\Gamma^{123}  + \frac{1}{2} \slashed{G}_{p-1}^a\epsilon^{abc} \Gamma^{bc}
 - \ha \slashed{G}_{p-1}^{a b} \epsilon^{abc} \G^c  -  \slashed{G}_{p-3}^{(3)} \mathbb{1} \ .
\ea
Putting these together, and sorting by form degree, we get
\be
  \slashed{F}_p \cdot   \Om^{-1}  =\tilde \slashed{ F}_{p-3} +  \tilde \slashed{ F}_{p-1} + \tilde  \slashed{ F}_{p+1} + \tilde  \slashed{ F}_{p+3}\ ,
\ee
where
\ba
\tilde  \slashed{ F}_{p-3} & = &  - A_0 \slashed{G}_{p-3}^{(3)}\  ,
\nonumber\\
\tilde  \slashed{ F}_{p-1} & = & A_a \slashed{G}_{p-1}^a -  {A_0\ov 2} \slashed{G}_{p-2}^{a b} \epsilon^{abc} \G^c
 -  A_a  \slashed{G}_{p-2}^{ab}  \G^b +  \frac{A_a}{2} \slashed{G}_{p-3}^{(3)} \e^{abc} \G^{bc}\ ,
 \nonumber \\
\tilde  \slashed{ F}_{p+1} & = & A_a \slashed{G}_p^{(0)} \G^a   +   \frac{A_0}{2} \slashed{G}_{p-1}^a\epsilon^{abc} \Gamma^{bc}
- A_a \slashed{G}_{p-1}^b  \G^{ab}  +  {A_a\ov 2} \slashed{G}_{p-2}^{bc}\epsilon^{abc} \G^{123} \ ,
\\
\tilde  \slashed{ F}_{p+3} & = & A_0 \slashed{G}_p^{(0)}\Gamma^{123}  \ .
\nonumber
\ea
 Then one reads off the expression for the T-dual forms which is similar to that in \eqn{fpfg}, i.e.
 \be
 \widehat F_p = \widehat G_p^{(0)} + \widehat G_{p-1}^{a}\wedge \hat e^a +  \ha \widehat G_{p-2}^{ab}\wedge \hat e^a \wedge \hat e^b
+ \widehat G_{p-3}^{(3)} \wedge \hat e^1 \wedge \hat e^2 \wedge\hat e^3 \ ,
\label{j3hg4or}
 \ee
 where
\ba
\widehat G_p^{(0)} & = &  e^{\Phi-\widehat \Phi} \left( - A_0 G_{p}^{(3)} + A_a G_{p}^a\right)\ ,
\nonumber\\
\widehat G_{p-1}^{a} & = &  e^{\Phi-\widehat \Phi} \left(-{A_0\ov 2} \e^{abc} G^{bc}_{p-1} +  A_b G^{ab}_{p-1} + A_a G^{(0)}_{p-1}\right)\ ,
\nonumber\\
\widehat G_{p-2}^{ab} & = &  e^{\Phi-\widehat \Phi} \left[  \e^{abc}\left(A_c G^{(3)}_{p-2} + A_0 G^{c}_{p-2}\right) -  (A_a G^b_{p-2} - A_b G^a_{p-2})\right]\ ,
\label{j3hg4}
\\
\widehat G_{p-3}^{(3)} & = &  e^{\Phi-\widehat \Phi} \left({A_a\ov 2} \e^{abc} G^{bc}_{p-3} + A_0 G^{(0)}_{p-3}\right)\ .
\nonumber
\ea
We would like to show that the non-Abelian T-duality transformation preserved the degrees of freedom associated to
the fact that not all of the RR $p$-forms are independent, by rather those with rank higher than five are related to
those with lower than five rank, as stated by \eqn{demoll1}. To proceed consider a $p$-form $X_p$ with legs in the seven-dimensional manifold.
Then note the identities
\ba
 \star X_p & = &  \star_7 X_p \wedge e^1 \wedge e^2 \wedge e^3 \ ,
\nonumber\\
 \star (X_p\wedge e^a) & = &  {(-1)^{p+1}\ov 2} \e^{abc} \star_7 X_p  \wedge e^b \wedge e^c\,
\nonumber\\
\star (X_p \wedge e^1 \wedge e^2 \wedge e^3) & = &  (-1)^{p+1} \star_7X_p \ ,
\\
 \star (X_p \wedge e^a \wedge e^b) &  = &  \e^{abc} \star_7 X_p \wedge e^c \ ,
\nonumber
\ea
valid for a Minkowskian seven-dimensional signature spacetime.
Then the condition \eqn{demoll1} that relates higher and lower forms gives the conditions
\ba
G^{(0)}_p & = &  (-1)^{p+[{p\ov 2}]} \star_7 G^{(3)}_{7-p}\ ,
\nonumber\\
G^a_{p-1} & = & {(-1)^{[{p\ov 2}]}} \e^{abc} \star_7 G^{bc}_{8-p}\ ,
\nonumber\\
G^{ab}_{p-2} & = & (-1)^{p+[{p\ov 2}]}  \e^{abc} \star_7 G^c_{9-p}\ ,
\label{cabp}
\\
G^{(3)}_{p-3} & = & (-1)^{[{p\ov 2}]} \star_7 G^{(0)}_{10-p}\ ,
\nonumber
\ea
which, as one may verify are self consistent. In particular, the first and the fourth as well as
the second and third are related by seven-dimensional Hodge duality.
Finally, using the above we can check
that the T-dual forms \eqn{j3hg4} preserve \eqn{cabp}, equivalently \eqn{demoll1}, as well. This is a non-trivial check of various
relevant factors in the T-duality transformations of the RR flux fields.

\section{SUGRA review and conventions}
\subsection{ Brief review of  IIB supergravity}
\ba
S_{\rm IIB} & = & {1\ov 2\kappa^2}\int_{M_{10}} \sqrt{g} \Bigg[e^{-2\Phi}
\left(R + 4 (\del\Phi)^2 -{H^2\ov 12}\right)  -\ha\left(F_1^2 + {F_3^2\ov 3!} +\ha {F_5^2\ov 5!}\right)\Bigg]
\nonumber\\
&&\ - \ha  C_4 \wedge H \wedge dC_2
\ ,
\label{act2}
\ea
where the potentials are
\be
H=dB \ , \quad F_1 = dC_0 \ , \quad F_3 =dC_2 - C_0 H \ , \quad  F_5 = dC_4 - H \wedge C_2 \ .
\ee
The field strength for $F_5$ is self dual (imposed by hand here). The Bianchi identities
\be
dH = 0 \ , \quad  dF_1 =  0 \ , \quad dF_3 = H\wedge F_1 \ ,  \quad dF_5 = H \wedge F_3.
\ee

\subsection{ Brief review of (massive) IIA supergravity}

In the conventions of \cite{Bergshoeff:2001pv}, the action of the massive type-IIA supergravity \cite{Romans:1985tz} is given by
\ba
S_{\rm Massive\ IIA} & = & {1\ov 2\kappa^2}\int_{M_{10}} \sqrt{g} \Bigg[e^{-2\Phi}
\left(R + 4 (\del\Phi)^2 -{H^2\ov 12}\right)  -\ha\left(m^2 + {F_2^2\ov 2} +{F_4^2\ov 4!}\right)
\nonumber\Bigg] \\
&&\ - \ha\left( dC_3 \wedge dC_3 \wedge B +  {m\ov 3} dC_3 \wedge B^3 + {m^2 \ov 20} B^5\right)
\ ,
\label{act22}
\ea
where the field strengths are defined as
\be
H= dB\ ,\qq  F_2 = dC_1 + m B \ ,\qq F_4 = dC_3 - H\wedge C_1 + {m\ov 2}B\wedge B\ ,
\label{hdbf2f4}
\ee
and where $m$ is the mass parameter. Note that, the presence of the $\ha m^2$ term in the action
reveals that $m$ plays the r\^ole of a zero-form $F_0$.
The field strengths are invariant under the gauge transformations
\be
\d B= d\L \ ,\qq \d C_1 = -m \L \ ,\qq \d C_3 = - m \L \wedge B\ ,
\label{gautr}
\ee
where $\L$ is a one-form.
The Bianchi identities are
\be
dH= 0 \ ,\qq dF_2= m H \ ,\qq dF_4 = H\wedge F_2\ .
\label{biiac}
\ee
The topological term in the action can be written as
\be
-\ha \int_{M_{10}}
dC_3 \wedge dC_3 \wedge B + {m\ov 3} dC_3 \wedge B^3 + {m^2 \ov 20} B^5 = -\ha \int_{M_{11}} F_4 \wedge F_4 \wedge H \ ,
\ee
where $\del M_{11}=M_{10}$, so that gauge invariance under \eqn{gautr} becomes manifest.

\no
The equations of motions that follow from varying the metric are
\ba
&& R_{\m\n}+2 D_\m D_\n\Phi -
{1\ov 4} H^2_{\m\n}
=e^{2\Phi}\Bigg[\ha (F_2^2)_{\m\n} + {1\ov 12} (F_4^2)_{\m\n}
\nonumber\\
&& \phantom{xxxxxxxxxxxxxxxxxxxxxxxxx} - {1\ov 4} g_{\m\n}
\left(\ha F_2^2 +{1\ov 24} F_4^2 + m^2\right)\Bigg]\ ,
\label{Einsteq}
\ea
whereas the dilaton equation is
\be
R + 4 D^2 \Phi - 4 (\del \Phi)^2 - {1\ov 12} H^2 = 0 \ .
\label{dilaeq}
\ee
From varying the fluxes we obtain
(after simplifying using Bianchi identities)
\ba
&& d\left(e^{-2\Phi}\star H\right) - F_2\wedge \star F_4 - \ha F_4\wedge F_4 = m \star F_2\ ,
\nonumber\\
&& d \star F_2 + H\wedge \star F_4 =0 \ ,
\label{fluxx}
\\
&&
 d \star F_4 + H\wedge  F_4 =0\ .
\nonumber
\ea
This set of equations is consistent with the Bianchi identities as it can be seen by applying
to each one of them the exterior derivative. In particular, we note the necessity of
the term proportional to $m$ in the right hand side of the first of \eqn{fluxx}.

The Bianchi identities and equations of motion can be recast in the following way:
\ba
0&=& d(F_2 - m B) \ , \nonumber  \\
0 &=& d(F_4 - B\wedge F_2 +\ha m B \wedge B)  \ , \nonumber  \\
0 &=& d( F_6 - B \wedge F_4 + \ha B^2 \wedge F_2 - \frac{1}{6} m B^3)  \ , \nonumber  \\
0 &=& d(F_8 - B\wedge F_6 +  \ha B^2\wedge F_4  - \frac{1}{6} B^3 \wedge F_2 + \frac{1}{24} mB^4)
\ea
in which we defined $F_6 = -\ast F_4$ and $F_8 = \ast F_2$.

\subsection{Supersymmetry}

\no
Our conventions for supersymmetry variations follow those of \cite{Hassan:1999bv}.
To package these variations we find it handy to introduce
a Killing spinor comprising of real Majorana--Weyl spinors
\be
\e = \left( \begin{array}{c} \e_+ \\ \e_- \end{array}\right).
\ee
In type-IIB we have $\G^{11} \e = \e$, while in type-IIA the conventions
are such that $\G^{11} \e_{\pm} = \mp \e_{\pm}$, that is:
\be
\textrm{IIB}: \quad \G_{11} \e = \mathbb{1}_2 \e \ , \quad \textrm{IIA}: \quad\G_{11} \e = - \s_3 \e  \ .
\ee

Using Pauli matrices, the type-IIA Killing spinor equations can be written as
\ba
&& \delta \lambda = \frac{1}{2} \slashed \del \Phi \e - \frac{1}{24} \slashed{H} \s_3 \e
+ \frac{1}{8} e^{\Phi} \biggl[  5 m \s^1 + {3\ov 2} \slashed{F}_2  (i \s^2)
+ {1\ov 24} \slashed{F}_4 \s^1 \biggr] \e\ ,
\nonumber\\
&& \delta \psi_{\mu} = D_{\mu} \e - \frac{1}{8} H_{\mu \nu \rho} \G^{\nu \rho} \s^3 \e
+ \frac{e^{\Phi}}{ 8} \biggl[ m  \s^1
+ {1\ov 2} \slashed{F}_2 (i \s^2) + {1\ov 24} \slashed{F}_{4} \s^1 \biggr]  \G_{\mu} \e\ ,
\label{killIIA}
\ea
where $\displaystyle D_\m \e = \del_\m \e + {1\ov 4} \om_\m^{ab} \G_{ab} \e $.
The Killing spinor of type-IIB are
\ba
&& \delta \lambda = \frac{1}{2} \slashed \del \Phi \e - \frac{1}{24} \slashed{H} \s_3 \e
+ \frac{1}{2} e^{\Phi} \biggl[ \slashed{F}_{1} (i \s^2) + \frac{1}{12}\slashed{F}_3  \s^1 \biggr] \e\ ,
\nonumber
\\
&& \delta \psi_{\mu} = D_{\mu} \e - \frac{1}{8} H_{\mu \nu \rho} \G^{\nu \rho} \s^3 \e
 - \frac{e^{\Phi}}{ 8} \biggl[ \slashed{F}_1 (i \s^2)
+ {1\ov 6} \slashed{F}_3 \s^1 +  \frac{1}{2 \times  5!} \slashed{F}_{5} (i \s^2) \biggr]  \G_{\mu}\e\ ,
\label{killIIB}
\ea
where as always we are using the notation $\displaystyle \slashed{F}_n \equiv  F_{i_1\dots i_n} \G^{i_1 \dots i_n}$.

For IIB it is helpful to combine the MW spinors into a complex $\e= \e_+ + i \e_-$.  For the simplest case where we only have $F_5$ turned on we have a trivial dilation variation and
\be
 \delta \psi_{\mu} = D_{\mu} \e + \frac{i e^\Phi }{2\times 8 \times 5!} F_{\mu_1\dots \mu_5} \G^{\mu_1 \dots \mu_5} \G_\mu \e = 0
\ee

\section{SUSY in the dual of KW }
In this appendix we evaluate explicitly the supersymmetry in the T-dual geometry of Klebanov-Witten.

Let us begin with the type IIB killing spinors of $T^{(1,1)}$ which obey
\be
\G_{12} \eta_1 = - \eta_2 \ , \quad \G_{45} \eta_1 =\eta_2 \ , \quad \G_{11} \eta_i = \eta_i  \ .
\ee
Let us define
\be
\e_1 = \eta_1 \ , \quad \e_2 = \Omega.\eta_2 \ ,
\ee
which have chiralities
\be
\G_{11} \e_1 = \e_1 \ ,  \quad \G_{11} \e_2 = - \e_2  \ ,
\ee
and thus these combine to give an ansatz for the Killing spinors of type IIA.

We begin with the dilatino equation
\ba
&& \delta \lambda_1 = \frac{1}{2} \slashed \del \Phi \e_1 - \frac{1}{24} \slashed{H}  \e_1
+ \frac{1}{8} e^{\Phi} \biggl[   {3\ov 2} \slashed{F}_2
+ {1\ov 24} \slashed{F}_4  \biggr] \e_2\ ,
\ea
Upon inserting the above ansatz we find this vanishes.
To see this first note that
\be
\e_2 = e^\Phi \left( \l \l_1^2 \G_3 \eta_1 - x_2\l \G_3 \eta_2 - x_1 \l_1 \G_1 \eta_1 \right) \ .
\ee
One may then use the projection conditions on the $\eta_i$ under $\G_{45}$ and $\G_{12}$ to simplify when contracting with the fluxes.

One finds
\ba
\frac{1}{2} \slashed{d} \Phi \e_1 &=&  \frac{e^{2 \Phi}}{2}  \left(- x_2 \l^3 \G_3\eta_1 + x_1 \l_1^3 \G_1\eta_2 -  \frac{x_1 x_2}{\l_1}(\l_1^2 - \l^2) \G_2\eta_2    \right) \\
 \frac{3}{8} e^{\Phi} \slashed{F}_2 \e_2 &=& \frac{e^{2\Phi} }{2}3  \left( \l^2 \l_1^4 \G_3 \eta_2 + x_2 \lambda^2 \lambda_1^2 \G_3 \eta_1 - x_1 \l \l_1^3 \G_1\eta_2 \right)  \\
 \frac{1}{8} e^{\Phi} \slashed{F}_4 \e_2 &=& \frac{e^{2\Phi} }{2} \left(( x_1^2 \l_1^2 + x_2^2\l^2 ) \G_3\eta_2 - x_2 \lambda^2 \lambda_1^2 \G_3 \eta_1 +   x_1 \l \l_1^3 \G_1\eta_2  \right)  \\
 -\frac{1}{24} \slashed{H} \e_1& =&  \frac{e^{2\Phi} }{2} \left( -\frac{1}{2} x_1 \l^2 \l_1  \G_1 \eta_2 + \frac{1}{2 \l_1 } x_1 x_2 \l^2 \G_2 \eta_2 +  \l \left( -\ha x_1^2 - x_2^2 + x_2^2 \l^2\l_1^{-2} - \l_1^4  \right)\G_3 \eta_2     \right)  \nonumber \\
 \ea

Combining terms we get
\ba
2 e^{-2\Phi} \delta \l_1 &=& x_2 \l^2 \left(  - \l + 2 \l_1^2\right) \G_3 \eta_1  -\frac{1}{2} x_1 \l_1 (\l^2 -2 \l_1^2 + 4 \l \l_1^2) \G_1\eta_2   +x_1 x_2 \left( \frac{3}{2} \l^2 \l_1^{-1} - \l_1  \right)  \G_2 \eta_2\nonumber  \\
    && \quad +\left( x_1^2 (- \ha \l + \l_1^2)  + x_2^2 \l \l_1^{-2} (\l^2 -\l_1^2 + \l \l_1^2 ) + \l \l_1^4 ( 3 \l - 1) \right) \G_3 \eta_2 \ ,
\ea
which vanishes as claimed when evaluated at the values of $\l^2 = \frac{1}{9}$ and  $\l_1^2 = \frac{1}{6}$.   The calculation of $\delta \l_2 $ proceeds in the same vein with the same result.

One may also show that the gravitino variations also vanish, however we need to also take into account the dependence on the AdS coordinates for the Killing spinors.   The full Killing spinors are then given by:
\ba
&&\e_{1+} = r^\ha \eta_{1+} \ ,  \quad  \e_{2+} = r^\ha \Omega.\eta_{2+} \ , \\
&& \e_{1-} = r^{-\ha} \eta_{1-}  + r^\ha \G_r (y^\mu \G_{y^\mu} )\e_{1 - }   \ , \quad  \e_{2-} = r^{-\ha} \Omega \eta_{2-}  + r^\ha \Omega \G_r (y^\mu \G_{y^\mu} )\e_{2 - }  \nonumber \\
\ea
In which the $\eta_{i \pm}$ are constant Majorana-Weyl spinors obeying
\be
\G_{(11)} \eta_{i \pm}= \eta_{i \pm} \ , \quad   \Gamma_{y^0 y^1 y^2 y^3} \eta_{1 \pm} = \pm \eta_{2 \pm }  \ ,  \quad \G_{12}\eta_{1 \pm} = -\eta_{2\pm} \ , \quad \G_{45}\eta_{1 \pm} = \eta_{2\pm}  \ .
\ee

We conclude that the dual preserves supersymmetry.

  \end{appendix}

\providecommand{\href}[2]{#2}\begingroup\raggedright\endgroup

\end{document}